\documentclass[useAMS, usenatbib, referee]{biom}

\def\bSig\mathbf{\Sigma}

\newcommand{\ba}{\mathbf{a}}

\newcommand{\bG}{\mathbf{G}}

\newcommand{\bX}{\mathbf{X}}

\newcommand{\bY}{\mathbf{Y}}
\newcommand{\bz}{\mathbf{z}}

\newcommand{\bbeta}{\bm{\beta}}
\newcommand{\bgamma}{\bm{\gamma}}
\newcommand{\btheta}{\bm{\theta}}
\newcommand{\blambda}{\bm{\lambda}}

\newcommand{\given}{\,|\,}
\newcommand{\trans}{^\text{T}}

\newcommand{\intd}{\text{d}}

\newcommand{\dnorm}{\text{N}}

\newcommand{\ddirich}{\text{Dirichlet}}

\newcommand{\dhalft}{\text{t}^+}
\newcommand{\preci}{\text{precision}}

\newcommand{\dbeta}{\text{Beta}}

\newcommand{\gwas}{\text{GWAS}}
\newcommand{\rnaseq}{\text{RNA-seq}}

\usepackage{amsmath}
\usepackage{amstext}
\usepackage{graphicx}
\usepackage{amsfonts}%
\usepackage{amssymb}
\usepackage[nooneline, FIGTOPCAP]{subfigure}
\usepackage{rotating}
\usepackage{url}
\usepackage{setspace}
\usepackage[usenames,dvipsnames]{xcolor}
\usepackage{longtable}
\usepackage{lscape}
\usepackage{ulem} 
\usepackage{multicol}

\title[A three-groups model for gene identification]{A Three-groups Non-local Model for Combining Heterogeneous Data Sources to Identify Genes Associated with Parkinson's Disease}

\author{Troy P. Wixson$^{1,*}$\email{twixson@rams.colostate.edu}, 
Benjamin A. Shaby$^{1}$, Daisy L. Philtron$^{2}$, \\ 
\textbf{International Parkinson Disease Genomics Consortium (IPDGC)$^{3}$,} \\
\textbf{Leandro A. Lima$^{4}$, Stacia K. Wyman$^{5}$, Julia A. Kaye$^{4}$, and Steven Finkbeiner$^{4,6,7}$} \\
$^{1}$Department of Statistics, Colorado State University, Fort Collins, Colorado, U.S.A. \\
$^{2}$Department of Applied Mathematics and Statistics, Colorado School of Mines, Golden, Colorado, U.S.A. \\
$^{3}$ https://pdgenetics.org \\
$^{4}$ Center for Systems and Therapeutics, Gladstone Institutes, San Francisco, CA, USA \\
$^{5}$Innovative Genomics Institute, University of California Berkeley, Berkeley, California, U.S.A.\\
$^{6}$ Taube/Koret Center for Neurodegenerative Disease, Gladstone Institutes, San Francisco, CA, USA \\
$^{7}$Department of Neurology and Physiology, University of California San Francisco, \\ San Francisco, California, U.S.A. }

\begin{document}

\date{{\it Received MONTH} YEAR. {\it Revised MONTH} YEAR. {\it Accepted MONTH} YEAR.}

\pagerange{\pageref{firstpage}--\pageref{lastpage}} 
\volume{xx}
\pubyear{2026}
\artmonth{XX}

\doi{xxx}

\label{firstpage}


\begin{abstract}
We seek to identify genes involved in Parkinson's Disease (PD) by combining information across different experiment types. Each experiment, taken individually, may contain too little information to distinguish some important genes from incidental ones. However, when experiments are combined using the proposed statistical framework, additional power emerges. The fundamental building block of the family of statistical models that we propose is a hierarchical three-groups mixture of distributions. Each gene is modeled probabilistically as belonging to either a null group that is unassociated with PD, a deleterious group, or a beneficial group. This three-groups formalism has two key features. By apportioning prior probability of group assignments with a Dirichlet distribution, the resultant posterior group probabilities automatically account for the multiplicity inherent in analyzing many genes simultaneously. By building models for experimental outcomes conditionally on the group labels, any number of data modalities may be combined in a single coherent probability model, allowing information sharing across experiment types. These two features result in parsimonious inference with few false positives, while simultaneously enhancing power to detect signals. Simulations show that our three-groups approach performs at least as well as commonly-used tools for GWAS and RNA-seq, and in some cases it performs better. We apply our proposed approach to publicly-available GWAS and RNA-seq datasets, discovering novel genes that are potential therapeutic targets.
\vspace{\baselineskip}
\end{abstract}

\begin{keywords}
GWAS; Non-local prior; RNA-seq; Variable selection.
\end{keywords}

\maketitle

\section{Introduction}
\label{sec:intro}

A substantial portion of the risk of developing even so-called ``sporadic disease'' is attributable to genetic variants harbored by an individual. For this reason, huge genetic studies have investigated many major human diseases to uncover genetic variants that either directly cause or modify disease, first with the analysis of single nucleotide variants (SNVs), then whole exomes, and now, increasingly, whole genomes. Conventionally, the analysis of genomic data focused on genome-wide association studies (GWAS), which have been used to identify associations in a gene variant and disease incidence or progression to determine if a genetic variant is correlated with a disease phenotype \citep{Uffelmann_2021}.

However, the variants identified from GWAS of complex traits, such as human disease, usually explain only a small fraction of the known heritability, and the effect sizes of the variants themselves are typically small. 
Some of the missing heritability appears to arise from common variants whose effect sizes are too small to detect using GWAS, even when the sample size is large \citep{boyle-2017a}.
Furthermore, the experiments required to functionally validate the role of a specific variant are resource intensive, and the number of potentially important variants that emerge from genetic studies vastly exceeds the capacity and resources available to evaluate them. 

This paper details an approach to begin to meet this need by integrating data sources across multiple experimental types to reliably detect weak genetic signals related to Parkinson's Disease (PD). 
Our approach probabilistically classifies genes as belonging to a null group, a deleterious group, or a beneficial group using joint models of the relationship of disparate data types to disease status. 
In this framework genes in the null group are unassociated with the disease, genes in the beneficial group are associated with either a better outcome or decreased incidence of a negative outcome, and genes in the deleterious group are associated with a worse outcome or increased incidence of a negative outcome. 
Prior probabilities of group assignments is apportioned with a Dirichlet distribution and thus posterior group probabilities automatically account for the multiplicity inherent in analyzing many genes simultaneously \citep{scott-2010a}. 

We consider a pair of experiments in this paper: an RNA-seq study in which biological samples of PD and control patients are assayed for differential expression, and a GWAS study where PD and control populations with no familial relationships are genotyped. 
Each data type has its own separately specified sub-model in which the response is modeled as conditionally independent given the group labels for each gene. 
These group labels are shared across (both) experiments. 
Conditional independence allows additional data types to easily be added when they become available and does not require the data types to be observed on a common set of individuals. 
Any number of data modalities can be combined in a single coherent probability model, as long as each collection of experimental outcomes can be formalized conditionally upon latent group labels, allowing information sharing across experiment types.
Let $\bY_1,\dots, \bY_M$ be response vectors for $M$ experimental data types, $\btheta = (\btheta_1\trans, \ldots, \btheta_M\trans)\trans$ the model parameters corresponding to sub-models for the $M$ experimental types, and $\bG$ the shared group labels, conditional independence allows the full data likelihood to be factorized as
$
\mathcal{L}(\bY_1,\dots,\bY_M \given \bG, \btheta) = \prod_{m=1}^M\mathcal{L}_m(\bY_m \given \bG,\btheta_m).
$

The statistical framework we propose enhances the power to detect weak signals by borrowing strength across different sources of information. This approach differs from the combination of multiple datasets or studies with meta-analyses. Traditional meta-analysis approaches for integrating data sources perform separate analyses for each data type and then combine the resulting summary measures like p-values or odds ratios (see, e.g., \citealp{Zeggini_Ioannidis_2009} for a review of meta-analyses in GWAS studies). Many techniques for $p$-value combination exist, including Fisher's or Stouffer's techniques \citep{fisher2-1929a,stouffer-1949a} as well as more modern approaches \citep{genovese-2004a,benjamini-2008a}. A critical weakness of meta-analysis is that the analysis of each individual dataset is siloed and, therefore, information can only be shared through the summary measures (e.g., p-values) rather than being shared among full datasets. Finally, traditional meta-analysis can suffer from selection bias, inconsistent analytic approaches, etc. \citep{Begum_2012}. In contrast, our proposed approach allows for consistent analytic treatment of each study and shares information across all data types to inform the analysis, increasing power to detect weak signals.

Several previous studies have emphasized the benefits of analyzing different datasets jointly, a strategy sometimes referred to as meta-dimensional methods \citep{ritchie-2015b} or multi-modal analysis \citep[see][for reviews]{richardson-2016a, li-2018a}. However, previous approaches to integrating multiple genomic data types are not directly applicable to our PD analysis, as they require either the same subjects in each experiment type, the same experiment types, or \textit{a priori} grouping of genes into sets \citep[see, e.g.][and references therein]{ding-2022a, holzinger-2013a, Tyekucheva_2011}.

Our hierarchical three-groups structure results in parsimonious inference with few false positives, while simultaneously enhancing power to detect signals. An additional benefit is that, like many hierarchical Bayesian formulations, it accommodates the situation where some genes are not measured in all data types. Genes that are included in some, but not all, experiment types are treated as missing data by iterative sampling from their posterior predictive distributions in the experiments from which they are absent. 
This flexibility is not available, for example, to $p$-value combination approaches which require genes to be measured in all data types.

Our primary scientific interest lies in the posterior probabilities of the group assignments. Genes with high posterior probability of being either beneficial or deleterious will be considered targets for our follow-up experiments, potentially as therapeutic targets.

\section{Model Details}
The proposed three-groups suite of statistical models requires the separate development of a sub-model for each experimental data type, conditional on the collection of genes belonging to null, deleterious, or beneficial groups. 
Each sub-model is tailored to its particular response type, but each is built upon the common three-groups structure, allowing pooling of information through the shared group labels. 
The shared group labels $G_j$ for gene $j \in 1, \ldots J$ are modeled with a Dirichlet-categorical distribution which induces an automatic multiplicity adjustment \citep{scott-2010a} (Section 2 of the Supplementary Materials).
Our complete model, including the RNA-seq sub-model, the GWAS sub-model, and shared components, is written out in the first section of the Supplementary Materials.

\subsection{Three-Groups Model for RNA-seq Data}\label{sec:rnaseq}

RNA-seq expression levels are measured as counts, necessitating statistical approaches that either model the data using discrete distributions such as Poisson or Negative Binomial like \texttt{edgeR} \citep{Robinson_2010} and \texttt{DESeq2} \citep{Love_2014} or normalize the data before applying statistical models for continuous responses like \texttt{limma+voom} \citep{Law_2014}. In either case, the goal of the analysis is to discover genes that are differentially expressed between the treatment (disease) and control (healthy) groups.

Let $Y_{ijk}^{\rnaseq}$ be the count for the $i^{\text{th}}$ replicate of the $j^{\text{th}}$ gene for treatment group $k\in \{0,1\}$. Similarly to \texttt{edgeR}, we model $Y_{ijk}^\rnaseq$ using the Negative Binomial distribution, using a mean and dispersion parametrization:
$
 Y_{ijk}^\rnaseq\sim \text{NegBin}(\mu_{ijk}, \phi_{j}),
$ 
where $\mu_{ijk}$ is the mean count from individual $i$ of gene $j$ from treatment group $k$, and $\phi_j$ is the dispersion parameter for gene $j$. The variance of $Y_{ijk}$ is then $\mu_{ijk}(1+\mu_{ijk}\phi_{j})$. We model the mean counts as
\[
 \log(\mu_{ijk}) = \alpha_j + \log(fc)_j*k + L_i + M_j + (\bX_i^\rnaseq)\trans\bbeta^\rnaseq.
\]
Here $\alpha_j$ is the gene-wise intercept, $(fc)_j$ is the fold change between expected counts in the treatment group compared to the control group, $L_i$ and $M_j$ are the natural logarithm of the library size for sample $i$ and the gene length for gene $j$ respectively (known), and $(\bX_i^\rnaseq)\trans$ is a row-vector of covariates associated with individual $i$ from the RNA-seq study. 
When $k=0$, indicating the control group, the mean function simplifies to $\log(\mu_{ijk}) = \alpha_j + L_i + M_j + (\bX_i^\rnaseq)\trans\bbeta^\rnaseq.$
Expected counts are proportional to both library size and gene length, in both groups, and thus these offsets are directly included in the model. This is an alternative to normalizing the data in a pre-processing step, which is common in standard analysis tools \citep{Robinson_2010,Love_2014,Law_2014}.
  
The fold change $(fc)_j$ of the counts associated with each gene is the focus of the inference, and hence is endowed with the three-groups structure: 
\begin{equation} \label{eqn:fold-change-prior}
\log(fc)_j\sim\begin{cases}
  \delta_0 & \text{if}\ G_j=1 \text{ (Null) }\\
  f^{\rnaseq^+} & \text{if}\ G_j=2\text{ (Deleterious) }\\
  f^{\rnaseq^-} & \text{if}\ G_j=3\text{ (Beneficial)}\\
\end{cases} 
\end{equation}
where $\delta_0$ is a point mass at zero, $f^{\rnaseq^+}$ is a distribution over the positive half-line, and $f^{\rnaseq^-}$ is a distribution over the negative half-line (see Section \ref{sec:piMOM_discussion}). 

We induce shrinkage on the gene-wise dispersion using a shared random effect scheme, with $\log(\phi_j) \sim \dnorm(\mu_{0}, \preci = \tau_0)$, $\mu_0 \sim \dnorm(0, \preci = 10^{-2})$, and $\tau_0 \sim \dhalft(\nu = 4)$, 
where $\dhalft$ refers to a $t$-distribution truncated to the positive half-line as in \citet{Gelman_2006}.

To complete the model, we assign the vague priors $\alpha_j \sim \dnorm(0, \preci = 10^{-3})$ and $\beta^\rnaseq_q \sim \dnorm(0, \preci = 10^{-3})$, independently for $j = 1, \ldots, J$ and $q = 1, \ldots, Q^\rnaseq$, where $Q^\rnaseq$ is the number of covariates considered.

\subsection{Three-Groups Model for GWAS Data with Binary Outcomes}\label{sec:gwas}

GWAS entails collecting genotypic (i.e. SNVs) and phenotypic data (disease status) from unrelated individuals with the goal of associating specific SNVs with disease status. In our model formulation for the RNA-seq data type, the unit of measurement is the gene; however, GWAS data are collected on the SNV level. Thus, integrating GWAS with the other models requires that SNVs be either summarized into their associated genes or identified as non-coding variants with no obvious parallel in our RNA-seq dataset. The need to aggregate SNV information to the gene level is not unique to our formulation and is also necessary, for example, for burden tests for association \citep{asimit-2012a,morgenthaler-2007a,li-2008a,madsen-2009a}. 
How best to formalize the association between SNVs and genes is an open question (see, e.g., \citealp{gazal2022combining} and references therein). The simplest approach, which we take here, is to collapse SNVs into genes in a binary fashion: if a gene contains a SNV in or near its coding region, it is assigned a value of 1, if it does not contain a SNV, it is assigned a value of 0. This binary collapse assumes a dominant model of disease (i.e., having one or more copies of the associated minor allele alters the risk). A simple alternative strategy is to sum the number of minor alleles in or near each gene's coding region. 
We report results from this alternative sum-mapping in Section 6 of the Supplementary Material.
Extensions include using weighted sums, and schemes allowing for different SNVs in a single region to act in opposite ways on the outcome.
Our model is agnostic with respect to the particular SNV-to-gene mapping, as different mappings only require corresponding changes the design matrix $\boldsymbol{X}_i^{GWAS}$ and thus the model itself is immediately applicable without modification.
 
Our three-groups model for GWAS is logistic regression, as the response is binary (1 if individual $i$ in the GWAS study has PD and 0 otherwise). The response $Y_i^\gwas$ is modeled as a Bernoulli random variable with probability $p_i$ of being in the PD group, with
\[
\text{logit}(p_i) = \bz_i\trans \bgamma + (\bX_i^\gwas)\trans \bbeta^\gwas,
\] 
where $\bgamma = (\gamma_1, \ldots, \gamma_J)\trans$ are gene effects and $\bbeta^{\gwas}$ is a vector of individual-level covariate effects. 
Here $\bz_i = (z_{i1}, \ldots, z_{iJ})\trans$ is a binary vector that indicates whether there was a SNV in genes $j= 1, \ldots, J$ for individual $i$ and $\bX_i^\gwas$ is the vector of covariates from individual $i$.
The gene effects are the focus of the inference, and thus the $\gamma_j$s, $j=1, \ldots, J$, are endowed with a three-groups prior structure analogous to \eqref{eqn:fold-change-prior} from Section \ref{sec:rnaseq}: 
\[
\gamma_j\sim\begin{cases}
  \delta_0 & \text{if}\ G_j=1 \text{ (Null) }\\
  f^{\gwas^+} & \text{if}\ G_j=2\text{ (Deleterious) }\\
  f^{\gwas^-} & \text{if}\ G_j=3\text{ (Beneficial)}.\\
\end{cases} 
\] 
where $\delta_0$, $f^{\gwas^+}$, and $f^{\gwas^-}$ are defined similarly to $\delta_0$, $f^{\rnaseq^+}$, and $f^{\rnaseq^-}$ in section \ref{sec:rnaseq}. The vague prior $\beta^\gwas_q \sim \dnorm(0, \preci = 10^{-3})$ for the individual level covariates $q = 1, \ldots, Q^\gwas$ completes the model.

\subsection{Gene Effect Priors}
\label{sec:piMOM_discussion}
Secondary scientific interest, after group assignment probabilities, lies in the magnitude of gene effects. 
These gene effects inform the group assignments and contain information regarding the potential benefit of clinical interventions. The three-groups framework allows for flexible modeling of these effect sizes through selection priors which may be asymmetric. That is, a point mass ($\delta_0$) at zero represents the null effect, and the distribution of beneficial effect sizes ($f^{(m)^-}$) may differ from that of deleterious effect sizes ($f^{(m)^+}$). This added flexibility above standard, symmetric, selection priors reflects the biological reality that genes with protective effects may behave very differently than genes with damaging effects.

Discontinuous priors which include null and non-null components (frequently termed ``spike and slab" priors) can be broadly categorized as \textit{local} and \textit{non-local} \citep{Johnson_2010}. 
A prior is said to be local if the density for its non-null component is non-zero in a neighborhood of the null value (i.e., very small effect sizes have prior mass). 
Conversely, a non-local prior has density values in its non-null component of exactly zero in a neighborhood of the null value. 
Intuitively, the benefit of non-local priors is that effects that might otherwise be estimated to be trivially small get pushed into the null group because very small effects have zero prior probability. 
We consider a non-local prior and and compare the results to local three-groups model which has half-normal priors for the non-null gene effects.

The non-local prior that we consider is a modification of the product inverse moment (piMOM) prior from \citet{Johnson_2010}, defined by the density $f(\bbeta | \tau, r) = [\tau^{r/2} / \Gamma(r/2)]^J$ $\prod_{j=1}^J |\beta_j|^{-(r+1)} \exp{(-\tau / \beta_j^2)}$. This is a two-parameter family, with $r$ controlling the tail decay (smaller $r$ gives heavier tails) and $\tau$ controlling the scale. To allow for asymmetry in the non-null gene effects, we truncate the density and use the positive component that has support on $\mathbb{R}^+$ for deleterious gene effects, and separately use a negative component that has support on $\mathbb{R}^-$ for the beneficial gene effects. Separate half-piMOM hyper-priors are placed on each of the $\tau$ parameters of these truncated densities. This allows us to consider the separate posterior distributions of beneficial and deleterious gene effects. Together with a point mass at zero, we thus arrive at a three-component mixture for the null, beneficial, and deleterious gene effects. Previous work on variable selection tasks using non-local priors \citep[e.g]{Johnson_2010, Johnson_2012,Nikooienejad_2016,LiWeibing_2022} found improved performance when using non-local priors relative to standard local selection priors. \citet{Johnson_2012} demonstrated that Bayesian model selection procedures based on the piMOM prior density with $r\geq 2$ results in consistent estimation of the true model. Here, we follow their recommendation and fix the tail decay rate at $r = 2$.

\section{Computation} 

The analysis was performed using reversible jump Markov chain Monte Carlo (RJMCMC) \citep{green-1995a} which improves runtime by accounting for the change of dimensionality when group assignment changes. A simpler conventional sampler without RJMCMC is possible but suffers from poor mixing because it requires needlessly sampling from gene effects that will just be multiplied by zero (as most genes are null).  RJMCMC avoids much of this wasted effort and is appreciably faster in our NIMBLE implementation \citep{nimble_2017_article, nimble_2023_package}.
Details on the RJMCMC are in Section 3 of the Supplementary Materials and code is available at \url{https://github.com/twixson/three_groups_simulations}.

\section{Simulations}
\label{sec:simulations}
In this section, we explore the behavior of our three-groups framework with asymmetric non-local gene effects relative to a local three-groups model with symmetric gene effects and standard analysis pipelines. Although we tried to replicate key features of real genomic data in our simulated datasets, the simulated data are illustrative only; thus, the total number of simulated genes is small ($J=250$) to allow for uncluttered visualizations. In each scenario, we assigned five genes to have beneficial effects and five genes to have deleterious effects. 
Each of the 300 simulated datasets includes a GWAS dataset and an RNA-seq dataset, which are independent conditional on the shared gene labels. To reflect the sample sizes in our actual datasets, we simulated 1,000 individuals for GWAS and 100 individuals for RNA-seq. Section 4 of the Supplementary Materials contains a full description of the simulated data.

\begin{figure}
  \centering
  \includegraphics[width=0.6\textwidth, clip = TRUE, trim = 5 20 5 5 ]{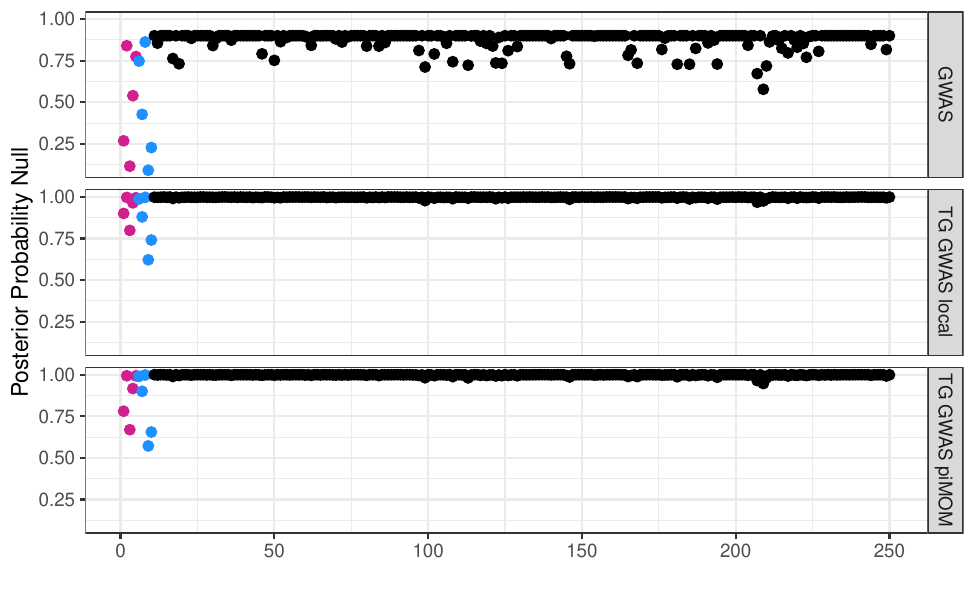}
  \caption{Posterior probability of inclusion in the null group for GWAS only methods, from a single simulated dataset. The top plot shows the individual logistic regression results, the middle plot shows our three-groups GWAS model with local priors on gene effects, and the bottom plot shows our three-groups GWAS model with piMOM priors on gene effects. Genes in blue were simulated to be deleterious, genes in purple were beneficial, and genes in black were null.}
  \label{fig:GWAS_one_run}
\end{figure}

\begin{figure}
\centering
\subfigure[RNA-seq only]{\label{subfig:RNA_one_run}\includegraphics[width=0.6\textwidth, clip = TRUE, trim = 5 20 5 5 ]{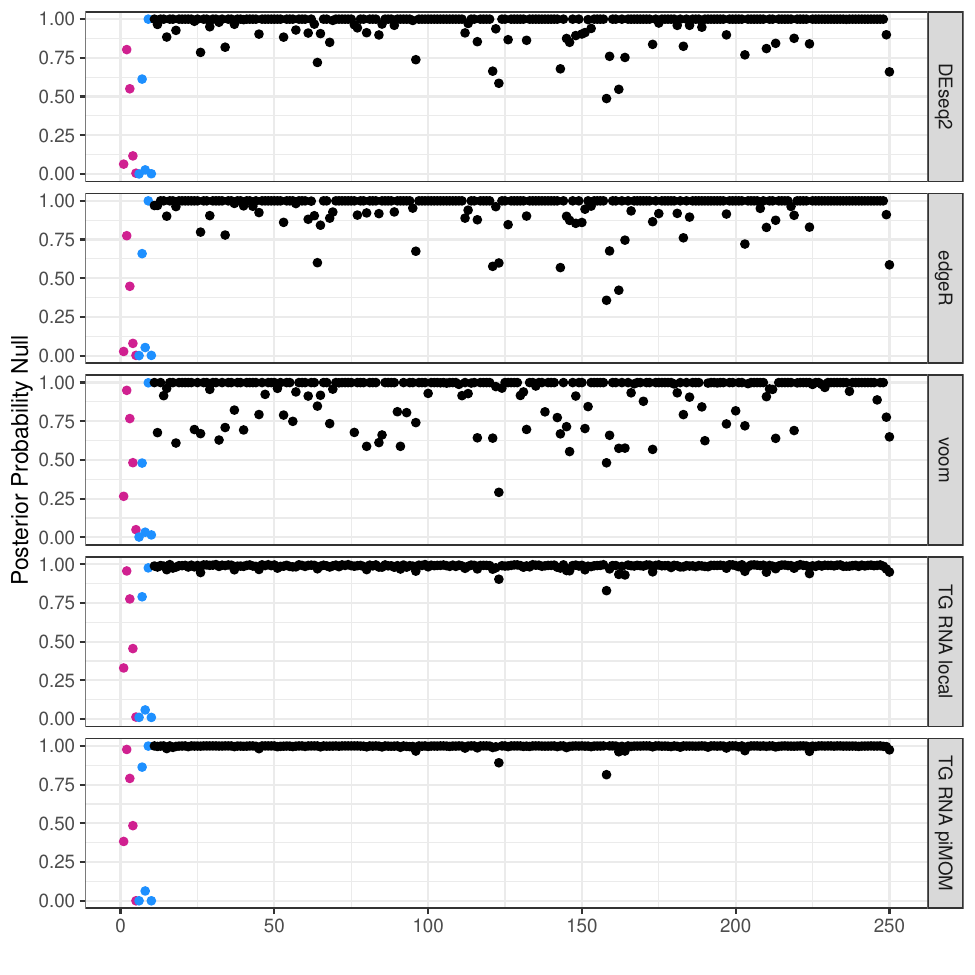}}
\subfigure[Combined models]{\label{subfig:Combined_one_run}\includegraphics[width=0.6\textwidth, clip = TRUE, trim = 5 20 5 5 ]{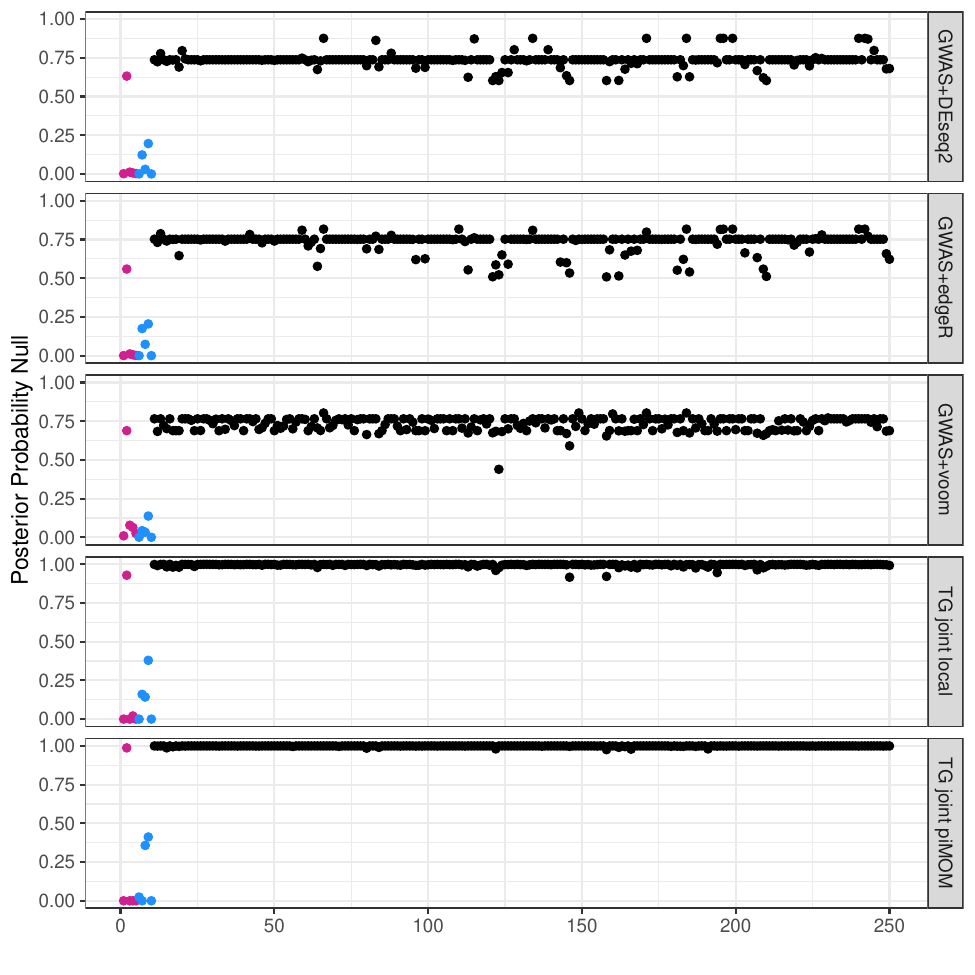}}
\caption{Panel \subref{subfig:RNA_one_run} shows posterior probabilities of inclusion in the null group for RNA-seq only methods, from a single simulated dataset. Panel \subref{subfig:Combined_one_run} shows the same, but for joint GWAS and RNA-seq methods. The top three plots in each panel are results from standard analysis tools, and the bottom two plots in each panel are results from our local and non-local three-groups models. The color scheme is the same as in Figure \ref{fig:GWAS_one_run}.}
\label{fig:one_run}
\end{figure}

We use each of the 300 simulated sets of joint GWAS and RNA-seq data to compare  conventional methods, a local symmetric three groups model, and a non-local asymmetric model for GWAS alone (Figure \ref{fig:GWAS_one_run}), RNA-seq alone (Figure~\ref{fig:one_run}\subref{subfig:RNA_one_run}), and joint models (Figure~\ref{fig:one_run}\subref{subfig:Combined_one_run}). 
In the joint three-groups models, the GWAS and RNA-seq models are combined into a single hierarchical model as described in the supplement. The conventional methods are combined using Fisher's $p$-value combination method \citep{fisher2-1929a} (the Cauchy combination method of \citet{liu2020cauchy} is explored in Section 5.2 of the Supplementary Materials). We chose competitor models for comparison based on what we take to be the most common analysis pipelines in the literature. Specifically, we used individual logistic regression (i.e. one logistic regression per gene, independently across genes) for GWAS; \texttt{DESeq2}, \texttt{edgeR}, and \texttt{limma+voom} for RNA-seq. Competitors methods were run with their default settings (including normalization) from the standard packages on Bioconductor. 

We control for multiple comparisons in the conventional methods using the local false discovery rate (lFDR) \citep{Efron_2001}. 
We chose lFDR over similar adjustment methods because of the comparable interpretation to posterior probabilities that result from Bayesian selection procedures. Specifically, the lFDR can be interpreted as the probability of a gene belonging to the null group conditional on the value of its test statistic.  
Additional comments concerning lFDR (including the odd clustering around 0.75 in Figure~\ref{fig:one_run}\subref{subfig:Combined_one_run}) are given in Section 5.3 of the Supplementary Materials.

As an illustration, Figures \ref{fig:GWAS_one_run} and \ref{fig:one_run} show the posterior probability of inclusion in the null group for each gene, from a single simulated dataset. Each dot represents a gene, and its height represents the posterior probability of being in the null group. The colored dots are true non-null genes, and the black dots are true null genes. The three-groups model further generates posterior probabilities of being in the beneficial and deleterious groups (not shown). 
These figures give some idea of how well each of the methods separates null from non-null genes. 
We investigate this more formally by considering performance across the 300 simulated joint datasets. Boxplots of run times for individual simulations are given in Section 5.8 of the Supplementary Materials.

\begin{figure}
\centering
\subfigure[Logarithmic Score (lower is better).]{\label{subfig:logscore}\includegraphics[width=0.45\textwidth, clip = TRUE, trim = 5 15 5 5 ]{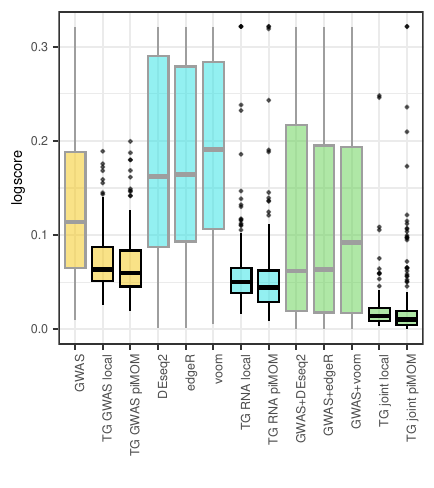}}
\subfigure[Brier Score (lower is better).]{\label{subfig:Brierscore}\includegraphics[width=0.45\textwidth, clip = TRUE, trim = 5 15 5 5 ]{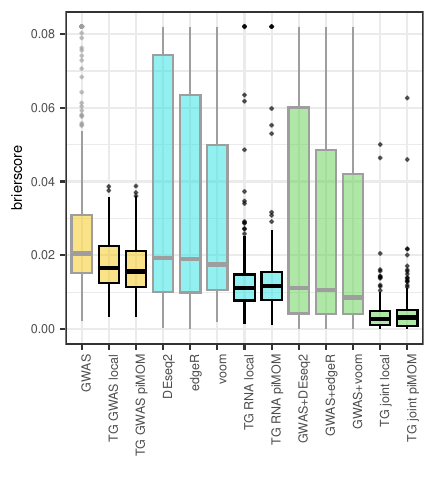}}
\subfigure[Area Under ROC Curve (higher is better).]{\label{subfig:AUC}\includegraphics[width=0.45\textwidth, clip = TRUE, trim = 5 15 5 5 ]{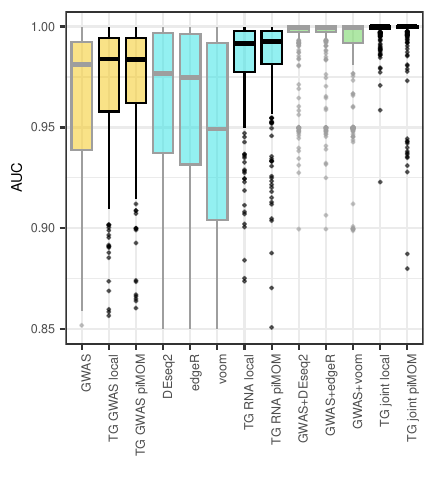}}
\subfigure[True positive rate at cutoffs (higher is better).]{\label{subfig:tpr}\includegraphics[width=0.45\textwidth, clip = TRUE, trim = 5 15 5 5 ]{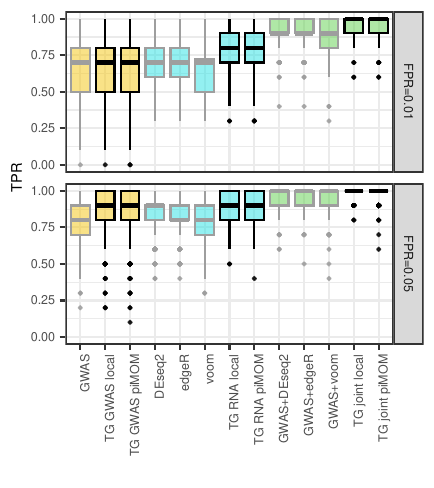}}
\caption{Boxplots of logarithmic scores \subref{subfig:logscore}, Brier scores \subref{subfig:Brierscore}, area under receiver operating characteristic curve \subref{subfig:AUC} computed from posterior probability of inclusion in the null group, and the true positive rates (i.e. power) computed at classification cutoffs that result in mean false positive rates of $0.01$ (top panel) and $0.05$ (bottom panel) in \subref{subfig:tpr} for each of 13 different models, based on 300 simulated datasets. Yellow indicates GWAS only models, blue indicates RNA-seq only models, and green indicates joint models. Boxplots for our three-groups models have black lines, while competitors have grey lines. Panels \subref{subfig:logscore}, \subref{subfig:Brierscore}, and \subref{subfig:AUC} show Winsorized boxplots to focus on the bulk of the distribution of points.}
\label{fig:proper_scoring}
\end{figure}

We assess the simulations using a variety of performance metrics. Each metric is computed on each model for each simulated dataset, and then the models are compared using boxplots of those metrics (further discussion of the metrics is in Section 5.1 of the Supplementary Materials). This allows us to consider the performance of each model across the many simulated datasets, which aids in understanding variability. 
First, we consider two proper scoring rules with attractive properties; the log score and Brier score \citep{Gneiting_2007}, oriented so that smaller values correspond to better performance (Figure \ref{fig:proper_scoring}). Both reward correct classifications more when they are made with higher confidence, and conversely penalize incorrect classifications more when they are made with higher confidence.

The results in Figure \ref{fig:proper_scoring} suggest that the three-groups versions of GWAS and RNA-seq, taken separately, are at least as good as standard analysis tools, and in some cases much better. Particularly striking is how much better the three-groups RNA-seq model performs in terms of log score, relative to \texttt{DESeq2}, \texttt{edgeR}, and \texttt{limma+voom}. This, again, is obtained without any normalizing schemes or elaborate tuning in the three-groups model. 

To assess performance of models at differing cutoffs, we compare the area under the receiver operating characteristic curve (Figure \ref{fig:proper_scoring}(c)), a popular metric in the classification literature. In addition, we show the true positive rate (i.e. power) computed at a cutoff for each model set such that each resulted in a fixed average false positive rate (Figure \ref{fig:proper_scoring}(d)). The joint models are superior to individual models in each metric and our three-groups family of models correctly identifies at least as many non-null genes as competitors while including very few false positives. 

These simulations (and others) demonstrate the value in the joint structure of our model (even when most of the signal comes from one data type). We investigate the performance of our model under scenarios with missing data, signal in one but not both data modalities, hyperprior sensitivity, and so forth with additional simulations in Sections 5 of the Supplementary Materials. Finally, a recurring theme in our investigation was the sparsity imposed by the three-groups family of models; Figures \ref{fig:GWAS_one_run} and \ref{fig:one_run} demonstrate that the three-groups models have a strong preference for fewer non-null genes, as explained in Section 2 of the Supplementary Materials, even with a uniform (i.e. non-informative) prior on the unit simplex for the group inclusion probabilities.

\section{Parkinson's Disease Analysis}
\label{sec:PD_analysis}

\subsection{Description of the Data} 

The data that we used used for the GWAS branch of this study came from the International Parkinson's Disease Genomics Consortium (IPDGC) NeuroX Dataset \citep{Nalls_2014}. The RNA-seq data that we used came from the Parkinson's Progression Markers Initiative (PPMI), obtained from PPMI upon request. Due to the intense computational burden of running the full MCMC, we analyzed a subset the full genome (1734 genes in the GWAS branch and 1697 in the RNA-seq branch). Details on the data and subsetting can be found in Section 7.1 of the Supplementary Materials.

\subsection{Description of the Analysis}

We analyzed the aforementioned data with the full joint three-groups model as well as with the individual sub-models separately with both symmetric local and asymmetric non-local gene effect priors. These six models allow for comparisons between our joint and individual models, as well as comparisons between the local and non-local versions of the models. In each case, we ran the MCMC for 20,000 iterations and threw out the first 10,000 as burn-in, as indicated by trace plots. 

\subsection{Results}
One of the benefits of the three-groups family of models is the sparsity it induces due to the built-in multiplicity adjustment, in addition to the borrowing of strength across both genes and sub-models. The induced sparsity is evident in this analysis: more than 1,650 of the 1,734 genes in the full analysis have posterior probability of being null which is greater than 0.99. We expect that, operationally, the $x$ genes with the smallest posterior probability of being null will be investigated further in mouse models or human iPSC cell models by knocking down the genes using siRNA or CRIPSRi/a etc., where $x$ is determined by available resources such as budget and personnel time. Here, we report ``interesting'' genes by setting $x = 20$. 
An alternative would be to classify genes according to their posterior probabilities, but this necessitates the somewhat arbitrary selection of a cutoff value. 
Instead, we report the 20 most interesting genes alongside the posterior probabilities and gene effect sizes (Table \ref{tab:results_table1}).

\begin{table}
\centering
\footnotesize{
\caption{Posterior probabilities and effect sizes for the 20 genes with the smallest posterior probability of inclusion in the null group in the joint three-groups models. The ``$+$'' symbols indicate genes that are in the top 20 most interesting genes in the respective model.
The ``--'' symbols indicate genes that are in the top 20 even though their posterior probability of inclusion in the null group is greater than 0.5. The single gene with a ``$\circ$'' is not in the top 20 list for the piMOM model but it has a posterior probability of inclusion in the null group that is less than 0.5. $P_\text{null}$, $P_\text{del.}$, and $P_\text{ben.}$ are the proportion of the 10,000 mcmc iterations that the gene was in the null, deleterious, and beneficial group respectively. Effect sizes are computed conditionally (i.e., the mean of the effect size when the gene was non-null). GWAS effect sizes are odds ratios; a value less than 1 represents a protective effect and a value greater than 1 represents a damaging effect. RNAseq effect sizes are fold changes and are interpreted similarly. Gene effects with ``*'' indicate that the gene was null $>99\%$ of the MCMC iterations in that model.}
\scalebox{0.85}{
\begin{tabular}{rl|lllll|rrrrrr|rrrrrr}
 & Gene & \begin{sideways} local \end{sideways} & \begin{sideways} piMOM \end{sideways} & \begin{sideways} GWAS + \texttt{edger} \end{sideways} & \begin{sideways} GWAS + \texttt{limma} \end{sideways} & \begin{sideways} GWAS + \texttt{DESeq2} \end{sideways} & \begin{sideways} $P_\text{null}$ local \end{sideways} & \begin{sideways} $P_\text{del.}$ local \end{sideways} & \begin{sideways} $P_\text{ben.}$ local \end{sideways} & \begin{sideways} $P_\text{null}$ piMOM \end{sideways} & \begin{sideways} $P_\text{del.}$ piMOM \end{sideways} & \begin{sideways} $P_\text{ben.}$ piMOM \end{sideways} & \begin{sideways} GWAS effect local \end{sideways} & \begin{sideways} RNA effect local \end{sideways} & \begin{sideways} Disp. local \end{sideways} & \begin{sideways} GWAS effect piMOM \end{sideways} & \begin{sideways} RNA effect piMOM \end{sideways} & \begin{sideways} Disp. piMOM \end{sideways} \\ 
 \hline
  1  & CHCHD6    & +  & +       & + & + & + & 0.00 & 0.00 & 1.00 & 0.00 & 0.00 & 1.00 & 0.81 & 0.90 & 0.16  & 0.80   & 0.86 & 0.16 \\ 
  2  & DUSP1     & +  & +       & + & + & + & 0.00 & 1.00 & 0.00 & 0.00 & 1.00 & 0.00 & 1.12 & 1.19 & 0.09  & 1.24   & 1.19 & 0.09 \\ 
  3  & SYTL3     & +  &         & + & + & + & 0.16 & 0.76 & 0.08 & 1.00 & 0.00 & 0.00 & 1.14 & 1.11 & 0.08  & *      & *    & 0.08 \\ 
  4  & CDIP1     & +  & $\circ$ & + & + & + & 0.26 & 0.09 & 0.65 & 0.49 & 0.13 & 0.38 & 0.36 & 0.93 & 0.14  & 0.41   & 0.87 & 0.14 \\ 
  5  & FGD4      & -- &         & + & + & + & 0.61 & 0.27 & 0.12 & 1.00 & 0.00 & 0.00 & 1.02 & 1.13 & 0.06  & *      & *    & 0.06 \\ 
  6  & TLR6      & -- &         & + & + & + & 0.90 & 0.05 & 0.05 & 1.00 & 0.00 & 0.00 & 1.16 & 1.10 & 0.05  & *      & *    & 0.05 \\ 
  7  & CNTNAP2   & +  & +       & + &   &   & 0.00 & 0.00 & 1.00 & 0.00 & 0.00 & 1.00 & 0.67 & 0.48 & 1.17  & 0.71   & 0.51 & 1.16 \\ 
  8  & IFRD1     & +  &         &   & + & + & 0.27 & 0.62 & 0.11 & 1.00 & 0.00 & 0.00 & 1.03 & 1.13 & 0.07  & *      & *    & 0.07 \\ 
  9  & RAP1GAP   & -- &         & + &   & + & 0.85 & 0.06 & 0.09 & 1.00 & 0.00 & 0.00 & 0.96 & 0.59 & 1.54  & *      & *    & 1.53 \\ 
  10 & FAM49B    & +  & +       &   &   &   & 0.00 & 1.00 & 0.00 & 0.00 & 1.00 & 0.00 & 2.19 & 1.06 & 0.02  & 8.67   & 1.09 & 0.02 \\ 
  11 & DPP10     & -- & +       &   &   &   & 0.58 & 0.29 & 0.13 & 0.00 & 1.00 & 0.00 & 2.20 & 1.15 & 3.08  & 2.56   & 1.23 & 3.06 \\ 
  12 & FCGR2A    &    & +       &   & + &   & 1.00 & 0.00 & 0.00 & 0.00 & 1.00 & 0.00 & *    & *    & 0.04  & 1.28   & 1.13 & 0.07 \\ 
  13 & CD180     & -- &         &   &   &   & 0.85 & 0.06 & 0.08 & 1.00 & 0.00 & 0.00 & 0.88 & 0.84 & 0.17  & *      & *    & 0.17 \\ 
  14 & PLA2G16   & -- &         &   &   &   & 0.87 & 0.05 & 0.07 & 0.83 & 0.07 & 0.10 & 0.56 & 0.68 & 25.24 & 0.64   & 0.73 & 1.06 \\ 
  15 & LINC00969 & -- &         &   &   &   & 0.89 & 0.05 & 0.06 & 0.88 & 0.06 & 0.06 & 1.62 & 1.67 & 25.31 & 1.38   & 1.14 & 1.08 \\ 
  16 & TMEM55A   & -- &         &   &   &   & 0.89 & 0.05 & 0.06 & 0.87 & 0.06 & 0.06 & 1.65 & 1.73 & 25.06 & 106.49 & 1.07 & 1.08 \\ 
  17 & MRPS21    & -- &         &   &   &   & 0.89 & 0.05 & 0.06 & 0.87 & 0.06 & 0.06 & 1.67 & 1.85 & 25.03 & 1.50   & 1.12 & 1.01 \\ 
  18 & C19orf70  & -- &         &   &   &   & 0.89 & 0.05 & 0.06 & 0.87 & 0.07 & 0.06 & 1.54 & 1.63 & 25.36 & 1.45   & 1.13 & 0.99 \\ 
  19 & C19orf66  & -- &         &   &   &   & 0.90 & 0.05 & 0.05 & 0.87 & 0.07 & 0.06 & 2.41 & 2.74 & 25.17 & 1.91   & 1.42 & 1.12 \\ 
  20 & C16orf52  & -- &         &   &   &   & 0.90 & 0.05 & 0.05 & 0.88 & 0.06 & 0.06 & 1.61 & 1.65 & 24.94 & 1.46   & 1.18 & 1.01 \\ 
  21 & ATP5J     & -- &         &   &   &   & 0.90 & 0.05 & 0.05 & 0.88 & 0.06 & 0.06 & 1.03 & 1.28 & 25.05 & 1.07   & 1.04 & 1.04 \\ 
  22 & CD82      &    & +       &   &   &   & 1.00 & 0.00 & 0.00 & 0.00 & 0.00 & 1.00 & 1.00 & 1.05 & 0.05  & 0.75   & 0.91 & 0.04 \\ 
  23 & CNTNAP4   &    & +       &   &   &   & 1.00 & 0.00 & 0.00 & 0.00 & 0.00 & 1.00 & 0.96 & 0.63 & 3.63  & 0.67   & 0.77 & 4.21 \\ 
  24 & CXCR4     &    & +       &   &   &   & 1.00 & 0.00 & 0.00 & 0.00 & 1.00 & 0.00 & 0.94 & 0.86 & 1.30  & 1.55   & 1.10 & 0.05 \\ 
  25 & DLGAP1    &    & +       &   &   &   & 1.00 & 0.00 & 0.00 & 0.00 & 1.00 & 0.00 & 0.99 & 0.80 & 2.89  & 1.49   & 1.19 & 1.53 \\ 
  26 & EFCAB6    &    & +       &   &   &   & 1.00 & 0.00 & 0.00 & 0.00 & 0.00 & 1.00 & *    & *    & 0.07  & 0.76   & 0.79 & 1.29 \\ 
  27 & FIGN      &    & +       &   &   &   & 1.00 & 0.00 & 0.00 & 0.00 & 1.00 & 0.00 & *    & *    & 4.26  & 1.51   & 1.18 & 1.83 \\ 
  28 & FRAS1     &    & +       &   &   &   & 1.00 & 0.00 & 0.00 & 0.00 & 1.00 & 0.00 & *    & *    & 1.53  & 1.59   & 1.21 & 2.89 \\ 
  29 & JARID2    &    & +       &   &   &   & 1.00 & 0.00 & 0.00 & 0.00 & 1.00 & 0.00 & *    & *    & 0.07  & 1.42   & 1.10 & 0.06 \\ 
  30 & LRFN5     &    & +       &   &   &   & 1.00 & 0.00 & 0.00 & 0.00 & 0.00 & 1.00 & *    & *    & 1.84  & 0.67   & 0.77 & 4.46 \\ 
  31 & PTPRN2    &    & +       &   &   &   & 1.00 & 0.00 & 0.00 & 0.00 & 1.00 & 0.00 & *    & *    & 0.06  & 1.67   & 1.16 & 0.19 \\ 
  32 & RIN3      &    & +       &   &   &   & 1.00 & 0.00 & 0.00 & 0.00 & 0.00 & 1.00 & *    & *    & 4.52  & 0.71   & 0.86 & 0.26 \\ 
  33 & VRK2      &    & +       &   &   &   & 1.00 & 0.00 & 0.00 & 0.00 & 0.00 & 1.00 & *    & *    & 0.20  & 0.75   & 0.90 & 0.09 \\ 
  34 & ATP8B4    &    & +       &   &   &   & 1.00 & 0.00 & 0.00 & 0.44 & 0.13 & 0.44 & *    & *    & 0.26  & 0.74   & 0.91 & 0.07 \\ 
  35 & C10orf90  &    & +       &   &   &   & 1.00 & 0.00 & 0.00 & 0.44 & 0.44 & 0.12 & *    & *    & 0.09  & 1.28   & 1.22 & 3.59 \\ 
  \hline
\end{tabular}
}
\label{tab:results_table1}
}
\end{table}

Our model identifies beneficial genes as well as deleterious genes (e.g., \textit{CHCHD6} and \textit{DUSP1} have posterior probability 1 of inclusion in the beneficial and deleterious groups, respectively, in both the local and non-local models). 
Five of the 20 most interesting genes from the local model are included in the 20 most interesting genes in the non-local model, and ten of the 35 genes identified by our three-groups models are in at least one of the lists of interesting genes from the $p$-value combinations of conventional methods.

Some identified genes have conflicting evidence between the two models.
Some genes (e.g., \textit{IFRD1} and \textit{SYTL3}) have very small estimated gene effects in the local model (Table \ref{tab:results_table1}). These genes are not identified by the non-local model because very small effects are shrunk to the null value by design. 
There are other genes which do not appear to fit this pattern, and we conjecture that the differences may be attributable to interactions among genes estimated as non-null, which alter the likelihood values. 
Finally, some genes are harder to categorize.
For example, \textit{CDIP1} would be classified in the beneficial group in the local model (posterior inclusion probability of 0.65) while the non-local model has posterior probabilities of 0.49, 0.13, and 0.38 for the null, deleterious, and beneficial groups, respectively.
The differences between models are not surprising when considering that the standard methods also report gene lists that are not consistent.

Our three-groups model identified several genes which have known links to PD in the literature. Several of the other identified genes are linked to pathways which have been implicated in PD, though we are are unaware of previous work
that directly links these genes to PD. Details of connections of the genes we identify to the PD literature are explored in Section 7 of the Supplement. Additional results including trace plots (Section 8), volcano plots (Section 9), and lists of genes indicated as interesting by the separate three-groups models and conventional models (Sections 10 and 11) are in the Supplementary Material.

\section{Discussion}
Structuring models for data conditional on the three-groups framework has advantages of modular incorporation of multiple experimental data types and automatic multiplicity adjustment. This increases power for improved prioritization of genes that show evidence of involvement in disease by combining information across genomic, transcriptomic, and potentially other data types. Furthermore, with the three-groups model as a platform, in the future we can combine functional data from future cell-based studies and screens that directly assess the impact of a gene on a cell's biological outcome, such as health, morphology, stress response, and proteostasis. In this way, we can harness the power of cell biology together with human genetics and gene expression. Additionally, the Bayesian formulation and use of MCMC allows for inclusion of genes that have observations in some but not all data types.

One challenge that we faced was the computational burden of running MCMC on this model; each run on the PD dataset used around 30Gb of RAM and took several days to finish. This is not a huge problem in the context of a years-long collaboration like ours, but it is inconvenient. \texttt{NIMBLE}'s RJMCMC features reduced the computational time, but not enough to make the sampler practical in casual settings. Previous implementations of non-local selection priors saved time by approximately marginalizing out the effect sizes with a Laplace approximation \citep{Johnson_2012,Nikooienejad_2016,shin-2018a}. This was unappealing to us because the effect sizes are important in our context. 

An alternative meta-analysis type approach uses conventional methods to estimate summary statistics and then treats those summaries as data in a hierarchical model. We adapted our  model to use summary statistics by replacing the experiment-specific models with a summary model. In particular, we treat the estimated gene effects, $\widehat{\log(fc)}_j$ and $\hat{\gamma}_j$, as approximately normal centered on the true effect sizes and use our  hyperprior on the true effect sizes. Preliminary assessment of this summary statistics model (Section 5.10 of the Supplement) suggests that a complete investigation (in future work) is warranted. The summary model is competitive in our standard simulation scenario but loses power, compared to our full model, with deviations from that scenario.

One issue that requires further attention is the effectiveness of our mapping between SNVs and genes in either GWAS or whole genome sequencing (WGS) datasets, particularly for SNVs in non-coding regions. The technique that we used (Section \ref{sec:gwas}) for mapping SNVs into genes is simplistic and could result in missed signals if SNVs within a single gene work in opposite directions. Other mapping techniques may enrich the procedure.

The strategy of combining genomic, transcriptomic, phenotypic, and potentially other sources of information using the three-groups framework can be applied to any heritable disease with multiple data types available. 
The PD datasets that we analyzed here are publicly available, but others like family pedigree WGS have been collected by our lab, and still others, including small interfering RNA screens, will result from follow-up experiments.

\backmatter

\section*{Acknowledgements}
This work was made possible by the NSF collaborative grant NSF-1761941/2309825. Additional support for this work came from NIH R01 LM013617, R01 NS124848, the Michael J Fox Foundation, and CIRM DISC0-16039. Data used in the preparation of this article were obtained 09/21/2018 from the Parkinson’s Progression Markers Initiative (PPMI) database (\url{www.ppmi-info.org/access-data-specimens/download-data}), RRID:SCR 006431. For up-to-date information on the study, visit \url{www.ppmi-info.org}. 
PPMI---a public-private partnership---is funded by the Michael J. Fox Foundation for Parkinson’s Research and funding partners, including 4D Pharma, Abbvie, AcureX, Allergan, Amathus Therapeutics, Aligning Science Across Parkinson's, AskBio, Avid Radiopharmaceuticals, BIAL, BioArctic, Biogen, Biohaven, BioLegend, BlueRock Therapeutics, Bristol-Myers Squibb, Calico Labs, Capsida Biotherapeutics, Celgene, Cerevel Therapeutics, Coave Therapeutics, DaCapo Brainscience, Denali, Edmond J. Safra Foundation, Eli Lilly, Gain Therapeutics, GE HealthCare, Genentech, GSK, Golub Capital, Handl Therapeutics, Insitro, Jazz Pharmaceuticals, Johnson \& Johnson Innovative Medicine, Lundbeck, Merck, Meso Scale Discovery, Mission Therapeutics, Neurocrine Biosciences, Neuron23, Neuropore, Pfizer, Piramal, Prevail Therapeutics, Roche, Sanofi, Servier, Sun Pharma Advanced Research Company, Takeda, Teva, UCB, Vanqua Bio, Verily, Voyager Therapeutics, the Weston Family Foundation and Yumanity Therapeutics. \vspace*{-8pt}

\section*{Supporting Information}
Supplementary Material, Tables, Figures, and simulation code referenced throughout the paper are available with this paper on arXiv.

\section*{Data Availability}
The data underlying this article were provided by the IPDGC (NeuroX GWAS data) and PPMI (RNA-seq data, obtained upon request). Data will be shared on request to the corresponding author with permission of the IPDGC and PPMI.

\bibliographystyle{biom_copy} 
\bibliography{2sources.bib}

@article{scott-2010a,
    AUTHOR = {Scott, James G. and Berger, James O.},
     TITLE = {Bayes and empirical-{B}ayes multiplicity adjustment in the
              variable-selection problem},
   JOURNAL = {The Annals of Statistics.},
  FJOURNAL = {The Annals of Statistics},
    VOLUME = {38},
      YEAR = {2010},
    NUMBER = {5},
     PAGES = {2587--2619},
}

@article{benjamini-2008a,
  title={Screening for partial conjunction hypotheses},
  author={Benjamini, Yoav and Heller, Ruth},
  journal={Biometrics},
  volume={64},
  number={4},
  pages={1215--1222},
  year={2008},
  publisher={Wiley Online Library}
}

@article{benjamini-1995a,
title={Controlling the false discovery rate: a practical and powerful approach to multiple testing},
author={Benjamini, Yoav and Hochberg, Yosef},
journal={Journal of the Royal Statistical Society Series B-Statistical Methodology},
pages={289--300},
year={1995},
volume = {57},
number = {1},
}

@article{fisher2-1929a,
  title={Tests of significance in harmonic analysis},
  author={Fisher, R.A.},
  journal={Proceedings of the Royal Society of London Series A-Mathematical and Physical Sciences},
  volume={125},
  number={796},
  pages={54--59},
  year={1929},
  publisher={JSTOR}
}

@article{stouffer-1949a,
  title={The {A}merican soldier: {A}djustment during army life, {V}ol. {I}},
  author={Stouffer, Samuel A. and Suchman, Edward A. and DeVinney, Leland C. and Star, Shirley A. and Williams Jr, Robin M.},
  journal={Studies in Social Psychology World War II},
  year={1949},
  publisher={Princeton University Press}
}

@article{genovese-2004a,
  title={A stochastic process approach to false discovery control},
  author={Genovese, Christopher and Wasserman, Larry},
  journal={The Annals of Statistics},
  volume={32},
  number={3},
  pages={1035--1061},
  year={2004},
  publisher={Institute of Mathematical Statistics}
}

@article{ritchie-2015b,
  title={Methods of integrating data to uncover genotype-phenotype interactions},
  author={Ritchie, Marylyn D. and Holzinger, Emily R. and Li, Ruowang and Pendergrass, Sarah A. and Kim, Dokyoon},
  journal={Nat Rev Genet},
  volume={16},
  number={2},
  pages={85},
  year={2015},
  publisher={Nature Publishing Group}
}

@article{holzinger-2013a,
  title={ATHENA: the analysis tool for heritable and environmental network associations},
  author={Holzinger, Emily R. and Dudek, Scott M. and Frase, Alex T. and Pendergrass, Sarah A. and Ritchie, Marylyn D.},
  journal={Bioinformatics},
  volume={30},
  number={5},
  pages={698--705},
  year={2013},
  publisher={Oxford University Press}
}

@article{asimit-2012a,
  title={{ARIEL} and {AMELIA}: testing for an accumulation of rare variants using next-generation sequencing data},
  author={Asimit, Jennifer L. and Day-Williams, Aaron G. and Morris, Andrew P. and Zeggini, Eleftheria},
  journal={Human Heredity},
  volume={73},
  number={2},
  pages={84--94},
  year={2012},
  publisher={Karger Publishers}
}

@article{morgenthaler-2007a,
  title={A strategy to discover genes that carry multi-allelic or mono-allelic risk for common diseases: a cohort allelic sums test (CAST)},
  author={Morgenthaler, Stephan and Thilly, William G.},
  journal={Mutation Research/Fundamental and Molecular Mechanisms of Mutagenesis},
  volume={615},
  number={1},
  pages={28--56},
  year={2007},
  publisher={Elsevier}
}

@article{li-2008a,
  title={Methods for detecting associations with rare variants for common diseases: application to analysis of sequence data},
  author={Li, Bingshan and Leal, Suzanne M.},
  journal={The American Journal of Human Genetics},
  volume={83},
  number={3},
  pages={311--321},
  year={2008},
  publisher={Elsevier}
}

@article{madsen-2009a,
  title={A groupwise association test for rare mutations using a weighted sum statistic},
  author={Madsen, Bo Eskerod and Browning, Sharon R.},
  journal={PLoS Genetics},
  volume={5},
  number={2},
  pages={e1000384},
  year={2009},
  publisher={Public Library of Science}
}

@article{boyle-2017a,
  title={An Expanded View of Complex Traits: From Polygenic to Omnigenic},
  author={Boyle, Evan A. and Li, Yang I. and Pritchard, Jonathan K.},
  journal={Cell},
  volume={169},
  number={7},
  pages={1177--1186},
  year={2017},
  publisher={Elsevier}
}

@book{gelman-bda3,
  author = {Gelman, Andrew},
  address = {Boca Raton},
  booktitle = {Bayesian data analysis},
  edition = {3rd},
  isbn = {9781439840955},
  language = {eng},
  lccn = {2013039507},
  publisher = {CRC Press},
  title = {Bayesian data analysis },
  year = {2014}
}

@article{Gerard_2020,
author = {Gerard, David},
address = {England},
copyright = {2020. This work is licensed under http://creativecommons.org/licenses/by/4.0/ (the “License”). Notwithstanding the ProQuest Terms and Conditions, you may use this content in accordance with the terms of the License.},
issn = {1471-2105},
journal = {BMC bioinformatics},
language = {eng},
number = {1},
pages = {206-206},
publisher = {BioMed Central},
title = {Data-based {RNA}-seq simulations by binomial thinning},
volume = {21},
year = {2020}
}

@Article{seqgendiff,
    title = {Data-based {RNA}-seq simulations by binomial thinning},
    year = {2020},
    journal = {BMC Bioinformatics},
    publisher = {BioMed Central Ltd},
    volume = {21},
    number = {1},
    pages = {206},
    issn = {1471-2105},
    doi = {10.1186/s12859-020-3450-9},
    author = {David Gerard}
  }

@article{Pickrell_2010U,
author = {Pickrell, Joseph K and Gilad, Yoav and Pritchard, Jonathan K and Marioni, John C and Pai, Athma A and Degner, Jacob F and Engelhardt, Barbara E and Nkadori, Everlyne and Veyrieras, Jean-Baptiste and Stephens, Matthew},
address = {LONDON},
copyright = {2015 INIST-CNRS},
issn = {0028-0836},
journal = {Nature (London)},
language = {eng},
number = {7289},
pages = {768-772},
publisher = {Springer Nature},
title = {Understanding mechanisms underlying human gene expression variation with RNA sequencing},
volume = {464},
year = {2010}
}

@article{Montgomery_2010,
author = {Montgomery, Stephen B and Dermitzakis, Emmanouil T and Sammeth, Micha and Gutierrez-Arcelus, Maria and Lach, Radoslaw P and Ingle, Catherine and Nisbett, James and Guigo, Roderic},
address = {LONDON},
copyright = {2015 INIST-CNRS},
issn = {0028-0836},
journal = {Nature (London)},
language = {eng},
number = {7289},
pages = {773-777},
publisher = {Springer Nature},
title = {Transcriptome genetics using second generation sequencing in a Caucasian population},
volume = {464},
year = {2010}
}

@article{Gneiting_2007,
author = {Gneiting, Tilmann and Raftery, Adrian E},
address = {PHILADELPHIA},
copyright = {American Statistical Association 2007},
issn = {0162-1459},
journal = {Journal of the American Statistical Association},
language = {eng},
number = {477},
pages = {359-378},
publisher = {Taylor & Francis},
title = {Strictly Proper Scoring Rules, Prediction, and Estimation},
volume = {102},
year = {2007}
}

@article{Johnson_2010,
issn = {1369-7412},author = {Johnson, Valen E. E. and Rossell, David},
address = {OXFORD},
journal = {Journal of the Royal Statistical Society. Series B, Statistical methodology},
language = {eng},
number = {2},
pages = {143-170},
publisher = {Oxford Univ Press},
title = {On the use of Non-Local Prior Densities in Bayesian Hypothesis Tests},
volume = {72},
year = {2010}
}

@article{Johnson_2012,
copyright = {Copyright Taylor & Francis Group, LLC 2012},
issn = {0162-1459},author = {Johnson, Valen E. and Rossell, David},
address = {PHILADELPHIA},
journal = {Journal of the American Statistical Association},
language = {eng},
number = {498},
pages = {649-660},
publisher = {Taylor & Francis Group},
title = {Bayesian Model Selection in High-Dimensional Settings},
volume = {107},
year = {2012}
}

@article{Nikooienejad_2016,
copyright = {The Author 2016. Published by Oxford University Press.},
issn = {1367-4803},
author = {Nikooienejad, Amir and Wang, Wenyi and Johnson, Valen E.},
address = {OXFORD},
journal = {Bioinformatics},
language = {eng},
number = {9},
pages = {1338-1345},
publisher = {Oxford Univ Press},
title = {Bayesian variable selection for binary outcomes in high-dimensional genomic studies using non-local priors},
volume = {32},
year = {2016}
}

@article{LiWeibing_2022,
copyright = {The Author(s), under exclusive licence to Springer-Verlag GmbH Germany, part of Springer Nature 2021},
issn = {0943-4062},
author = {Li, Weibing and Chekouo, Thierry},
address = {Berlin/Heidelberg},
journal = {Computational statistics},
language = {eng},
number = {1},
pages = {287-302},
publisher = {Springer Berlin Heidelberg},
title = {Bayesian group selection with non-local priors},
volume = {37},
year = {2022}
}

@article{Nalls_2014,
copyright = {info:eu-repo/semantics/restrictedAccess},
issn = {1061-4036},
author = {Nalls, Mike A and Pankratz, Nathan and Lill, Christina M and Do, Chuong B and Hernandez, Dena G and Saad, Mohamad and DeStefano, Anita L and Kara, Eleanna and Bras, Jose and Sharma, Manu and Schulte, Claudia and Keller, Margaux F and Arepalli, Sampath and Letson, Christopher and Edsall, Connor and Stefansson, Hreinn and Liu, Xinmin and Pliner, Hannah and Lee, Joseph H and Cheng, Rong and Ikram, M. Arfan and Ioannidis, John P. A and Hadjigeorgiou, Georgios M and Bis, Joshua C and Martinez, Maria and Perlmutter, Joel S and Goate, Alison and Marder, Karen and Fiske, Brian and Sutherland, Margaret and Xiromerisiou, Georgia and Myers, Richard H and Clark, Lorraine N and Stefansson, Kari and Hardy, John A and Heutink, Peter and Chen, Honglei and Wood, Nicholas W and Houlden, Henry and Payami, Haydeh and Brice, Alexis and Scott, William K and Gasser, Thomas and Bertram, Lars and Eriksson, Nicholas and Foroud, Tatiana and Singleton, Anew B},
address = {NEW YORK},
journal = {Nature genetics},
language = {eng},
number = {9},
organization = {GenePD},
pages = {989-+},
publisher = {Springer Nature},
title = {Large-scale meta-analysis of genome-wide association data identifies six new risk loci for {P}arkinson's disease},
volume = {46},
year = {2014},
}

@article{Agarwal_2020,
copyright = {The Author(s) 2020. This work is published under http://creativecommons.org/licenses/by/4.0/ (the “License”). Notwithstanding the ProQuest Terms and Conditions, you may use this content in accordance with the terms of the License.},
issn = {2041-1723},
author = {Agarwal, Devika and Sandor, Cynthia and Volpato, Viola and Caffrey, Tara M and Monzón-Sandoval, Jimena and Bowden, Rory and Alegre-Abarrategui, Javier and Wade-Martins, Richard and Webber, Caleb},
address = {England},
journal = {Nature communications},
language = {eng},
number = {1},
pages = {4183-4183},
publisher = {Nature Publishing Group},
title = {A single-cell atlas of the human substantia nigra reveals cell-specific pathways associated with neurological disorders},
volume = {11},
year = {2020}
}

@article{Efron_2001,
author = {Efron, Bradley and Tibshirani, Robert and Storey, John D and Tusher, Virginia},
address = {ALEXANDRIA},
copyright = {American Statistical Association 2001},
issn = {0162-1459},
journal = {Journal of the American Statistical Association},
language = {eng},
number = {456},
pages = {1151-1160},
publisher = {Taylor & Francis},
title = {Empirical Bayes Analysis of a Microarray Experiment},
volume = {96},
year = {2001},
}

@Article{nimble_2017_article,
  title = {Programming with models: writing statistical algorithms for general model structures with {NIMBLE}},
  journal = {Jour. of Computational and Graphical Statistics},
  volume = {26},
  issue = {2},
  pages = {403-413},
  year = {2017},
  author = {Perry {de Valpine} and Daniel Turek and Christopher Paciorek and Cliff Anderson-Bergman and Duncan Temple {Lang} and Ras Bodik},
  doi = {10.1080/10618600.2016.1172487},
}

@Manual{nimble_2023_package,
  title = {{NIMBLE}: {MCMC}, Particle Filtering, and Programmable Hierarchical Modeling},
  author = {Perry {de Valpine} and Christopher Paciorek and Daniel Turek and Nick Michaud and Cliff Anderson-Bergman and Fritz Obermeyer and Claudia {Wehrhahn Cortes} and Abel Rodrìguez and Duncan {Temple Lang} and Sally Paganin},
  url = {https://cran.r-project.org/package=nimble},
  year = {2023},
  version = {1.0.1},
  note = {{R} package version 1.0.1},
  doi = {10.5281/zenodo.1211190},
}

@article{Barbieri_2004,
author = {Barbieri, Maria Maddalena and Berger, James O.},
address = {CLEVELAND},
copyright = {Copyright 2004 Institute of Mathematical Statistics},
issn = {0090-5364},
journal = {The Annals of statistics},
language = {eng},
number = {3},
pages = {870-897},
publisher = {Institute of Mathematical Statistics},
title = {Optimal Predictive Model Selection},
volume = {32},
year = {2004},
}

@article{ding-2022a,
  title={Cooperative learning for multiview analysis},
  author={Ding, Daisy Yi and Li, Shuangning and Narasimhan, Balasubramanian and Tibshirani, Robert},
  journal={Proceedings of the National Academy of Sciences},
  volume={119},
  number={38},
  pages={e2202113119},
  year={2022},
  publisher={National Acad Sciences}
}

@article{li-2018a,
  title={A review on machine learning principles for multi-view biological data integration},
  author={Li, Yifeng and Wu, Fang-Xiang and Ngom, Alioune},
  journal={Briefings in bioinformatics},
  volume={19},
  number={2},
  pages={325--340},
  year={2018},
  publisher={Oxford University Press}
}

@article{richardson-2016a,
  title={Statistical methods in integrative genomics},
  author={Richardson, Sylvia and Tseng, George C and Sun, Wei},
  journal={Annual review of statistics and its application},
  volume={3},
  pages={181--209},
  year={2016},
  publisher={Annual Reviews}
}

@article{Tyekucheva_2011,
author = {Tyekucheva, Svitlana and Marchionni, Luigi and Karchin, Rachel and Parmigiani, Giovanni},
address = {LONDON},
copyright = {Copyright ©2011 Tyekucheva et al.; licensee BioMed Central Ltd. 2011 Tyekucheva et al.; licensee BioMed Central Ltd.},
issn = {1465-6906},
journal = {Genome biology},
language = {eng},
number = {10},
pages = {R105-R105},
publisher = {Springer-Verlag},
title = {Integrating diverse genomic data using gene sets},
volume = {12},
year = {2011},
}

@article{Love_2014,
author = {Love, Michael I. and Huber, Wolfgang and Anders, Simon},
address = {LONDON},
copyright = {Love et al.; licensee BioMed Central. 2014},
issn = {1474-760X},
journal = {Genome biology},
language = {eng},
number = {12},
pages = {550-550},
publisher = {Springer Nature},
title = {Moderated estimation of fold change and dispersion for RNA-seq data with DESeq2},
volume = {15},
year = {2014},
}

@article{Robinson_2010,
author = {Robinson, Mark D. and McCarthy, Davis J. and Smyth, Gordon K.},
address = {OXFORD},
copyright = {The Author(s) 2009. Published by Oxford University Press. 2009},
issn = {1367-4803},
journal = {Bioinformatics},
language = {eng},
number = {1},
pages = {139-140},
publisher = {Oxford University Press},
title = {{edgeR}: a Bioconductor package for differential expression analysis of digital gene expression data},
volume = {26},
year = {2010},
}

@article{Law_2014,
author = {Law, Charity W and Chen, Yunshun and Shi, Wei and Smyth, Gordon K},
address = {LONDON},
copyright = {Copyright © 2014 Law et al.; licensee BioMed Central Ltd. 2014 Law et al.; licensee BioMed Central Ltd.},
issn = {1465-6906},
journal = {Genome biology},
language = {eng},
number = {2},
pages = {R29-R29},
publisher = {Springer-Verlag},
title = {voom: precision weights unlock linear model analysis tools for RNA-seq read counts},
volume = {15},
year = {2014},
}

@article {shin-2018a,
    AUTHOR = {Shin, Minsuk and Bhattacharya, Anirban and Johnson, Valen E.},
     TITLE = {Scalable {B}ayesian variable selection using nonlocal prior
              densities in ultrahigh-dimensional settings},
   JOURNAL = {Stat Sin.},
  FJOURNAL = {Statistica Sinica},
    VOLUME = {28},
      YEAR = {2018},
    NUMBER = {2},
     PAGES = {1053--1078},
      ISSN = {1017-0405,1996-8507},
   MRCLASS = {62F15 (62H15 62J05)},
  MRNUMBER = {3791100},
}

@article {green-1995a,
    AUTHOR = {Green, Peter J.},
     TITLE = {Reversible jump {M}arkov chain {M}onte {C}arlo computation and
              {B}ayesian model determination},
   JOURNAL = {Biometrika},
  FJOURNAL = {Biometrika},
    VOLUME = {82},
      YEAR = {1995},
    NUMBER = {4},
     PAGES = {711--732},
      ISSN = {0006-3444,1464-3510},
   MRCLASS = {62F15},
  MRNUMBER = {1380810},
       DOI = {10.1093/biomet/82.4.711},
       URL = {https://doi.org/10.1093/biomet/82.4.711},
}

@article{Uffelmann_2021,
  title={Genome-wide association studies},
  author={Uffelmann, Emil and Huang, Qin Qin and Munung, Nchangwi Syntia and De Vries, Jantina and Okada, Yukinori and Martin, Alicia R and Martin, Hilary C and Lappalainen, Tuuli and Posthuma, Danielle},
  journal={Nature Reviews Methods Primers},
  volume={1},
  number={1},
  pages={59},
  year={2021},
  publisher={Nature Publishing Group UK London}
}

@article {Gelman_2006,
    AUTHOR = {Gelman, Andrew},
     TITLE = {Prior distributions for variance parameters in hierarchical
              models (comment on article by {B}rowne and {D}raper)},
   JOURNAL = {Bayesian Anal.},
  FJOURNAL = {Bayesian Analysis},
    VOLUME = {1},
      YEAR = {2006},
    NUMBER = {3},
     PAGES = {515--533},
      ISSN = {1936-0975,1931-6690},
   MRCLASS = {99-01},
  MRNUMBER = {2221284},
       DOI = {10.1214/06-BA117A},
       URL = {https://doi.org/10.1214/06-BA117A},
}

@article{bose2019mitochondrial,
  title={Mitochondrial dysfunction and oxidative stress in induced pluripotent stem cell models of {P}arkinson's disease},
  author={Bose, Anindita and Beal, M Flint},
  journal={Eur J Neurosci.},
  volume={49},
  number={4},
  pages={525--532},
  year={2019},
  publisher={Wiley Online Library}
}

@article{zaltieri2015mitochondrial,
  title={Mitochondrial dysfunction $\alpha$-{S}ynuclein synaptic pathology in {P}arkinson’s disease: who’s on first?},
  author={Zaltieri, Michela and Longhena, Francesca and Pizzi, Marina and Missale, Cristina and Spano, PierFranco and Bellucci, Arianna and others},
  journal={Parkinson's disease},
  volume={2015},
  year={2015},
  publisher={Hindawi}
}

@article{moon2015mitochondrial,
  title={Mitochondrial dysfunction in {P}arkinson's disease},
  author={Moon, Hyo Eun and Paek, Sun Ha},
  journal={Experimental neurobiology},
  volume={24},
  number={2},
  pages={103},
  year={2015},
  publisher={Korean Society for Brain and Neural Science}
}

@article{clayton1998synucleins,
  title={The synucleins: a family of proteins involved in synaptic function, plasticity, neurodegeneration and disease},
  author={Clayton, David F and George, Julia M},
  journal={Trends in neurosciences},
  volume={21},
  number={6},
  pages={249--254},
  year={1998},
  publisher={Elsevier}
}

@article{bagetta2010synaptic,
  title={Synaptic dysfunction in {P}arkinson's disease},
  author={Bagetta, Vincenza and Ghiglieri, Veronica and Sgobio, Carmelo and Calabresi, Paolo and Picconi, Barbara},
  journal={Biochemical Society Transactions},
  volume={38},
  number={2},
  pages={493--497},
  year={2010},
  publisher={Portland Press Ltd.}
}

@article{morais2009parkinson,
  title={Parkinson's disease mutations in {PINK1} result in decreased Complex {I} activity and deficient synaptic function},
  author={Morais, Vanessa A and Verstreken, Patrik and Roethig, Anne and Smet, Jo{\'e}l and Snellinx, An and Vanbrabant, Mieke and Haddad, Dominik and Frezza, Christian and Mandemakers, Wim and Vogt-Weisenhorn, Daniela and others},
  journal={EMBO molecular medicine},
  volume={1},
  number={2},
  pages={99--111},
  year={2009},
  publisher={WILEY-VCH Verlag Weinheim}
}

@article{lev2003apoptosis,
  title={Apoptosis and {P}arkinson's disease},
  author={Lev, Nirit and Melamed, Eldad and Offen, Daniel},
  journal={Progress in Neuro-Psychopharmacology and Biological Psychiatry},
  volume={27},
  number={2},
  pages={245--250},
  year={2003},
  publisher={Elsevier}
}

@article{tatton2003apoptosis,
  title={Apoptosis in {P}arkinson's disease: signals for neuronal degradation},
  author={Tatton, William G and Chalmers-Redman, Ruth and Brown, David and Tatton, Nadine},
  journal={Ann Neurol.},
  volume={53},
  number={S3},
  pages={S61--S72},
  year={2003},
  publisher={Wiley Online Library}
}

@article{mochizuki1996histochemical,
  title={Histochemical detection of apoptosis in {P}arkinson's disease},
  author={Mochizuki, Hideki and Goto, Keigo and Mori, Hideo and Mizuno, Yoshikuni},
  journal={Journal of the neurological sciences},
  volume={137},
  number={2},
  pages={120--123},
  year={1996},
  publisher={Elsevier}
}

@article{zhao2010stress,
  title={Stress-sensitive regulation of {IFRD1} {mRNA} decay is mediated by an upstream open reading frame},
  author={Zhao, Chenyang and Datta, Shyamasree and Mandal, Palash and Xu, Shuqing and Hamilton, Thomas},
  journal={Journal of Biological Chemistry},
  volume={285},
  number={12},
  pages={8552--8562},
  year={2010},
  publisher={ASBMB}
}

@article{chang2020ifrd1,
  title={{IFRD1} regulates the asthmatic responses of airway via {NF-$\kappa$B} pathway},
  author={Chang, Ming and Zhang, Yali and Hui, Zhenggang and Wang, Dan and Guo, Huanli},
  journal={Molecular Immunology},
  volume={127},
  pages={186--192},
  year={2020},
  publisher={Elsevier}
}

@article{liu2008dusp1,
  title={{DUSP1} is controlled by p53 during the cellular response to oxidative stress},
  author={Liu, Yu-Xin and Wang, Jianli and Guo, Jianfen and Wu, Jingjing and Lieberman, Howard B and Yin, Yuxin},
  journal={Molecular Cancer Research},
  volume={6},
  number={4},
  pages={624--633},
  year={2008},
  publisher={AACR}
}

@article{wang2016role,
  title={Role for {DUSP1} (dual-specificity protein phosphatase 1) in the regulation of autophagy},
  author={Wang, Juan and Zhou, Jun-Ying and Kho, Dhonghyo and Reiners Jr, John J and Wu, Gen Sheng},
  journal={Autophagy},
  volume={12},
  number={10},
  pages={1791--1803},
  year={2016},
  publisher={Taylor \& Francis}
}

@article{singh2020parkinson,
  title={Parkinson's: a disease of aberrant vesicle trafficking},
  author={Singh, Pawan Kishor and Muqit, Miratul MK},
  journal={Annual review of cell and developmental biology},
  volume={36},
  pages={237--264},
  year={2020},
  publisher={Annual Reviews}
}

@article{perrett2015endosomal,
  title={The endosomal pathway in {P}arkinson's disease},
  author={Perrett, Rebecca M and Alexopoulou, Zoi and Tofaris, George K},
  journal={Molecular and Cellular Neuroscience},
  volume={66},
  pages={21--28},
  year={2015},
  publisher={Elsevier}
}

@article{esposito2012synaptic,
  title={Synaptic vesicle trafficking and {P}arkinson's disease},
  author={Esposito, Giovanni and Ana Clara, Fernandes and Verstreken, Patrik},
  journal={Developmental neurobiology},
  volume={72},
  number={1},
  pages={134--144},
  year={2012},
  publisher={Wiley Online Library}
}

@article{brehm2015genetic,
  title={A genetic mouse model of {P}arkinson’s disease shows involuntary movements and increased postsynaptic sensitivity to apomorphine},
  author={Brehm, N and Bez, Francesco and Carlsson, T and Kern, B and Gispert, S and Auburger, G and Cenci, MA},
  journal={Molecular neurobiology},
  volume={52},
  pages={1152--1164},
  year={2015},
  publisher={Springer}
}

@article{usenko2021comparative,
  title={Comparative transcriptome analysis in monocyte-derived macrophages of asymptomatic GBA mutation carriers and patients with {GBA}-associated {P}arkinson’s disease},
  author={Usenko, Tatiana and Bezrukova, Anastasia and Basharova, Katerina and Panteleeva, Alexandra and Nikolaev, Mikhail and Kopytova, Alena and Miliukhina, Irina and Emelyanov, Anton and Zakharova, Ekaterina and Pchelina, Sofya},
  journal={Genes},
  volume={12},
  number={10},
  pages={1545},
  year={2021},
  publisher={MDPI}
}

@article{gazal2022combining,
  title={Combining SNP-to-gene linking strategies to identify disease genes and assess disease omnigenicity},
  author={Gazal, Steven and Weissbrod, Omer and Hormozdiari, Farhad and Dey, Kushal K and Nasser, Joseph and Jagadeesh, Karthik A and Weiner, Daniel J and Shi, Huwenbo and Fulco, Charles P and O’Connor, Luke J and others},
  journal={Nature genetics},
  volume={54},
  number={6},
  pages={827--836},
  year={2022},
  publisher={Nature Publishing Group US New York}
}

@article{Gu_2023,
  title={Expanding causal genes for {P}arkinson’s disease via multi-omics analysis},
  author={Gu, Xiao-Jing and Su, Wei-Ming and Dou, Meng and Jiang, Zheng and Duan, Qing-Qing and Yin, Kang-Fu and Cao, Bei and Wang, Yi and Li, Guo-Bo and Chen, Yong-Ping},
  journal={npj Parkinson's Disease},
  volume={9},
  number={1},
  year={2023},
  publisher={Nature Publishing Group UK London}
}

@article{schilder_2022fine,
  title={Fine-mapping of {P}arkinson’s disease susceptibility loci identifies putative causal variants},
  author={Schilder, Brian M and Raj, Towfique},
  journal={Human Molecular Genetics},
  volume={31},
  number={6},
  pages={888--900},
  year={2022},
  publisher={Oxford University Press}
}

@article{hu_2024early,
  title={Early B Cell Factor 3 ({EBF3}) attenuates {P}arkinson's disease through directly regulating contactin-associated protein-like 4 ({CNTNAP4}) transcription: An experimental study},
  author={Hu, Wentao and Wang, Menghan and Sun, Guifang and Zhang, Limin and Lu, Hong},
  journal={Cellular Signalling},
  volume={,},
  pages={111-139},
  year={2024},
  publisher={Elsevier}
}

@article{zhang_2020cntnap4,
  title={{CNTNAP4} deficiency in dopaminergic neurons initiates {P}arkinsonian phenotypes},
  author={Zhang, Wenlong and Zhou, Miaomiao and Lu, Weiye and Gong, Junwei and Gao, Feng and Li, Yuanquan and Xu, Xuandong and Lin, Yuwan and Zhang, Xiaokang and Ding, Liuyan and others},
  journal={Theranostics},
  volume={10},
  number={7},
  pages={3000},
  year={2020},
  publisher={Ivyspring International Publisher}
}

@article{ma_2023cxcr4,
  title={{CXCR4} knockout induces neuropathological changes in the {MPTP}-lesioned model of {P}arkinson's disease},
  author={Ma, Jianjun and Dong, Linrui and Chang, Qingqing and Chen, Siyuan and Zheng, Jinhua and Li, Dongsheng and Wu, Shaopu and Yang, Hongqi and Li, Xue},
  journal={Biochimica et Biophysica Acta (BBA)-Molecular Basis of Disease},
  volume={1869},
  number={2},
  pages={166597},
  year={2023},
  publisher={Elsevier}
}

@article{bonham2018cxcr4,
  title={CXCR4 involvement in neurodegenerative diseases},
  author={Bonham, Luke W and Karch, Celeste M and Fan, Chun C and Tan, Chin and Geier, Ethan G and Wang, Yunpeng and Wen, Natalie and Broce, Iris J and Li, Yi and Barkovich, Matthew J and others},
  journal={Translational psychiatry},
  volume={8},
  number={1},
  pages={73},
  year={2018},
  publisher={Nature Publishing Group UK London}
}

@article{li2022long,
  title={Long noncoding RNA BACE1-antisense transcript plays a critical role in {P}arkinson's disease via {microRNA}-214-3p/Cell death-inducing p53-target protein 1 axis},
  author={Li, Lina and Wang, Hongjuan and Li, Huicang and Lu, Xin and Gao, Yanxiang and Guo, Xiaofeng},
  journal={Bioengineered},
  volume={13},
  number={4},
  pages={10889--10901},
  year={2022},
  publisher={Taylor \& Francis}
}

@article{strobbe2018distinct,
  title={Distinct mechanisms of pathogenic {DJ}-1 mutations in mitochondrial quality control},
  author={Strobbe, Daniela and Robinson, Alexis A and Harvey, Kirsten and Rossi, Lara and Ferraina, Caterina and De Biase, Valerio and Rodolfo, Carlo and Harvey, Robert J and Campanella, Michelangelo},
  journal={Frontiers in Molecular Neuroscience},
  volume={11},
  pages={68},
  year={2018},
  publisher={Frontiers Media SA}
}

@article{kochmanski2022parkinson,
  title={Parkinson's disease-associated, sex-specific changes in {DNA} methylation at {PARK7} ({DJ}-1), {SLC17A6} ({VGLUT2}), {PTPRN2} ({IA}-2$\beta$), and {NR4A2} ({NURR1}) in cortical neurons},
  author={Kochmanski, Joseph and Kuhn, Nathan C and Bernstein, Alison I},
  journal={npj Parkinson's Disease},
  volume={8},
  number={1},
  pages={120},
  year={2022},
  publisher={Nature Publishing Group UK London}
}

@article{chuang2019longitudinal,
  title={Longitudinal epigenome-wide methylation study of cognitive decline and motor progression in {P}arkinson's disease.}, 
  journal={J Parkinsons Dis.}, 
  volume={9}, 
  pages={389--400},
  author={Chuang, YH and Lu, AT and Paul, KC and Folle, AD and Bronstein, JM and Bordelon, Y and Horvath, S and Ritz, B},
  doi={doi. org/10.3233/JPD-18154},
  year={2019}
}

@article{tan2020parkinson,
  title={Parkinson disease and the immune system—associations, mechanisms and therapeutics},
  author={Tan, Eng-King and Chao, Yin-Xia and West, Andrew and Chan, Ling-Ling and Poewe, Werner and Jankovic, Joseph},
  journal={Nature Reviews Neurology},
  volume={16},
  number={6},
  pages={303--318},
  year={2020},
  publisher={Nature Publishing Group UK London}
}

@article{albert2020map,
  title={{MAP/ERK} signaling in developing cognitive and emotional function and its effect on pathological and neurodegenerative processes},
  author={Albert-Gasc{\'o}, H{\'e}ctor and Ros-Bernal, Francisco and Castillo-G{\'o}mez, Esther and Olucha-Bordonau, Francisco E},
  journal={Int J Mol Sci.},
  volume={21},
  number={12},
  pages={4471},
  year={2020},
  publisher={MDPI}
}

@article{toskas2022prc2,
  title={{PRC2}-mediated repression is essential to maintain identity and function of differentiated dopaminergic and serotonergic neurons},
  author={Toskas, Konstantinos and Yaghmaeian-Salmani, Behzad and Skiteva, Olga and Paslawski, Wojciech and Gillberg, Linda and Skara, Vasiliki and Antoniou, Irene and S{\"o}dersten, Erik and Svenningsson, Per and Chergui, Karima and others},
  journal={Science Advances},
  volume={8},
  number={34},
  pages={eabo1543},
  year={2022},
  publisher={American Association for the Advancement of Science}
}

@article{liu2021aberrant,
  title={Aberrant mitochondrial morphology and function associated with impaired mitophagy and {DNM1L-MAPK/ERK} signaling are found in aged mutant {P}arkinsonian {LRRK2R1441G} mice},
  author={Liu, Huifang and Ho, Philip Wing-Lok and Leung, Chi-Ting and Pang, Shirley Yin-Yu and Chang, Eunice Eun Seo and Choi, Zoe Yuen-Kiu and Kung, Michelle Hiu-Wai and Ramsden, David Boyer and Ho, Shu-Leong},
  journal={Autophagy},
  volume={17},
  number={10},
  pages={3196--3220},
  year={2021},
  publisher={Taylor \& Francis}
}

@article{Zeggini_Ioannidis_2009,
    author = {Eleftheria Zeggini and John PA Ioannidis},
    title = {Meta-Analysis in Genome-Wide Association Studies},
    journal = {Pharmacogenomics},
    volume = {10},
    number = {2},
    pages = {191--201},
    year = {2009},
    publisher = {Taylor \& Francis},
    doi = {10.2217/14622416.10.2.191},
    note ={PMID: 19207020},
    URL = {https://doi.org/10.2217/14622416.10.2.191},
}

@article{Begum_2012,
    author = {Begum, Ferdouse and Ghosh, Debashis and Tseng, George C. and Feingold, Eleanor},
    title = {Comprehensive literature review and statistical considerations for GWAS meta-analysis},
    journal = {Nucleic Acids Research},
    volume = {40},
    number = {9},
    pages = {3777-3784},
    year = {2012},
    month = {01},
    issn = {0305-1048},
    doi = {10.1093/nar/gkr1255},
    url = {https://doi.org/10.1093/nar/gkr1255},
    eprint = {https://academic.oup.com/nar/article-pdf/40/9/3777/16954210/gkr1255.pdf},
}

@article{liu2020cauchy,
  title={Cauchy combination test: a powerful test with analytic p-value calculation under arbitrary dependency structures},
  author={Liu, Yaowu and Xie, Jun},
  journal={Journal of the American Statistical Association},
  volume={115},
  number={529},
  pages={393--402},
  year={2020},
  publisher={Taylor \& Francis}
}

\label{lastpage}

\newpage


\appendix

\section{1: Full model}
\label{sec:full_model}

\small{
\begin{multicols}{2}
    \noindent
    $Y_{ijk}^{RNA}$ is the RNA count from individual $i$ for gene $j$. \\
    \begin{align*}
        Y_{ijk}^{RNA} \sim \text{Neg}&\text{Bin}(\text{mean } = \mu_{ijk}, \text{dispersion } = \phi_j)\\
        \log(\mu_{ijk}) &= \alpha_j + \log(fc)_j * k + L_i + M_j\\
        & \hspace{0.2in} + (\mathbf{X}_i^{RNA})^T \beta^{RNA} \\
        & \hspace{-0.3in} \alpha_j \sim N(0, \text{precision} = 10^{-3})\\
        & \hspace{-0.3in}\log(fc)_j \sim 
        \begin{cases}
            0 & \text{if } G_j = 1\\
            f^{RNA^+} & \text{if } G_j = 2\\
            f^{RNA^-} & \text{if } G_j = 3\\
        \end{cases}\\
        & \hspace{-0.3in} k = \mathbb{I}\{\text{individual } i \text{ has PD} \}.\\
        & \hspace{-0.3in} L_i = \log(\text{library size of individual i}).\\
        & \hspace{-0.3in} M_j = \log(\text{gene length of gene } j).\\
        & \hspace{-0.3in} \mathbf{X}_i^{RNA}: \text{ covariates for individual i}.\\
        & \hspace{-0.3in} \beta_q^{RNA} \sim N(0, \text{precision} = 10^{-3})\\
        & f^{RNA^+} \sim \text{half-piMOM}(t = t^{RNA^+}, r = 2)\\
        -& f^{RNA^-} \sim \text{half-piMOM}(t = t^{RNA^-}, r = 2)\\
        & \hspace{0.2in} t^{RNA^+} \sim \text{half-piMOM}(t = 0.05, r = 1)\\
        & \hspace{0.2in} t^{RNA^-} \sim \text{half-piMOM}(t = 0.05, r = 1)\\
        \log(\phi_j) \sim& N(\mu_0, \text{precision} = \tau_0)\\
        & \hspace{-0.3in} \mu_0 \sim N(0, \text{precision} = 10^{-2})\\
        & \hspace{-0.3in} \tau_0 \sim t^+ (\nu = 4)\\
    \end{align*}
    $Y_{i}^{GWAS}$ is $1$ if the $i^{th}$ individual in the GWAS study has PD and $0$ otherwise. 
    \begin{align*}
        Y_{i}^{GWAS} \sim& \text{Bern}(p_i)\\
        \text{logit}(p_i) &= \mathbf{z}_i^T \mathbf{\gamma} + (\mathbf{X}_i^{GWAS})^T \mathbf{\beta}^{GWAS}\\
        & \hspace{-0.3in} z_{ij} = \mathbb{I}\{ \text{Individual } i \text{ has a SNV in gene } j \}\\
        & \hspace{-0.3in} \gamma_j \sim
        \begin{cases}
            0 & \text{if } G_j = 1\\
            f^{GWAS^+} & \text{if } G_j = 2\\
            f^{GWAS^-} & \text{if } G_j = 3\\
        \end{cases}\\
        & \hspace{-0.3in} \mathbf{X}_i^{GWAS}: \text{ covariates for individual i}.\\
        & \hspace{-0.3in} \beta_q^{GWAS} \sim N(0, \text{precision} = 10^{-3})\\
        &f^{RNA^+} \sim \text{half-piMOM}(t = t^{GWAS^+}, r = 2)\\
        -&f^{RNA^-} \sim \text{half-piMOM}(t = t^{GWAS^-}, r = 2)\\
        & \hspace{0.2in} t^{GWAS^+} \sim \text{half-piMOM}(t = 0.05, r = 1)\\
        & \hspace{0.2in} t^{GWAS^-} \sim \text{half-piMOM}(t = 0.05, r = 1)\\
    \end{align*}
\end{multicols}
\vspace{-0.75in}
\begin{align*}
    G_j \sim& \text{ Multinomial}\big(n = 1, p_m = (\lambda_1, \lambda_2, \lambda_3)\big)\\
    & (\lambda_1, \lambda_2, \lambda_3) \sim \text{Dirichlet}(1,1,1)
\end{align*}
}

\section{2: A Three-component Mixture and Automatic Multiplicity Adjustment}

Our three-groups model adjusts for multiple comparisons by assigning a prior distribution to the vector of group assignment probabilities $\blambda$ that depends on the number of comparisons $J$.
The unknown group label $G_j$ for gene $j$ has a categorical distribution with probability vector denoted as $\blambda = (\lambda_1, \lambda_2, \lambda_3)\trans$, exchangeably across all genes $1, \ldots, J$.  
The common structure across the sub-models induces sharing of information among the disparate data types.  
Placing a prior distribution on $\blambda$ results in a penalty for large numbers of non-null genes that acts as an automatic adjustment for multiple comparisons, and hence results in few false positives. 
This automatic multiplicity adjustment is well known in the beta-binomial case \citep{scott-2010a} and is an alternative to the common practice of performing many independent tests and adjusting $p$-values \textit{post hoc} \citep[][e.g.]{benjamini-1995a}.

\begin{figure}[ht]
\centering
\includegraphics[width=0.35\textwidth]{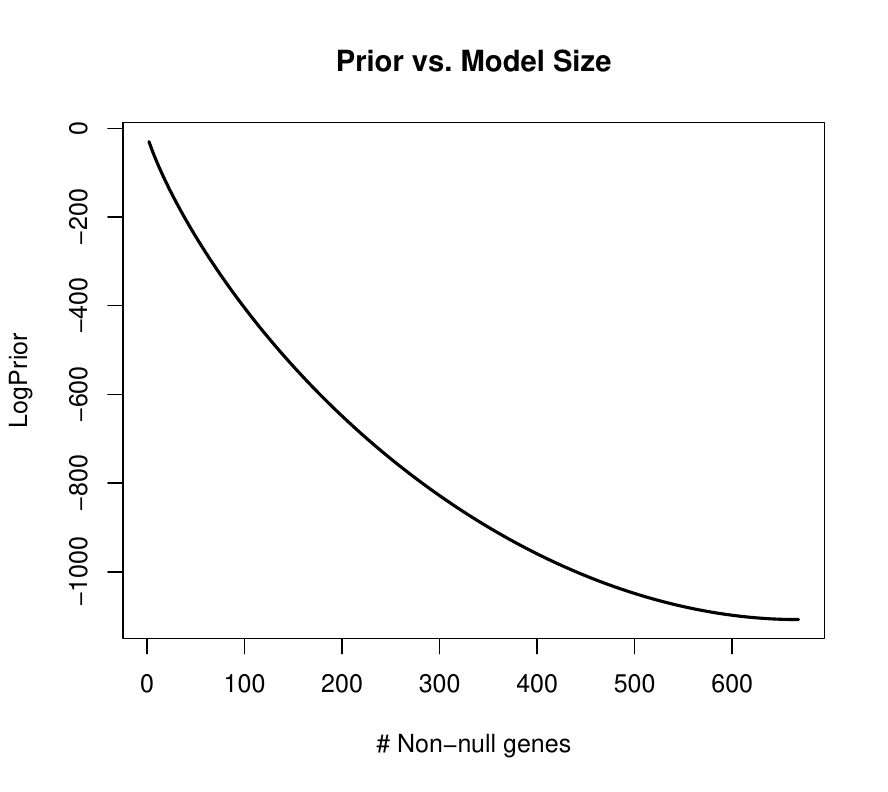}
\caption{The log prior probability mass function for models $\mathcal{M_\bG}$ corresponding to group assignments $\bG$, as a function of the number of non-null elements in $\bG$.  The penalty for including more non-null genes in the model is strong, resulting in an aggressive multiple comparisons adjustment.}
\label{fig:log-prior}
\end{figure}

The apportionment of prior mass on $\blambda$ depends on the number of comparisons $J$ and induces automatic multiplicity adjustment \citep{scott-2010a}. 
Let $\mathcal{M_\bG}$ be the model with group assignments $\bG = (G_1, \dots, G_J)$ where $G_j \in 1,2,3$ for all $j \in 1, \dots, J$. 
Assign the hyper-prior distribution $\blambda \sim \ddirich(\kappa\ba)$.  
Define $(j_1, j_2, j_3)\trans$ as the number of genes in groups 1, 2, and 3 as determined by $\bG$ (so that $j_1 + j_2 + j_3 = J)$. 
Then the prior probability mass function (pmf) for each model $\mathcal{M_\bG}$ given $\blambda$ is 
$
  P(\mathcal{M_\bG} \given \blambda) = \lambda_1^{J-j_2-j_3} \lambda_2^{j_2} \lambda_3^{j_3}
$ 
and the marginal prior pmf for each model $\mathcal{M_\bG}$ is
\[
  p(\mathcal{M}_\bG) = \int_{\blambda} P(\mathcal{M}_\bG \given 
  \blambda)\pi(\blambda)\intd \blambda = 
   \frac{\Gamma\left(\sum_{i=1}^3\kappa a_i \right)\prod_{i=1}^3 \Gamma(\kappa a_i + j_i)}
        {\Gamma\left(\sum_{i=1}^3\kappa a_i + j_i\right) \prod_{i=1}^3 \Gamma(\kappa a_i)}.
\]
For simplicity of visualization, assume that $j_2 = j_3$ so that $j_2 + j_3$ is the number of non-null genes and $j_1 = J - 2*j_2$ is the number of null genes. Also assume $\kappa=1$, and $\ba = (1,1,1)\trans$. If  $J = 1,000$ genes, then the log marginal prior pmf for each model $\mathcal{M_\bG}$, as a function of the number of non-null genes $j_2 + j_3$ is displayed in Figure~\ref{fig:log-prior}. 

Figure~\ref{fig:log-prior} shows a strong prior preference for models with few genes classified as non-null. 
In this simple example with $J = 1000$ genes, the model with no non-null genes is many thousands of times more likely \textit{a priori} than each model with 50 non-null genes. 
This strong prior preference results in posterior inference with few false positives.

To understand this prior preference, let the model size vector $(j_1, j_2, j_3)$ denote a model with $j_1$ genes in group 1, $j_2$ genes in group 2, and $j_3$ genes in group 3. 
Observe that there are ${{J}\choose{j_1, j_2, j_3}}$ models of size $(j_1, j_2, j_3)$ and thus when we add the prior mass of all models of size $(j_1, j_2, j_3)$ we get

\begin{align*}
    {{J}\choose{j_1, j_2, j_3}} \frac{\Gamma\left(\sum_{i=1}^3\kappa a_i \right)\prod_{i=1}^3 \Gamma(\kappa a_i + j_i)}
        {\Gamma\left(\sum_{i=1}^3\kappa a_i + j_i\right) \prod_{i=1}^3 \Gamma(\kappa a_i)} 
        &= {{J}\choose{j_1, j_2, j_3}} \frac{\Gamma\left(3\right)\prod_{i=1}^3 \Gamma(1 + j_i)}
        {\Gamma\left(3 + \sum_{i=1}^3j_i\right) \prod_{i=1}^3 \Gamma(1)} \\
        &= {{J}\choose{j_1, j_2, j_3}} \frac{2 j_1! j_2! j_3!}
        {(J+2)!} \\
        &= \frac{2}{(J+1)(J+2)}. 
\end{align*}
Note that (1) this doesn't depend on the model size $(j_1, j_2, j_3)$ and (2) this is the inverse of the size of the set of all possible model sizes from $J$ genes. 
Finally, note that there are many more models of size $(334, 333, 333)$ than models of size $(998, 1, 1)$ and thus each individual model of size $(998, 1, 1)$ is much more likely \textit{a priori} than each model of size $(334, 333, 333)$. 
Changing the Dirichlet hyper-prior values changes the specific mass but we find this explanation remains useful in understanding the sparsity inducing mechanism which relies on the prior exchangability. 

\subsection{Computational Efficiency}
For computational efficiency, we use a stick-breaking representation for the Dirichlet-multinomial portion of the model. That is, for $\blambda \sim \ddirich(\kappa\ba)$ the marginal distribution of each component follows a $\dbeta$ distribution (e.g. $\lambda_1 \sim \dbeta(\kappa a_1, \kappa a_2 + \kappa a_3 )$) which allows us to model the prior probabilities according to two $\dbeta$ distributions: one controlling the prior probability of being a null gene and the other controlling the conditional probability of being beneficial given the gene is not in the null group \citep[pg. 585]{gelman-bda3}. Correspondingly, the marginal and conditional distributions of categorical random variables are Bernoulli.

\section{3: RJMCMC comments}

Our numerical implementation used \texttt{NIMBLE}'s RJMCMC which required some additional customization. 
Genes in the null group have effect sizes of zero, but our numerical implementation requires beneficial and deleterious effect sizes to be updated at every MCMC iteration. 
In a traditional sampler, the effect sizes for genes in the null group are multiplied by an indicator of inclusion into one of the non-null groups in an intermediate step in each iteration.  
We used \texttt{NIMBLE}'s RJMCMC in order to cut out this unnecessary sampling and multiplication by zero. 
Unfortunately, the RJMCMC does not actually shrink the parameter space but instead fixes the excluded parameters at some pre-specified value (zero by default). 
In our case, this sets the effect sizes for null genes at zero. 

Updating the hyper-parameter values (e.g., $\tau$) requires a likelihood calculation which includes all of these fixed values. 
This fixing of effect sizes at zero results in errors because the non-local effect-size distributions have no mass at zero. 
To overcome this challenge, we wrote custom distributions which set the log-likelihood at zero for all effects which are set at zero by the RJMCMC toggler.  
This results in equivalent likelihood calculations to those from the true shrunk parameter space.

\section{4: Data Generation for Simulations}
\label{sec:simulations}

We simulated GWAS datasets from a standard logistic regression model. This scenario is idealistic in that we fit the exact data-generating response model; however the standard GWAS pipeline also fits this data-generating response model, so comparisons between our three-groups approach and standard GWAS are on an even footing. To generate the binary predictor variables, which indicate the presence (or absence) of any minor alleles, we simulated a gene-wise minor allele frequency from a $\dbeta(20, 35)$ distribution, which has most of its mass between $0.2$ and $0.5$ (i.e., for $X\sim \dbeta(20, 35)$, $P\{X \in (0.2, 0.5)\} = 0.977$). Then, conditional on the minor allele frequency, we simulated the binary predictor for each individual for that gene from a Bernoulli distribution with success probability equal to the minor allele frequency. Next, the probability of being in the treatment group is simulated from a Bernoulli distribution with success probability equal to the inverse logit of a linear combination of a normally distributed intercept and the predictors, multiplied by the fixed gene effects. The sign of the gene effects indicates group membership, with positive effects being deleterious and negative effects being beneficial. 

To make things as realistic as possible, we generated the RNA-seq data from subsets of a real RNA-seq dataset. This involved selecting 250 genes from an RNA-seq dataset and adding signal to the genes which should be included in the beneficial and deleterious groups. We started with the combined data from \citet{Pickrell_2010U} and \citet{Montgomery_2010} (\url{https://bowtie-bio.sourceforge.net/recount/}), as this dataset has a large number of biological replicates (129) and many genes (11,107). We added signal using a binomial thinning scheme, as in \citet{Gerard_2020}, using the referenced \texttt{R} package \texttt{seqgendiff}. This method allows the simulated data to retain the characteristics of real RNA-seq data and does not bias results towards one method or another.

\section{5: Simulation Comments and Additional Simulations}
\label{sec:app_sim}

\subsection{Metrics used to compare models}

We give a few more details of the metrics that were discussed in section 4 of the main text. 
To compute the logarithmic and Brier scores we let $p_j^{null}$ be the posterior probability that gene $j$ is null and let $x_j$ be an indicator which is $1$ if gene $j$ is null. The log score is a binomial log-likelihood and we negate it so that smaller values indicate better performance:
\[
\text{log-score} = - \frac{1}{J}\sum_{j=1}^J \bigg[x_j \log(p_j^{null}) + (1 - x_j) \log(1 - p_j^{null})  \bigg].
\]
The Brier score squares the distance between the prediction and the truth:
\[
\text{Brier-score} = \frac{1}{J}\sum_{j=1}^J (x_j - p_j^{null})^2.
\]
The receiver operating characteristic curve plots the false positive rate vs the true positive rate. 
We report the area under this curve as this is a one-number summary of the performance of the classifier. 
A random classifier will have an average AUC of 0.5 and a perfect classifier will have an AUC of 1. 
Our final comparison metric is the true positive rate at a fixed false positive rate. 
For this metric we determine the cutoff at which the average false positive rate (across 300 simulations) is at the fixed value (0.01 and 0.05) and compute the true positive rate for each simulation.

\subsection{$p$-value Combination}
\begin{figure}
    \centering
    \subfigure[Logarithmic Score (lower is better).]{\label{subfig:logscore_caucy}\includegraphics[width=0.45\textwidth, clip = TRUE, trim = 5 15 5 5 ]{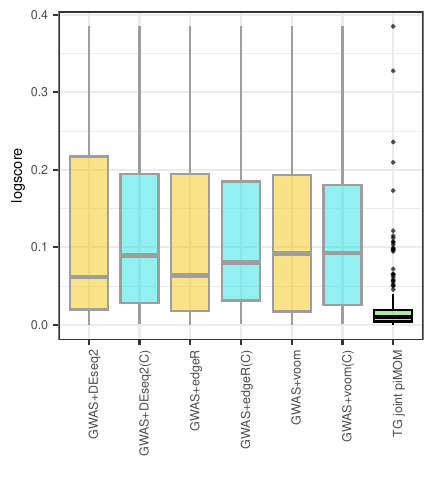}}
    \subfigure[Brier Score (lower is better).]{\label{subfig:Brierscore_cauchy}\includegraphics[width=0.45\textwidth, clip = TRUE, trim = 5 15 5 5 ]{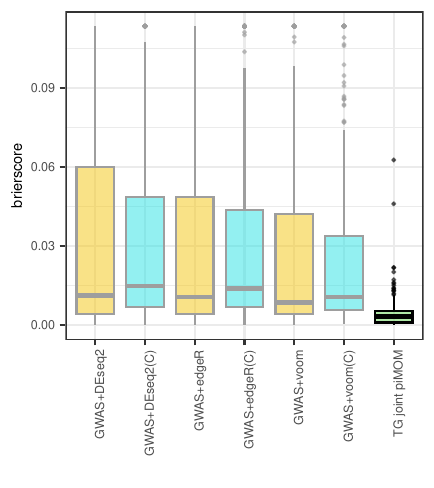}}
    \caption{Comparison of $p$-value combination methods. Yellow filled boxplots (i.e., without ``(C)") are results using Fisher's combination method and are the same as in the main manuscript. Blue filled boxplots (labeled with ``(C)") are results using the Cauchy combination method was used. Included is our model for comparison (filled in green).}
    \label{fig:cauchy_plots}
\end{figure}

The simulation in the main text compares our joint model with $p$-value combinations of competitors models. There are many $p$-value combination methods. The plots created for the main text used Fisher's $p$-value combination which computes the test statistic $p_{f,comb} = -2\sum_{i = 1}^k \log(p_i)$ where $p_i$ is the $p$-value for the $i^{th}$ test \citep{fisher2-1929a}. This test statistic has a chi-squared distribution under the null. A different method constructs the test statistic $p_{c, comb} = \sum_{i = 1}^k w_i \tan \{(0.5 - p_i) \pi\}$ which, under the null, can be approximated by a Cauchy distribution \citep{liu2020cauchy}. Figure~\ref{fig:cauchy_plots} displays the difference in simulation results for these two combination methods. These plots show that, while the particular scores are different, the character of the results is the same with both combination methods because, unlike our model, $p$-value combination methods cannot share information across datasets.

\subsection{lFDR Comments}
\begin{figure}
\centering
\subfigure[GWAS only]{\label{subfig:GWAS_new_lfdr_one}\includegraphics[width=0.45\textwidth, clip = TRUE, trim = 5 20 5 5  ]{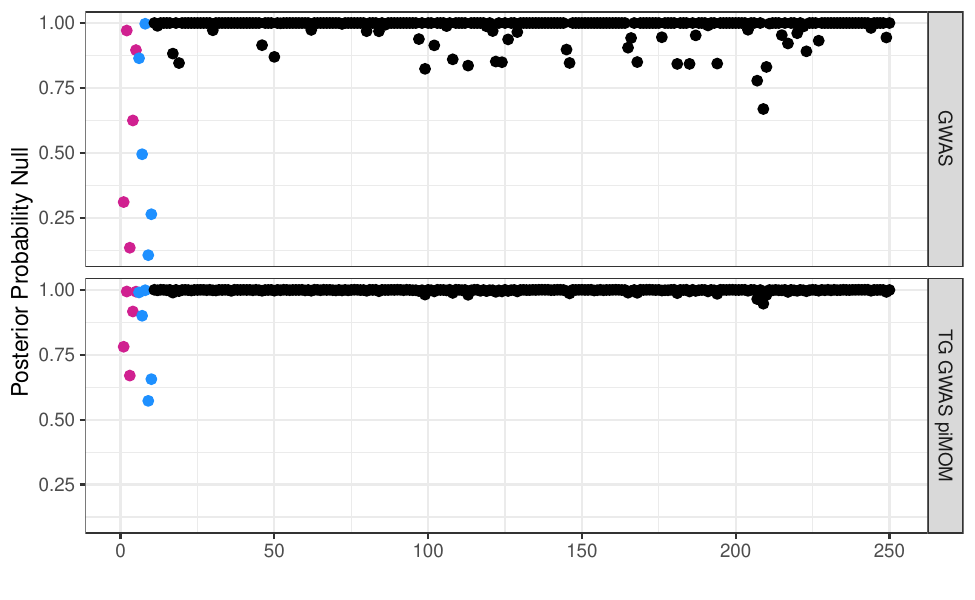}}
\newline
\subfigure[RNA-seq only]{\label{subfig:RNA_new_lfdr_one}\includegraphics[width=0.45\textwidth, clip = TRUE, trim = 5 20 5 5  ]{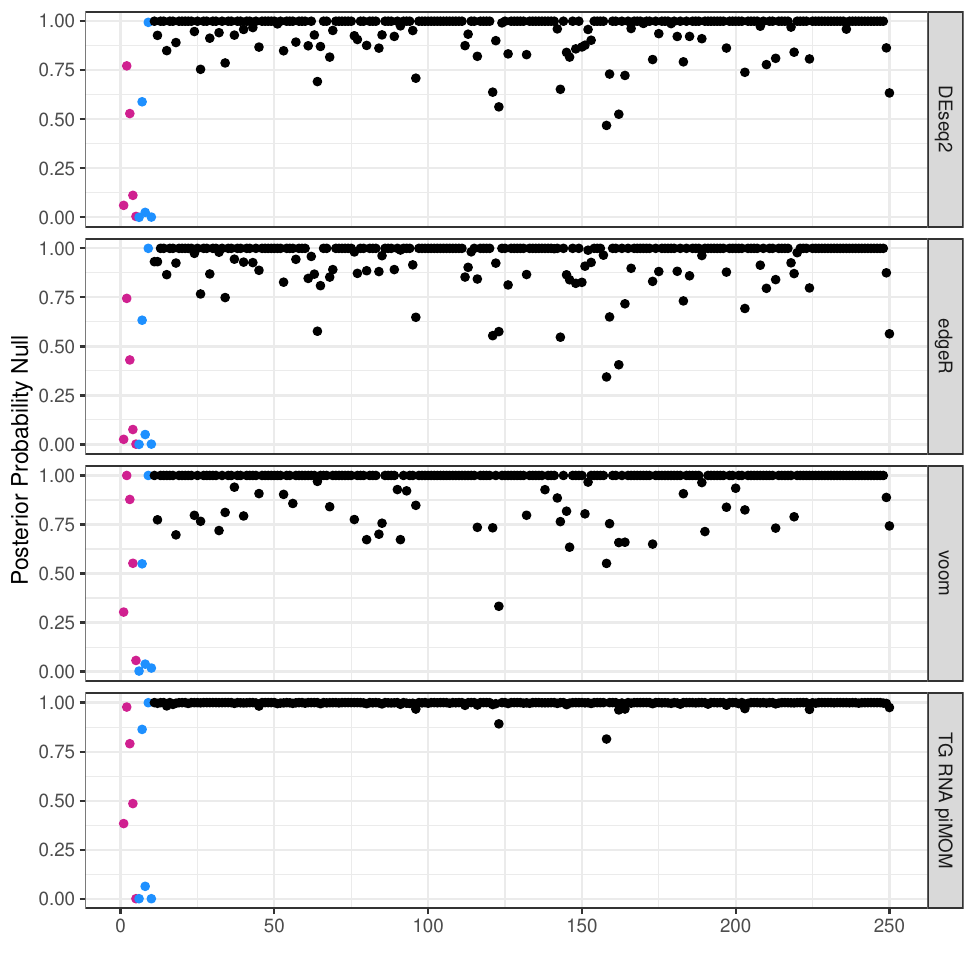}}
\subfigure[Combined models]{\label{subfig:Combined_new_lfdr_one}\includegraphics[width=0.45\textwidth, clip = TRUE, trim = 5 20 5 5  ]{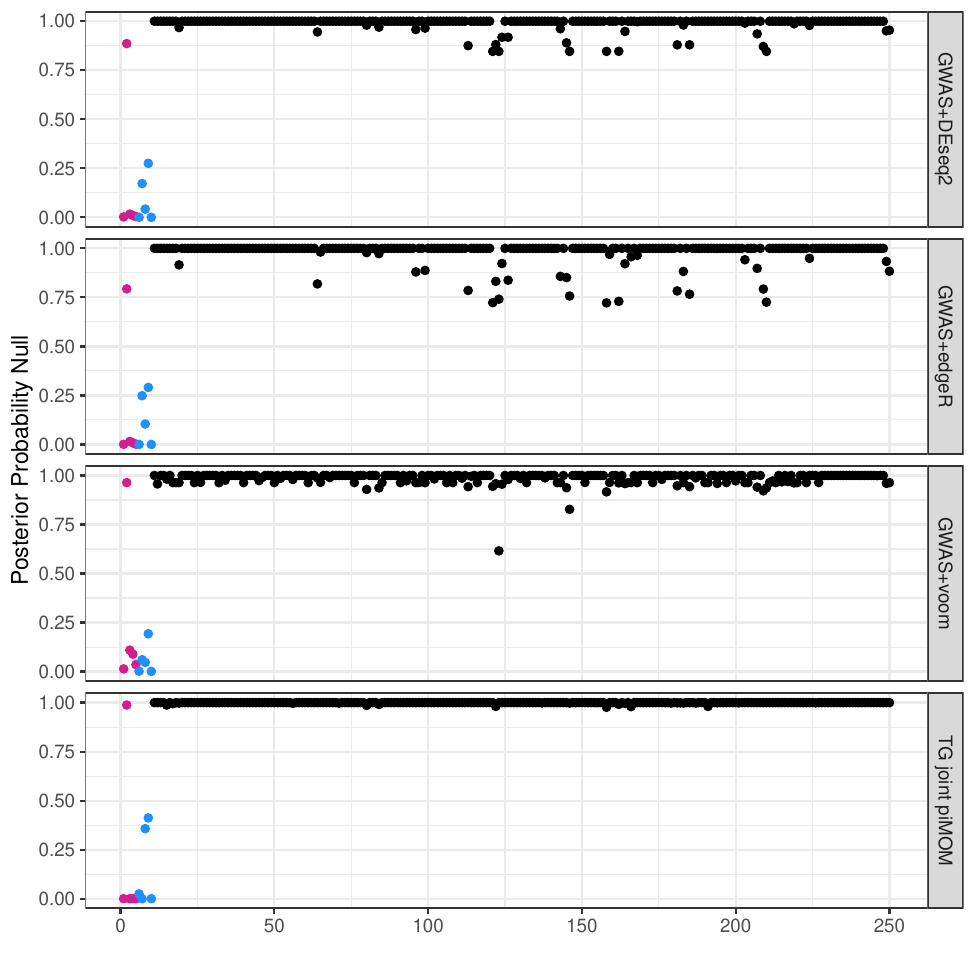}}
\caption{Posterior probability of inclusion in the null group for one simulation when true proportion of null genes is used in the LFDR computation for competitors.}
\label{fig:new_lfdr_one_run}
\end{figure}

\begin{figure}
    \centering
    \subfigure[Logarithmic Score (lower is better).]{\label{subfig:logscore_new_lfdr}\includegraphics[width=0.45\textwidth, clip = TRUE, trim = 5 15 5 5 ]{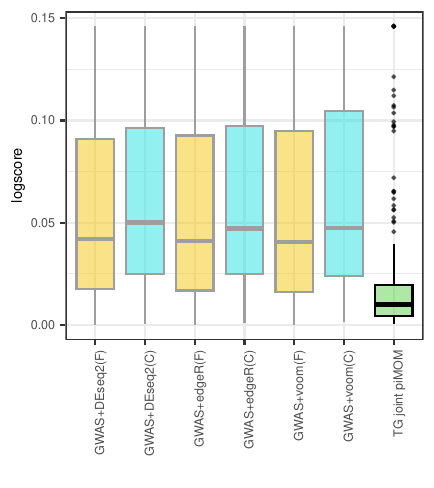}}
    \subfigure[Brier Score (lower is better).]{\label{subfig:Brierscore_new_lfdr}\includegraphics[width=0.45\textwidth, clip = TRUE, trim = 5 15 5 5 ]{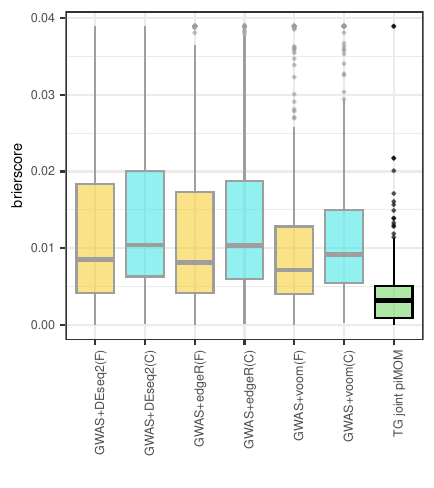}}
    \caption{Comparison of models when the true proportion of non-null genes is used in the computation of the LFDF for competitors methods. Labels with ``(C)" indicate that the Cauchy combination method was used (see previous section). Labels with ``(F)" indicate that Fisher's combination method was used.}
    \label{fig:new_lfdr_boxplots}
\end{figure}

Simulation results in the main text show the lFDR when computed using the default values in the \texttt{qvalue} package. 
Inherent in any estimation of lFDR is an estimation of $p_0$ which is the true proportion of null variables (genes in our case). 
When we allowed the software to estimate $p_0$ we noticed that the posterior probability plots for competitors had most of the null genes around some value that was less than one (e.g., in Figure 2b competitors models have null genes all around 0.75).
This strange artifact is do to poor estimation of $p_0$. 
We investigate the effect of this estimation by specifying the true $p_0$ in this section. 
Figure~\ref{fig:new_lfdr_one_run} shows plots analogous to Figures 1 and 2 in the manuscript. 
We note that now, in the competitors models, the majority of the null genes have posterior probability of inclusion in the null group of 1 which is similar to our model. 
While the poor estimation of $p_0$ alters plots of a single run quite a bit, we note that the character of the full results does not change.  
Figure~\ref{fig:new_lfdr_boxplots} plots the boxplots of logscores and Brier scores from the same simulation as was done in the main text but with the lFDR computed with the true $p_0$ with both Fisher's $p$-value combination and the Cauchy combination method. 
This additional information (which is not available in a real data context) noticeably improves the scores of the competitors' methods but our model still outperforms the competitors in these metrics.

\subsection{Benefits of the Dirichlet hyperprior}
\begin{figure}
    \centering
    \subfigure[Logarithmic Score (lower is better).]{\label{subfig:logscore_fixedeff}\includegraphics[width=0.45\textwidth, clip = TRUE, trim = 5 15 5 5 ]{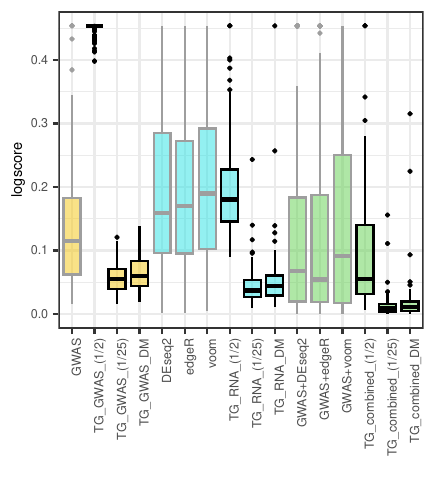}}
    \subfigure[Brier Score (lower is better).]{\label{subfig:Brierscore_fixedeff}\includegraphics[width=0.45\textwidth, clip = TRUE, trim = 5 15 5 5 ]{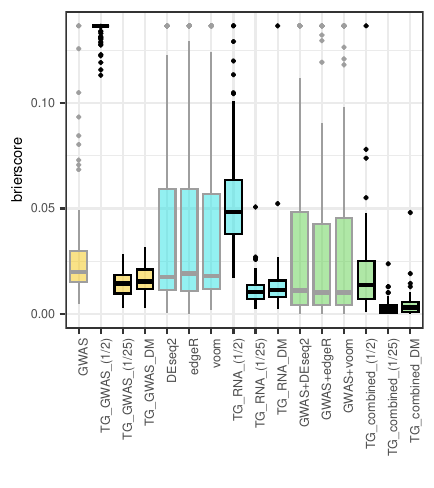}}
    \caption{Comparison of models. The three TG models all have piMOM hyperpriors on gene effects but differ in the structure for inclusion in non-null groups. The first two, which are labeled with ``\ldots(1/2)" and ``\ldots(1/25)", have probabilities of being included in one of the non-null groups fixed at 1/2 and 1/25. The models labeled with ``DM" have the Dirichlet-Multinomial prior setup to adaptively learn these probabilities.}
    \label{fig:fixed_inclProbs}
\end{figure}

The inclusion of the Dirichlet-Multinomial hyperprior complicates our model and thus it may be useful to breifly explore the benefits. We do so by performing simulations where the probability of inclusion in a non-null group is fixed at 1/2 and 1/25 where 1/25 is the true inclusion probability. Results from these simulations are in Figure \ref{fig:fixed_inclProbs}. These figures demonstrate that performance is similar when the inclusion probability is fixed at the truth but is substantially worse when inclusion probabilities are fixed at non-informative values. This differs from the Dirichlet-Multinomial structure where our Dirichlet hyperprior is uniform (and thus non-informative) and yet is adaptive to differing sample sizes. The adaptive nature of the Dirichlet-Multinomial structure for automatic multiplicity correction is what allows us to perform simulations with a small number of genes and suggest that performance will be similar when many more genes are included. These plots also demonstrate that, even when the inclusion probabilities are much too loose, there are benefits to borrowing strength across data types. 

\subsection{Hyperprior sensitivity}

We ran several additional versions of the three-groups model and report results here. The main text includes an asymmetric non-local three-groups model which has half-piMOM effect size distributions ($f^{(m)^-}$ and $f^{(m)^+}$) with separate half-piMOM hyper-priors placed on the scaling parameter $\tau$. This model is compared to a symmetric local three-groups model which has half-normal effect size distributions that are fixed. In this appendix we compare these models to another asymmetric non-local model that instead has inverse gamma hyper-priors on the $\tau$ parameters, a symmetric non-local model that has fixed $\tau$ values, and an asymmetric local model that has separate inverse gamma hyper-priors on the means and separate half-piMOM hyper-priors on the standard deviations of the half-normal effect size priors. Table \ref{tab:table_appendix_models_sim} demonstrates the distinctions between these models. We do not include results from an asymmetric local model which had separate inverse gamma hyper-priors on the means and standard deviations as they were uniformly worse and made the plots unreadable.

\begin{table}
\centering
\caption{Additional three-groups models}
\begin{tabular}{lrrrr}
Label &  Effect size & Hyper-parameters & Is local? & Is symmetric? \\ 
 & distributions & & & \\
 \hline
nonL piMOM & half-piMOM & $\tau \sim$ half-piMOM & non-local & asymmetric \\
nonL invG & half-piMOM & $\tau \sim$ inverse Gamma & non-local & asymmetric \\
nonL fixed & half-piMOM & fixed & non-local & symmetric \\
local piMOM & half-Normal & $\mu \sim$ inverse Gamma & local & asymmetric \\
 & & $\sigma \sim$ half-piMOM & & \\
local fixed & half-Normal & fixed & local & symmetric \\
   \hline
\label{tab:table_appendix_models_sim}
\end{tabular}
\end{table}

\begin{figure}
\centering
\subfigure[Logarithmic Score (lower is better).]{\label{subfig:logscore_app}\includegraphics[width=0.45\textwidth, clip = TRUE, trim = 5 15 5 5 ]{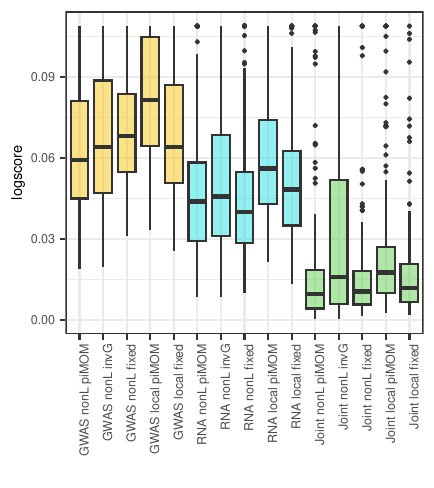}}
\subfigure[Brier Score (lower is better).]{\label{subfig:Brierscore_app}\includegraphics[width=0.45\textwidth, clip = TRUE, trim = 5 15 5 5 ]{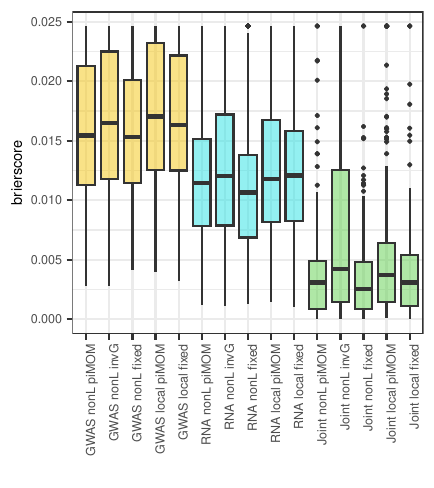}}
\subfigure[Area Under ROC Curve (higher is better).]{\label{subfig:AUC_app}\includegraphics[width=0.45\textwidth, clip = TRUE, trim = 5 15 5 5 ]{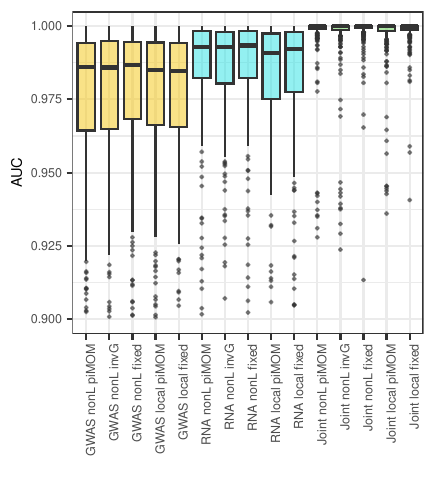}}
\subfigure[True positive rate at cutoffs]{\label{subfig:tpr_app}\includegraphics[width=0.45\textwidth, clip = TRUE, trim = 5 15 5 5 ]{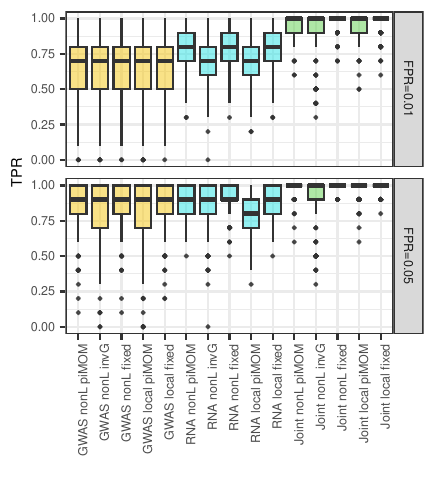}}
\caption{Boxplots of logarithmic scores \subref{subfig:logscore_app}, Brier scores \subref{subfig:Brierscore_app}, area under receiver operating characteristic curve \subref{subfig:AUC_app} computed from posterior probability of inclusion in the null group, and the true positive rates (i.e. power) computed at classification cutoffs that result in mean false positive rates of $0.01$ (top panel) and $0.05$ (bottom panel) in \subref{subfig:tpr_app} as in Figure 4 in the main text. 
The first term in the labels (``GWAS", ``RNA", ``Joint") indicates which sub-models are run. 
The second term in each label (``nonL" or ``local") indicates whether the gene effect size distributions are non-local or local. 
the last term (``piMOM", ``invG", ``fixed") indicates the hyper-prior distribution on the scale parameter of the gene effect size distributions (e.g., $t^{RNA+}$ in Appendix \ref{sec:full_model}).} 
\label{fig:app_sim_fig}
\end{figure}

Figure \ref{fig:app_sim_fig} demonstrates a range of choices (for modeling the effect sizes) which provide comparable results.
This range illustrates that most of the benefit of our model comes from the borrowing of strength across data types using three-groups structure.

While there are many similarities between these plots, we think that it is worth highlighting one important difference. The non-local model with inverse gamma hyper-priors required tuning of the hyper-parameter values; we had to move the mass away from zero to ensure the posterior distributions for $\tau$ were not nearly zero. This suggests to us that the sparsity invoked by the automatic multiplicity correction of the Dirichlet-categorical inclusion scheme can become swamped by many very small empirical effect sizes. In other words, some engineering is required to ensure that the effect size distributions do not become additional ``spikes" at trivially small values.

\begin{figure}
\centering
\subfigure[Logarithmic Score (lower is better).]{\label{subfig:logscore_df}\includegraphics[width=0.45\textwidth, clip = TRUE, trim = 5 15 5 5 ]{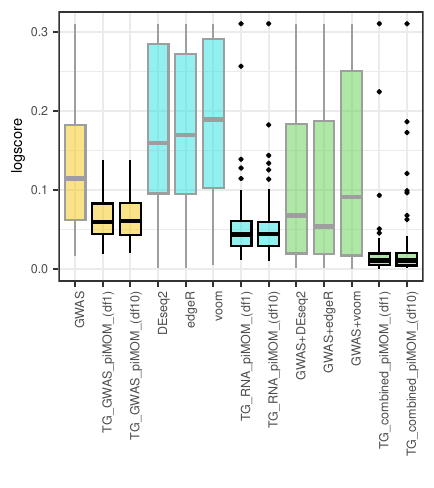}}
\subfigure[Brier Score (lower is better).]{\label{subfig:Brierscore_df}\includegraphics[width=0.45\textwidth, clip = TRUE, trim = 5 15 5 5 ]{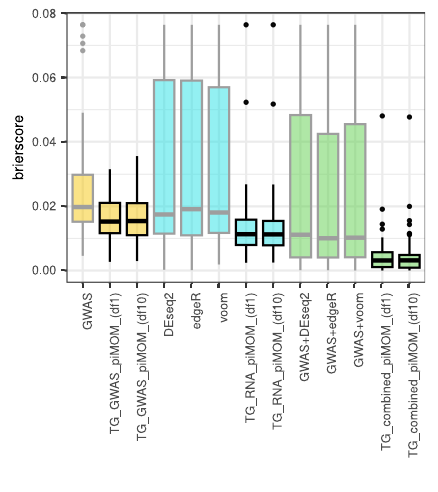}}
\caption{Boxplots of logarithmic scores \subref{subfig:logscore_df} and Brier scores \subref{subfig:Brierscore_df}. 
The last term in the ``TG" models indicates the degrees of freedom placed on the hyper-prior for the dispersion parameter of the RNA-seq model.} 
\label{fig:df_sim_fig}
\end{figure}

Many hyperparameters in the model are set to make the hyperpriors approximately flat in the region of interest (e.g., $N(0, 10^{-3})$) or tuned to the region that is scientifically meaningful (e.g., the half-piMOM excludes very small values and the mass is concentrated away from extremely large odds ratios). 
An exception to this is the hyperparameter which determines the degrees of freedom of the hyper-prior placed on the precision for the log-Normal prior of the negative binomial dispersion. 
We performed an additional set of simulations to test the sensitivity to this hyperparameter. 
The hyper-prior is a half-$t$ distribution with degrees of freedom $\nu = 4$ in all results in the manuscript. Figure \ref{fig:df_sim_fig} displays results of simulation studies with $\nu = 1$ and $\nu = 10$. 
The results appear to be insensitive to this change in hyper-prior, with MCMC output seeming to differ only up to Monte Carlo error.

\subsection{Genes with effects in only one experiment type}

\begin{figure}
    \centering
    \includegraphics[width=0.85\linewidth]{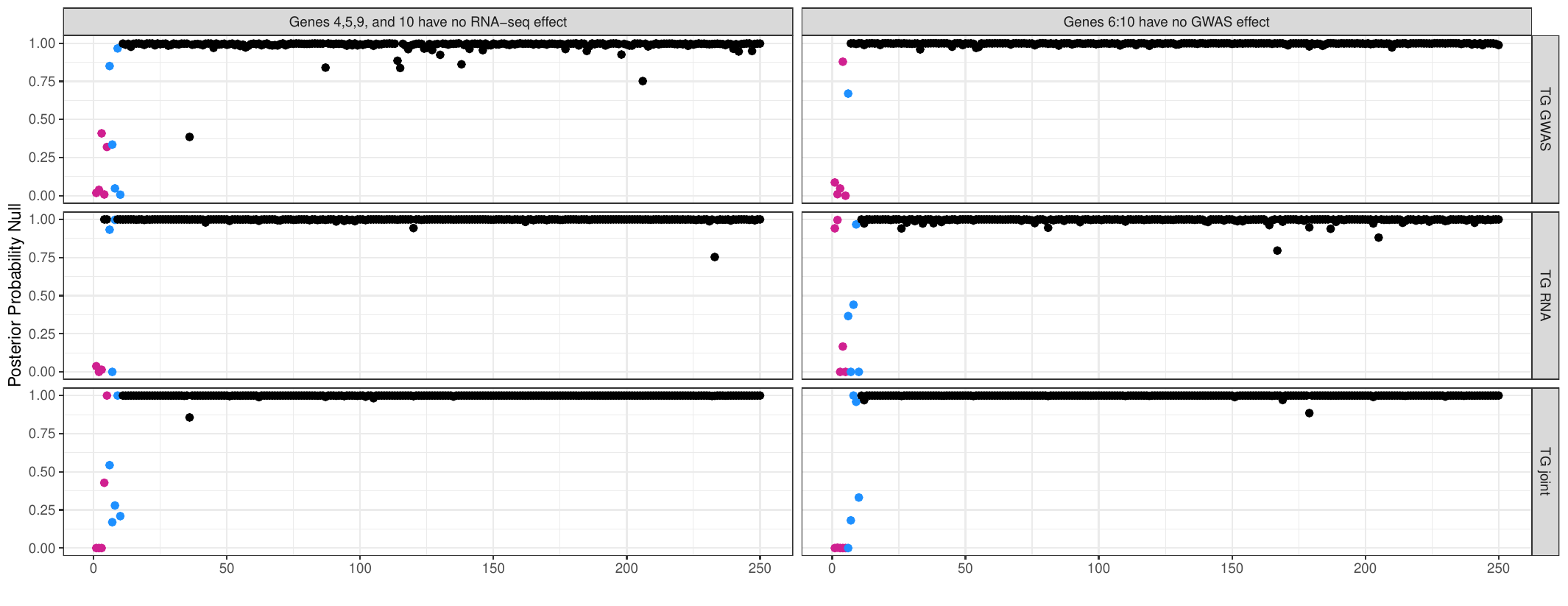}
    \caption{Simulation results with genes that have effect in a single experiment type. Each column demonstrates our model's performance on one simulated set of 250 genes. In the first column, the first 10 genes are generated with an effect in the GWAS branch and six of these first 10 are generated with an effect in the RNA-seq branch. In the second column, six of the first 10 genes are simulated with an effect in the GWAS branch and the first 10 are simulated with an effect in the RNA-seq branch.}
    \label{fig:noEffect}
\end{figure}

Our model was designed to identify genes with weak signal in multiple data types through borrowing of strength. The simulation in the manuscript demonstrates that the model performs better than competitors at this task. 
We think that it is biologically reasonable that some some genes are associated with disease status but have signal in a subset of the included data types (e.g., a gene may exhibit differential expression between the PD and control groups but the minor allele proportions may be the same). 
We explore whether our model can detect these genes through an additional simulation study. 
In this set of simulations, data are generated such that some genes are non-null in one experiment type only. 
Figure \ref{fig:noEffect} displays the results of two simulations with 250 genes where six genes were simulated with effects in both branches and an additional four genes are simulated with an effect only in one branch. 
The first two rows of the figure demonstrate the sub-models performance which indicates that genes without effects are not identified in these sub-model runs. 
The joint model (bottom row) indicates two important features of our model; first, that some genes that have signal in one model only are identified in the joint model and, second, that some genes that have relatively weak signal in one of the sub-models are identified by the joint model. 
Boxplots of comparison metrics like in Figure \ref{fig:app_sim_fig} tell the same story as in other simulations; our joint model performs better than p-value combinations of competing models. 

\subsection{Missing genes}
\label{subsec:missing_genes}
Our model seamlessly handles genes which are missing in one of the data types in a standard, fully Bayesian, fashion that treats missing data as parameters to be estimated. 
Here the missing values are simply drawn from their posterior predictive distributions within the MCMC sampler. 
This imputation approach means that any signal from genes that are missing in one data type must be captured from the data in a single study. 
This missingness makes it more challenging for the model to identify these non-null genes. 

\begin{figure}
    \centering
    \includegraphics[width=0.95\linewidth]{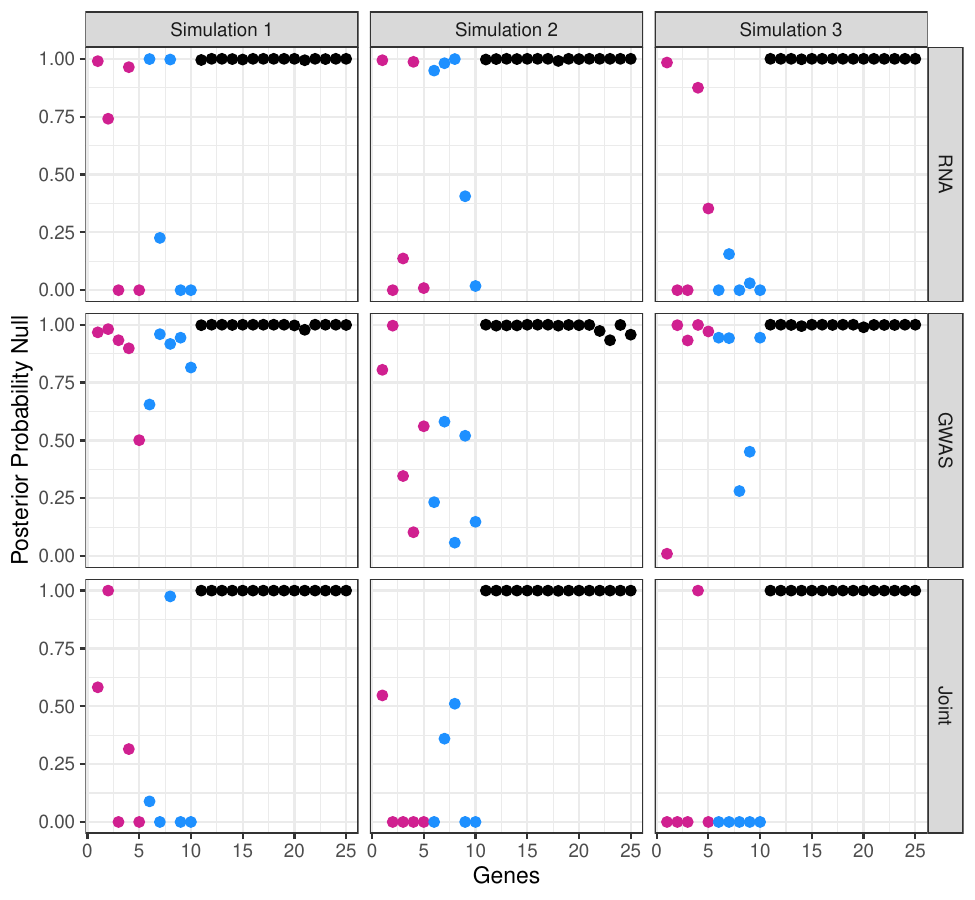}
    \caption{Simulation results with a gene that only has data in one experiment type. Each column demonstrates our model's performance on one simulated set of 100 genes (only 25 are shown for clarity). In all three simulations gene number 1 is missing from the RNA-seq dataset.}
    \label{fig:missing}
\end{figure}

The challenge of identifying non-null genes which are missing in one data type is not unlike the challenge in identifying genes which are simulated with effects in only one data type, a scenario which was discussed in the previous section. 
In this section we briefly show that our model is still capable of identifying non-null genes when the data are missing in one experiment. 
Figure~\ref{fig:missing} shows the posterior probabilities of inclusion in the null group from three simulations which differ only in the sampling variability in the data generation. 
In each of these simulations gene number 1 is missing in the RNA-seq dataset (we generated data just as we had for other experiments but replaced this gene's data with NAs). 
The posterior probability of inclusion in the null group is 1.0 in the RNA-only model. 
In the first two simulations we see that the small signal that is detectable in the GWAS-only model is enough so that the posterior probability of inclusion in the null group is near 0.5 in the joint model. 
In the third experiment (column 3) we see that this gene has a strong signal in the GWAS only model and thus it is clearly identified as non-null by the joint model.

\subsection{Simulation Run times}
\begin{figure}
    \centering
    \includegraphics[width=0.85\linewidth]{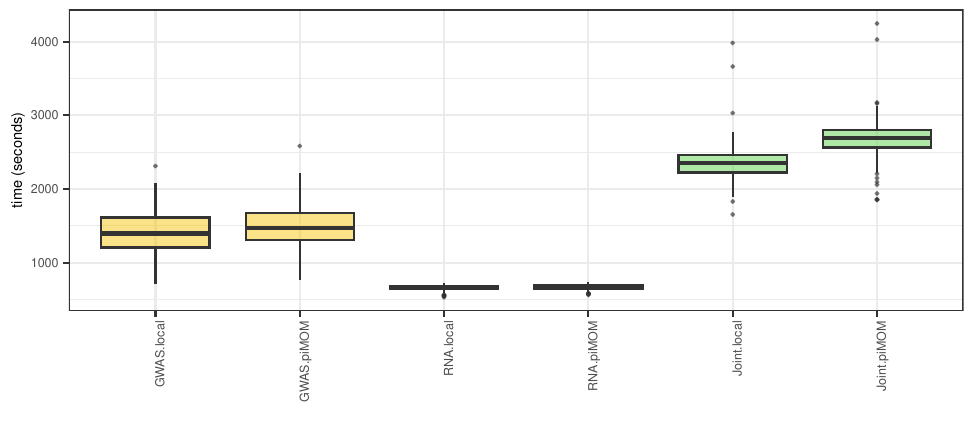}
    \caption{Boxplots which show the computation time (in seconds) of each of our models for each of the 300 simulation in the main text.}
    \label{fig:run_times}
\end{figure}

Computation is always a challenge with MCMC methods. In this section we share boxplots of run times for the simulation in the main text (Figure~\ref{fig:run_times}). 
This figure shows that the RNA only model is quite fast when compared to the GWAS model. We note that the piMOM models are slower than the local models as we had to write custom distributions instead of taking advantage of previous work that speeds up computation with the normal distribution.

\subsection{Sensitivity to Conditional Independence Assumption}
Our model implicitly assumes that RNA counts are independent conditional on PD status. This is more lenient than the full independence assumed in the conventional models used for comparison. It is plausible that both independence assumptions are violated through, for example, differences in the cellular composition of each sample. We assessed our model through simulations with RNA-seq data which are generated by adding signal to a real RNA-seq dataset. We expect these synthetic data to reflect potential violations of our assumptions that may also be present in the real data. In the following, we simulate dependent RNA-seq data to assess the sensitivity of our model to known violations of conditional independence.

\begin{figure}
    \centering
    \subfigure[Surrogate variables]{\label{subfig:surrogate_logscore}\includegraphics[width=0.45\textwidth, clip = TRUE, trim = 5 15 5 5 ]{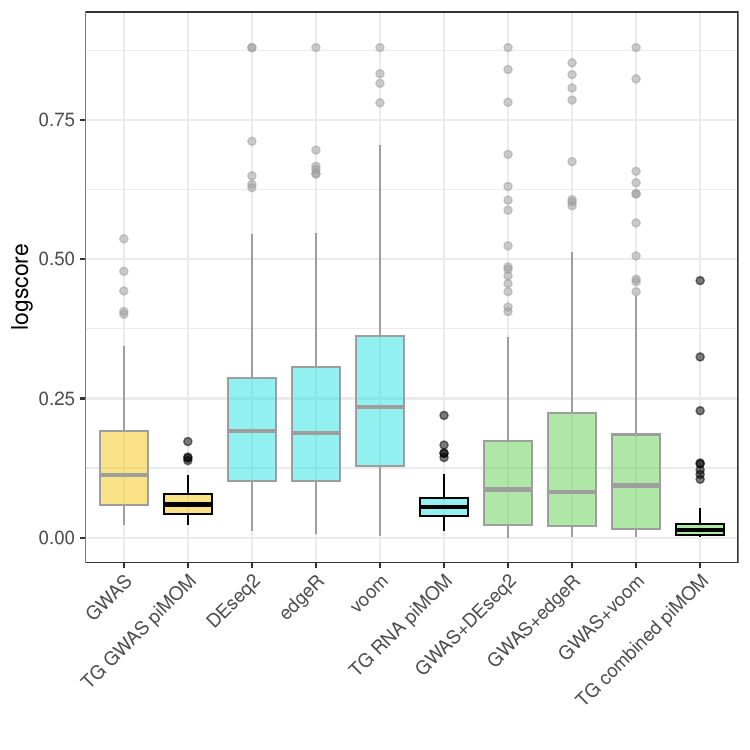}}
    \subfigure[Two-stage, moderate effect]{\label{subfig:twostage_14}\includegraphics[width=0.45\textwidth, clip = TRUE, trim = 5 15 5 5 ]{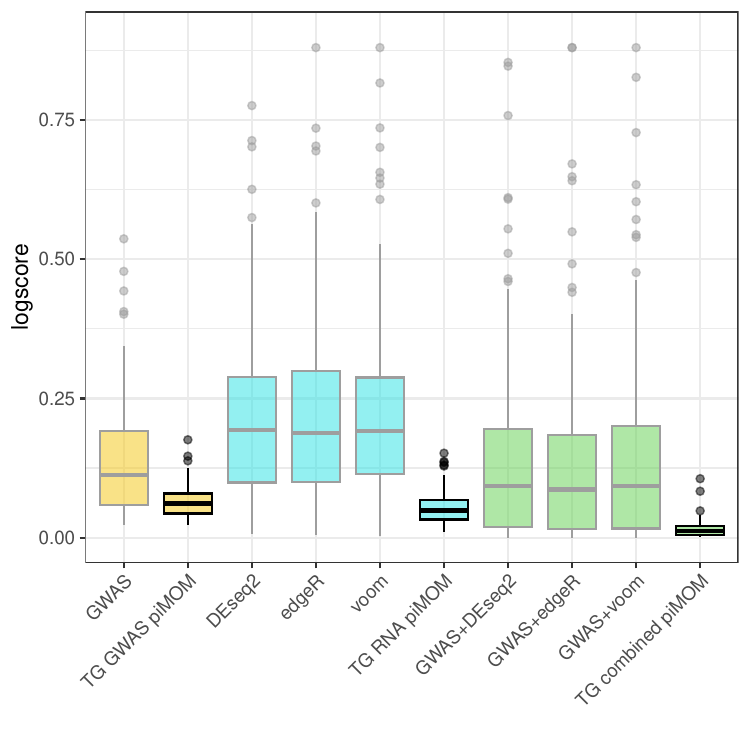}}
    \\
    \subfigure[Two-stage, large effect]{\label{subfig:twostage_5}\includegraphics[width=0.45\textwidth, clip = TRUE, trim = 5 15 5 5 ]{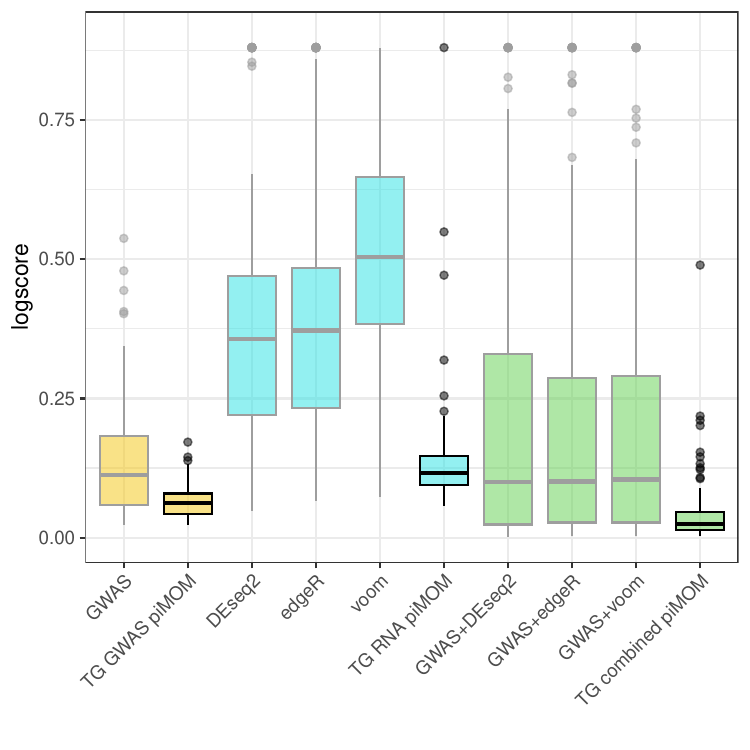}}
    \subfigure[Two-stage, extreme effect]{\label{subfig:twostage_75}\includegraphics[width=0.45\textwidth, clip = TRUE, trim = 5 15 5 5 ]{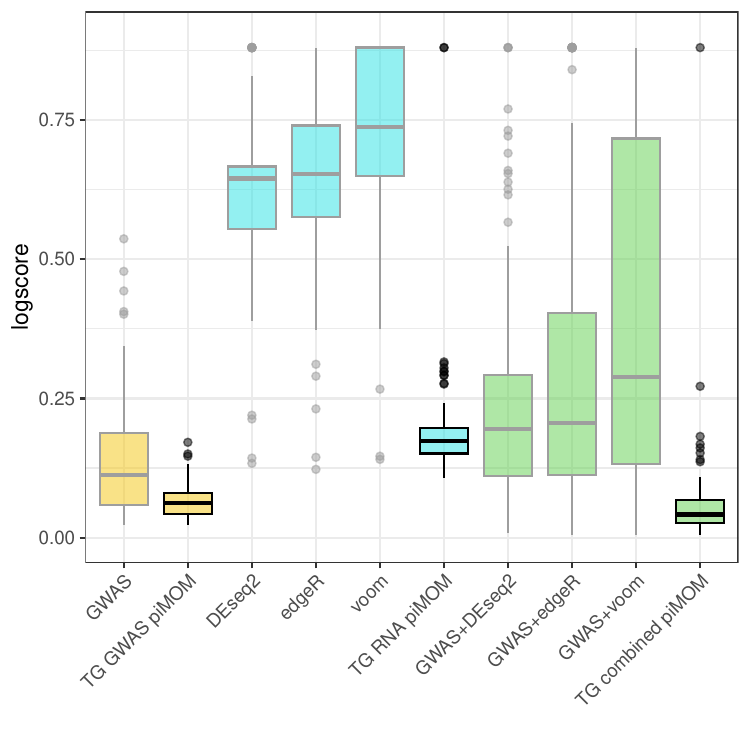}}
    \caption{Simulation results when the assumption of conditional independence of RNA counts is broken. RNA-seq data for Subfigure~\ref{fig:breakIndep_plot}\subref{subfig:surrogate_logscore} were simulated using the surrogate variables feature in \texttt{seqgendiff}. Synthetic data for simulations in the other three plots were made by adding spurious signal to the underlying RNA-seq counts before adding the PD signal. The "moderate" effect is the same magnitude as the PD signal.}
    \label{fig:breakIndep_plot}
\end{figure}

The package used for RNA-seq data generation in our simulations (\texttt{seqgendiff}) has a method to add surrogate variables. The inclusion of these surrogates imposes dependence across genes. The surrogate variable information is ignored in the analysis of the data. Subfigure~\ref{fig:breakIndep_plot}\subref{subfig:surrogate_logscore} shows results of one set of simulations in which the RNA-seq data are generated with three surrogates of differing strength (weaker, equal to, and stronger than the PD signal) and prevalence (affecting 15\% to 50\% of the genes).

A second method for generating data which breaks the conditional independence assumption is to add signal to the RNA-seq data twice; the first signal is a synthetic covariate that will be ignored in the subsequent analysis and the second second signal is the PD signal. This introduces dependence between genes that the model does not account for. To accomplish this we perform the binomial thinning of \citet{Gerard_2020} twice. This two-stage approach ensures that the PD signal is added to RNA-seq data which are known to be dependent. We performed simulations using this two-stage data generation with three different levels of effects for the simulated covariate. The lowest level of dependence involves signal that is the same strength as the gene-effects for the genes with differential expression in PD (a coefficient of 1.4 which is a multiplicative effect of 0.48). The other two levels of dependence include coefficients of 5 and 75 (multiplicative effects of 2.32 and 6.23 respectively). Results in Figure \ref{fig:breakIndep_plot} demonstrate that with increasing dependence comes decreasing performance (larger log-scores) of all methods that include the RNA-seq data (the GWAS results are identical as those data are the same in all simulations).

The results from these simulations with known violations of assumptions indicate that the performance of all models degrades as the strength of the conditional dependence strengthens. This set of simulations suggests that a moderate level of dependence does not degrade performance dramatically as Subfigures~\ref{fig:breakIndep_plot}\subref{subfig:surrogate_logscore} and ~\ref{fig:breakIndep_plot}\subref{subfig:twostage_14} are quite similar to simulation results without intentionally breaking the conditional independence. Furthermore, our three-groups RNA-seq model appears more robust to violations of conditional independence than conventional models. Finally, we highlight that the increased power of our model, demonstrated in previous simulations, remains across these increasing violations of the assumption of conditional independence.

\subsection{An alternative joint model with summary statistics}
In this section we compare our full joint three-groups model to a version of the model adapted for summary statistics. Our three-groups model analyzes the data from a GWAS and RNA-seq experiment. This is unlike meta-analysis methods which analyze summary statistics computed from these experiments. Summary statistics are convenient because they are often more broadly available than the raw data and of smaller dimensions, so the models that use them as inputs are less computationally demanding. These models may be less powerful than models developed for the raw data.

We adapted our three-groups model for use with summary statistics. We assume that the gene effects estimated from conventional methods (we used GWAS and DEseq2) are approximately Normal random variables that are centered on the true gene effect with standard errors approximated by the respective conventional methods. Let $\hat{\beta}_j^{RNA}$ be the $j$th estimated gene effect and $\text{se}_j^{RNA}$ be the associated standard error as estimated by one of the conventional RNA-seq models. Let $\hat{\beta}_j^{GWAS}$ be the $j$th estimated gene effect and $\text{se}_j^{GWAS}$ be the associated standard error as estimated by one-at-a-time logistic regression. Our summary model is:

\small{
\vspace{-20pt}
\begin{multicols}{2}
    \noindent
    \begin{align*}
        \hat{\beta}_j^{RNA} \sim &N(\beta_j^{RNA}, \text{se}_j^{RNA})\\
        \beta_j^{RNA} & \sim 
        \begin{cases}
            0 & \text{if } G_j = 1\\
            f^{RNA^+} & \text{if } G_j = 2\\
            f^{RNA^-} & \text{if } G_j = 3\\
        \end{cases}\\
        f&^{RNA^+} \sim \text{half-piMOM}(t = t^{RNA^+}, r = 2)\\
        -f&^{RNA^-} \sim \text{half-piMOM}(t = t^{RNA^-}, r = 2)\\
        & \hspace{0.15in} t^{RNA^+} \sim \text{half-piMOM}(t = 0.05, r = 1)\\
        & \hspace{0.15in} t^{RNA^-} \sim \text{half-piMOM}(t = 0.05, r = 1)\\
    \end{align*}
    \begin{align*}
        \hat{\beta}_j^{GWAS} \sim &N(\beta_j^{GWAS}, \text{se}_j^{GWAS})\\
        \beta_j^{GWAS} & \sim 
        \begin{cases}
            0 & \text{if } G_j = 1\\
            f^{GWAS^+} & \text{if } G_j = 2\\
            f^{GWAS^-} & \text{if } G_j = 3\\
        \end{cases}\\
        f&^{GWAS^+} \sim \text{half-piMOM}(t = t^{GWAS^+}, r = 2)\\
        -f&^{GWAS^-} \sim \text{half-piMOM}(t = t^{GWAS^-}, r = 2)\\
        & \hspace{0.15in} t^{GWAS^+} \sim \text{half-piMOM}(t = 0.05, r = 1)\\
        & \hspace{0.15in} t^{GWAS^-} \sim \text{half-piMOM}(t = 0.05, r = 1)\\
    \end{align*}
\end{multicols}
\vspace{-0.75in}
\begin{align*}
    G_j \sim& \text{ Multinomial}\big(n = 1, p_m = (\lambda_1, \lambda_2, \lambda_3)\big)\\
    & (\lambda_1, \lambda_2, \lambda_3) \sim \text{Dirichlet}(1,1,1).
\end{align*}
}

This summary statistics version of the model is considerably simplified and has very few observations for parameter estimation. The true gene effects $\beta_j^{m}$ (where $m$ is either RNA or GWAS) each have a single observation of the estimated gene effect $\hat{\beta}_j^{m}$. The group membership $G_j$ must be estimated from the two estimated gene effects ($\hat{\beta}_j^{RNA}$ and $\hat{\beta}_j^{GWAS}$) and their associated standard errors. We expect that the summary model would perform better in cases where there are multiple experiments of each type. In addition, an analysis with multiple experiments in each data type would benefit even more from the computational advantages of the summary model. 

\begin{figure}
    \centering
    \subfigure[Simulated as in manuscript]{\label{subfig:logscore_summary_stat}\includegraphics[width=0.45\textwidth, clip = TRUE, trim = 5 15 5 5 ]{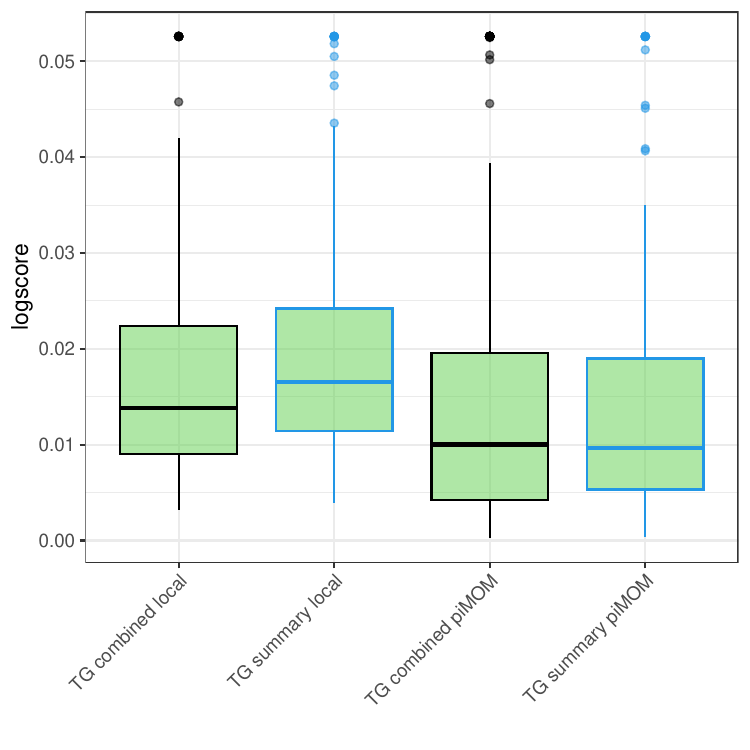}}
    \subfigure[Simulated with missing data]{\label{subfig:missing_summary_stat}\includegraphics[width=0.45\textwidth, clip = TRUE, trim = 5 15 5 5 ]{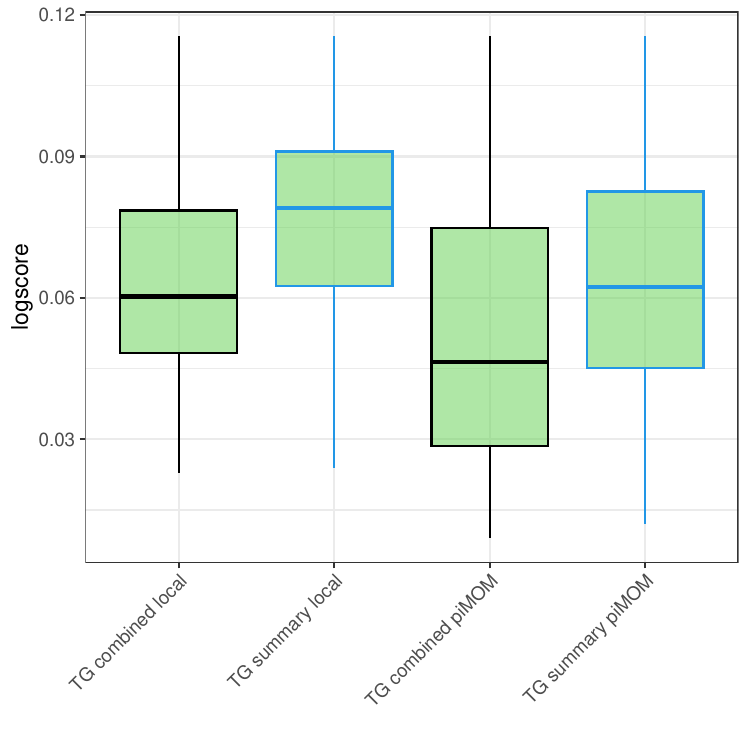}}
    \caption{Simulation results comparing our full three-groups model and our three-groups model adapted to summary statistics. Subfigure~\ref{fig:summary_stats_plot}\subref{subfig:logscore_summary_stat} has the same results for our full three-groups model as the main results in the manuscript. Subfigure~\ref{fig:summary_stats_plot}\subref{subfig:missing_summary_stat} is from simulations where a few genes have been generated with data only in the GWAS branch of the study.}
    \label{fig:summary_stats_plot}
\end{figure}

Figure \ref{fig:summary_stats_plot} displays results from two simulation studies that compare our raw-data three-groups model to this summary statistics model. The first study (Subfigure~\ref{fig:summary_stats_plot}\subref{subfig:logscore_summary_stat}) uses the same simulation data that were used to assess our full three-groups model in the manuscript. These results demonstrate that this summary model handily outperforms $p$-value combinations and is quite competitive with our raw data model. The real data have 34 genes which are only measured in the GWAS experiment. For this reason we also performed a simulation study to assess the summary model with data that were generated with missingness (Subfigure~\ref{fig:summary_stats_plot}\subref{subfig:missing_summary_stat}). In this second study, a handful of genes were generated with no information in the RNA-seq experiment (as in Section \ref{subsec:missing_genes} of this Supplement). Here we see that the raw data model more clearly outperforms the summary model.

The stark computational advantage of the summary model may be enough of a benefit that it overrides a decrease in power compared to the full model. For this reason, we analyzed the real data using the summary model. Results from this analysis were disappointing; all genes but one (\textit{CHCHD6}) have a posterior probability of inclusion in the null group that is greater than 0.5. Two other genes (\textit{CDIP1} and \textit{DUSP1}) have posterior probability of inclusion in the null group that is between 0.5 and 0.95.

The results of our simulations and real data analysis suggest that a complete investigation into the summary model is needed. The model appears to do well on simulated data but power degrades more rapidly than the raw data model when we move away from perfect simulation scenarios. One area for future exploration has to do with the distinction between estimating gene effects in a joint model versus one-at-a-time models. The gene effects in the GWAS branch of our summary model are estimated using one-at-a-time logistic regression. By contrast, our raw data model jointly estimates gene effects and thus it is capable of borrowing strength across genes. We conjecture that the one-at-a-time GWAS data generation used in our simulations obscures this advantage of the raw data model.

\section{6: An alternative SNV-to-gene mapping}

Our GWAS data are collected on the SNV level but our model is on the gene level.
This discrepancy requires us to map SNVs to genes. In the manuscript we have reported results using a binary mapping which indicates whether or not there is a minor allele in each gene. 
Here we report results of an alternative approach that sums the number of SNVs in or near each coding region. 
Our model is immediately applicable to this setting when the counts are centered and scaled as part of the preprocessing. 
When the sum-mapped GWAS data are analyzed with both the GWAS-only TG model and the joint TG model (with half-piMOM hyperpriors on the gene effects) our model finds two non-null genes.
The two genes are \textit{CHCHD6} (which was found in the indicator-mapped data) and \textit{NSF} (which was found by both DESeq2 and the joint GWAS-DESeq2 models with the indicator mapping).
In the GWAS-only model these genes had empirical posterior inclusion probability of $1.0$ and $0.9997$ respectively. 
The next highest inclusion probability was $0.1122$.
The joint model included both of these interesting genes in all 10000 of the MCMC iterations (post warm-up) and the next largest inclusion probability was $0.0407$.

\section{7: Parkinson's Disease Data Analysis}
\subsection{Description of the Data} 

The data that we used used for the GWAS branch of this study came from the International Parkinson's Disease Genomics Consortium (IPDGC) NeuroX Dataset \citep{Nalls_2014}. 
Due to the intense computational burden of running the full MCMC, we analyzed a subset the full genome.
We chose the 2,000 genes that exhibited the largest differential expression between the cortex and the substantia nigra (the region most affected in PD) in \citet{Agarwal_2020}, as well as 19 additional genes which seemed promising in exploratory analyses. 
We associated each allele with a particular gene if it was annotated within that gene in the GWAS data (including intronic and 5' or 3' UTR plus 3kbp). This resulted in a list of 53,559 variants associated with the 2,019 genes. 
The NeuroX dataset included 1,734 of the 2,019 genes on our list, sequenced from 11,402 individuals. 
We then summarized the SNV data to the gene level using an indicator function for whether a given gene was mapped to at least one SNV. 
In our analysis, we also included the subject-specific covariates age and sex from the NeuroX dataset, and used PD status as the response for all individuals.

The RNA-seq data that we used came from the Parkinson's Progression Markers Initiative (PPMI), obtained from PPMI upon request. 
We included individuals who were identified as healthy controls or untreated PD cases, resulting in data for 370 individuals. 
We extracted RNA fragment counts for each individual for genes that appeared in the same list of 2,019 genes, resulting in data on 1,697 genes (37 genes only had data in the NeuroX dataset). 
Age, years of education, race, sex, and the phase of the PPMI study are included as covariates.

\subsection{Results}
In the following we classify genes as deleterious or beneficial using the so-called median probability model, hereafter MM, wherein genes are included in a group if the corresponding posterior inclusion probability is greater than 0.5 \citep{Barbieri_2004}.

The joint three-groups model with local priors identifies three beneficial and four deleterious genes. The three beneficial genes are \textit{CDIP1}, \textit{CHCHD6}, and \textit{CNTNAP2}. These genes are known to be involved in dysregulated pathways in PD; mitochondrial function \citep{bose2019mitochondrial, zaltieri2015mitochondrial, moon2015mitochondrial}
, synaptic function \citep{clayton1998synucleins, morais2009parkinson, bagetta2010synaptic}, 
and apoptosis \citep{tatton2003apoptosis, mochizuki1996histochemical, lev2003apoptosis}. 
The four deleterious genes are \textit{DUSP1}, \textit{FAM49B}, \textit{IFRD1}, and \textit{SYTL3} also all have functions in previously implicated PD pathways such as mitochondrial function \citep{bose2019mitochondrial, zaltieri2015mitochondrial, moon2015mitochondrial}, 
stress responses \citep{zhao2010stress, chang2020ifrd1}, autophagy \citep{liu2008dusp1, wang2016role}, vesicular trafficking and endocytosis \citep{singh2020parkinson, 
perrett2015endosomal, esposito2012synaptic}. Three of these genes (\textit{CDIP1}, \textit{CNTNAP2}, \textit{DUSP1}) have connections with PD in the literature \citep{li2022long, brehm2015genetic, usenko2021comparative}. The other genes collectively play roles in pathways previously associated with Parkinson's disease, rendering them plausible candidates for further investigation. The estimated effect sizes (conditional on inclusion in the respective non-null groups) are reported in Table 1 of the main manuscript which contains the union of the top 20 genes identified as non-null by the three-groups joint models.

The non-local, joint three-groups model identifies eight genes as beneficial and ten genes as deleterious groups. These 18 genes were in their respective groups for all 10,000 post burn-in MCMC iterations. The two other genes in the top 20 non-null genes were in each of the three groups for at least 10\% of the MCMC iterations. Five of the 20 interesting genes are also identified in the local model (\textit{CHCHD6}, \textit{CNTNAP2}, \textit{DUSP1}, \textit{FAM49B}, and \textit{DPP10}) and the 21st most interesting gene in this model (\textit{CDIP1}) was also identified by the local model. Several genes identified by this non-local model are linked to the PD literature. \textit{CNTNAP4}, \textit{CXCR4}, \textit{FCGR2A}, and \textit{PTPRN2} are directly linked to PD in the literature \citep[and references therein]{hu_2024early, zhang_2020cntnap4, ma_2023cxcr4, bonham2018cxcr4, Gu_2023, schilder_2022fine, chuang2019longitudinal, kochmanski2022parkinson}. \textit{EFCAB6} has been studied in connection with a mutation known to be associated with early-onset PD in \citet{strobbe2018distinct}. Several others can be linked to pathways which have been studied in relationship with PD though we are are unaware of work which directly links these other genes to PD. These pathway links include \textit{ATP8B4}'s association with the innate immune system, \textit{FRAS1}'s link to ERK signaling, and \textit{JARID2}'s interaction with the Polycomb repressive complex 2, all of which have been implicated or have relation to PD in \citet{liu2021aberrant, tan2020parkinson, albert2020map, toskas2022prc2} respectively.

\section{8: MCMC trace plots}

\begin{figure}
    \centering
    \includegraphics[width=1\linewidth]{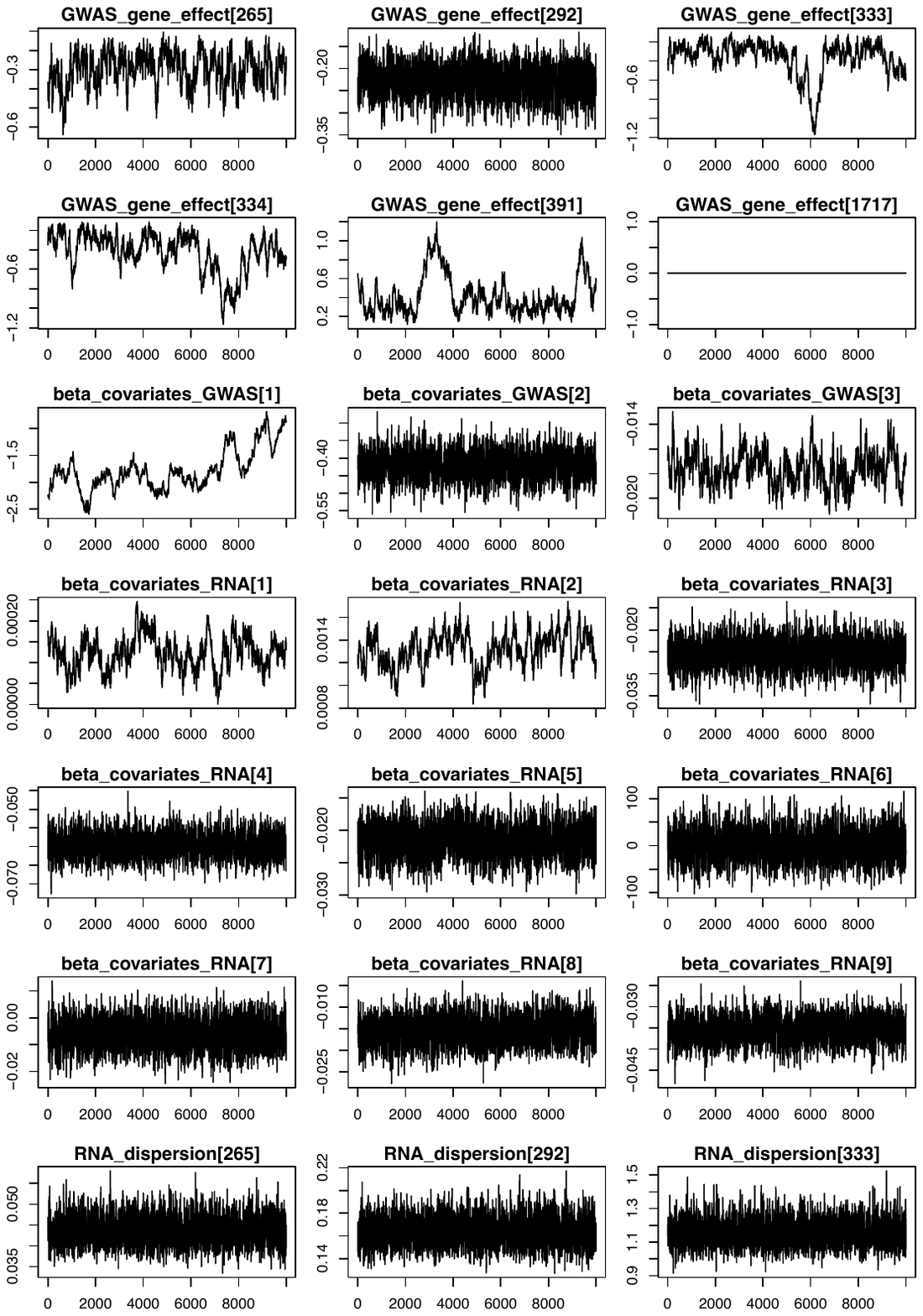}
    \caption{Coda plots 1}
    \label{fig:coda1}
\end{figure}

\begin{figure}
    \centering
    \includegraphics[width=1\linewidth]{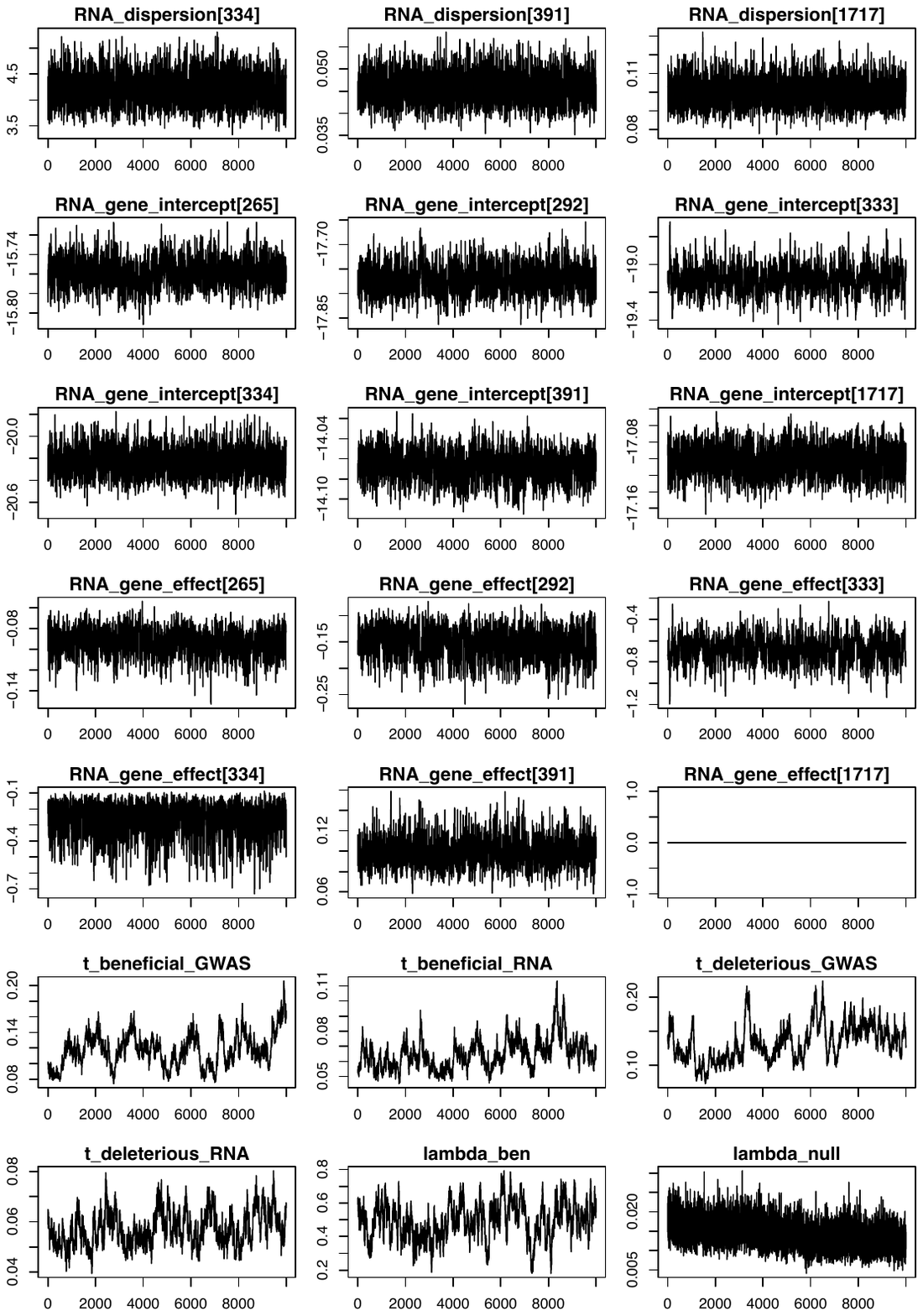}
    \caption{Coda plots 2}
    \label{fig:coda2}
\end{figure}

Example trace plots from our joint three-groups model with piMOM priors on gene effects are given in Figures~\ref{fig:coda1} and \ref{fig:coda2}. Included are trace plots for parameters from the six genes with the lowest posterior probability of inclusion in the null group and one gene that is in null in all mcmc samples. Trace plots for covariates and other hyper-parameters are also included.

\section{9: Volcano Plots}

\begin{figure}
    \centering
    \includegraphics[width=0.95\textwidth]{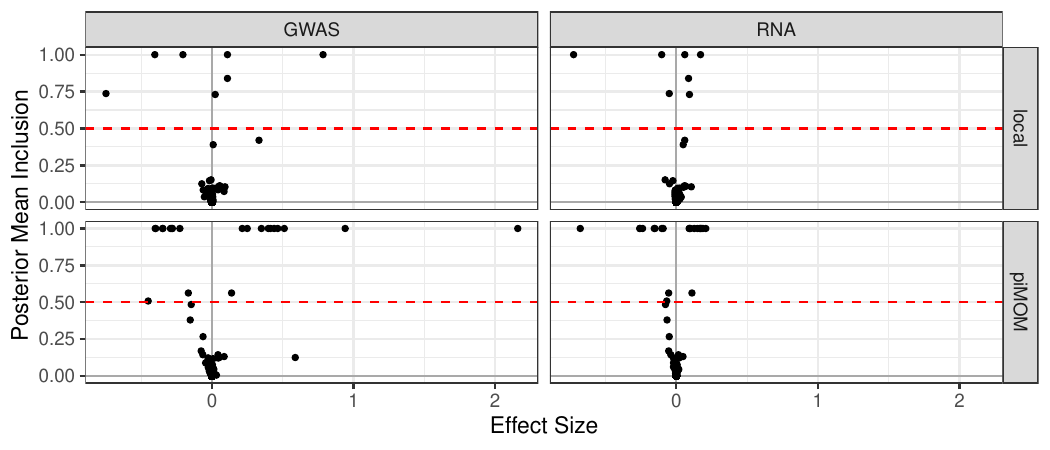}
    \caption{Volcano plots for the local model (top row) and piMOM model (bottom row). Effect sizes are the marginal log odds ratio (GWAS) and log fold change (RNA). The red dashed line indicates the MM cutoff.}
    \label{fig:Volcano}
\end{figure}

It is common to use volcano plots to identify the most promising genes in RNA-seq analyses. These plots conventionally have the log fold change on the horizontal axis and the negative log p-value on the vertical axis which allows investigators to quickly subset genes which are both statistically significant and have large effect sizes. Our model does not benefit from these plots in the same manner due to the inherent sparsity. Nevertheless, we have created a version of these plots in Figure \ref{fig:Volcano} with the marginal log effect size on the horizontal axis and posterior mean of inclusion on the vertical axis. The vast majority of genes are not visible because they are piled up exactly at the origin. These plots highlight the non-local nature of the piMOM model; no included genes have small effect sizes. 

\section{10: Non-null genes from all models}
\label{sec:app_nonnull_all_models}

Table \ref{tab:table_appendix_nonnull_all} displays a list of all of the genes that can be considered non-null by any of the models we ran on the real data. Any gene with a posterior inclusion probability in the null group that was less than 0.5 is considered a non-null gene. This median model (MM) cutoff is applied to our three-groups models except the local GWAS model (even with a cutoff of 0.0001 there were 94 non-null genes). Genes are considered non-null in the conventional models if their lFDR value is below 0.05. Once again, this cutoff is different because the MM cutoff finds too many non-null genes to be useful (the MM for GWAS combined with \texttt{edgeR}, \texttt{limma+voom}, and \texttt{DESeq2} results in the inclusion of 418, 220, and 533 genes respectively whereas the 0.05 cutoff results in 14, 8, and 27 non-null genes respectively).

{\footnotesize
\begin{spacing}{1.0}
\begin{longtable}{rl|llllll|lllllll}
 & Gene & \begin{sideways} TG RNA local \end{sideways} & \begin{sideways} TG RNA piMOM \end{sideways} & \begin{sideways} TG GWAS local \end{sideways} & \begin{sideways} TG GWAS piMOM \end{sideways} & \begin{sideways} TG joint local \end{sideways} & \begin{sideways} TG joint piMOM \end{sideways} & \begin{sideways} \texttt{edgeR} \end{sideways} & \begin{sideways} \texttt{voom} \end{sideways} & \begin{sideways} \texttt{DESeq2} \end{sideways} & \begin{sideways} GWAS \end{sideways} & \begin{sideways} GWAS+\texttt{edgeR} \end{sideways} & \begin{sideways} GWAS+\texttt{voom} \end{sideways} & \begin{sideways} GWAS+\texttt{DESeq2} \end{sideways} \\ 
  \hline
\endhead
\hline
\multicolumn{15}{l}{\footnotesize Continued on next page}
\endfoot
\endlastfoot
 \hline
1 & ACTR10 &  &  & * &  &  &  &  &  &  &  &  &  &  \\ 
  2 & ADAMTS19 &  &  & * &  &  &  &  &  &  &  &  &  &  \\ 
  3 & AFF3 &  &  &  &  &  &  &  &  & * &  &  &  &  \\ 
  4 & ALOX5AP &  &  & * &  &  &  &  &  & * &  &  &  &  \\ 
  5 & ANK3 &  &  & * &  &  &  &  &  &  &  &  &  &  \\ 
  6 & ANKFN1 &  &  &  & * &  &  &  &  &  &  &  &  &  \\ 
  7 & AOAH &  &  & * &  &  &  &  &  &  &  &  &  &  \\ 
  8 & APOD & * &  &  &  &  &  & * &  &  &  & * &  &  \\ 
  9 & ARF1 &  &  & * &  &  &  &  &  &  &  &  &  &  \\ 
  10 & ARHGEF26 &  &  & * &  &  &  &  &  &  &  &  &  &  \\ 
  11 & ATP2B1 &  &  & * &  &  &  &  &  &  &  &  &  &  \\ 
  12 & ATP2C1 &  &  & * &  &  &  &  &  &  &  &  &  &  \\ 
  13 & ATP5A1 &  &  & * &  &  &  &  &  &  &  &  &  &  \\ 
  14 & ATP6V0B &  &  &  &  &  &  &  & * & * &  &  &  &  \\ 
  15 & ATP8B4 &  &  &  &  &  & * &  &  &  &  &  &  &  \\ 
  16 & AZI2 &  &  &  & * &  &  &  &  &  &  &  &  &  \\ 
  17 & BDH1 &  &  & * &  &  &  &  &  &  &  &  &  &  \\ 
  18 & C10orf90 &  &  &  &  &  & * &  &  &  &  &  &  &  \\ 
  19 & C2CD5 &  &  & * &  &  &  &  &  &  &  &  &  &  \\ 
  20 & C6orf136 &  &  & * &  &  &  &  &  &  &  &  &  &  \\ 
  21 & CAMLG &  &  & * &  &  &  &  &  &  &  &  &  &  \\ 
  22 & CAMTA1 &  &  &  & * &  &  &  &  &  &  &  &  &  \\ 
  23 & CANX &  &  & * &  &  &  &  &  &  &  &  &  &  \\ 
  24 & CCDC136 &  &  & * &  &  &  &  &  &  &  &  &  &  \\ 
  25 & CCPG1 &  &  & * &  &  &  &  &  & * &  &  &  &  \\ 
  26 & CD180 &  &  &  &  &  &  &  &  & * &  &  &  &  \\ 
  27 & CD200 & * & * &  &  &  &  & * &  &  &  &  &  &  \\ 
  28 & CD82 &  &  &  &  &  & * &  &  &  &  &  &  &  \\ 
  29 & CD83 & * &  &  &  &  &  & * &  &  &  & * &  &  \\ 
  30 & CDH7 &  &  & * &  &  &  &  &  &  &  &  &  &  \\ 
  31 & CDIP1 &  &  &  & * & * & * &  &  &  & * & * & * & * \\ 
  32 & CDK14 &  &  &  &  &  &  &  &  & * &  &  &  &  \\ 
  33 & CHCHD3 &  &  & * &  &  &  &  &  &  &  &  &  &  \\ 
  34 & CHCHD6 &  &  & * & * & * & * &  &  &  & * & * & * & * \\ 
  35 & CHD6 &  &  & * &  &  &  &  &  &  &  &  &  &  \\ 
  36 & CLEC7A &  &  &  &  &  &  &  &  & * &  & * & * & * \\ 
  37 & CNTNAP2 & * & * &  &  & * & * & * &  &  &  & * &  &  \\ 
  38 & CNTNAP4 &  &  &  &  &  & * &  &  &  &  &  &  &  \\ 
  39 & CNTNAP5 &  &  & * & * &  &  &  &  &  &  &  &  &  \\ 
  40 & CREM &  &  &  &  &  &  &  &  & * &  &  &  &  \\ 
  41 & CSMD1 &  &  &  & * &  &  &  &  &  &  &  &  &  \\ 
  42 & CTSB &  &  & * &  &  &  &  &  &  &  &  &  &  \\ 
  43 & CXCR4 &  &  &  &  &  & * &  &  &  &  &  &  &  \\ 
  44 & CYB5R1 &  &  & * &  &  &  &  &  &  &  &  &  &  \\ 
  45 & DACH1 &  &  &  & * &  &  &  &  &  &  &  &  &  \\ 
  46 & DDIT3 &  &  & * &  &  &  &  &  &  &  &  &  &  \\ 
  47 & DLGAP1 &  &  &  &  &  & * &  &  &  &  &  &  &  \\ 
  48 & DOCK4 &  &  &  &  &  &  & * &  & * &  &  &  &  \\ 
  49 & DPP10 &  &  &  &  &  & * &  &  &  &  &  &  &  \\ 
  50 & DPYSL5 &  &  & * &  &  &  &  &  &  &  &  &  &  \\ 
  51 & DUSP1 & * &  & * &  & * & * & * & * & * &  & * & * & * \\ 
  52 & EDIL3 &  &  &  & * &  &  &  &  &  &  &  &  &  \\ 
  53 & EFCAB6 &  &  &  &  &  & * &  &  &  &  &  &  &  \\ 
  54 & EFEMP1 &  &  & * &  &  &  &  &  &  &  &  &  &  \\ 
  55 & ENPP2 &  &  & * &  &  &  &  &  &  &  &  &  &  \\ 
  56 & EVL &  &  & * &  &  &  &  &  &  &  &  &  &  \\ 
  57 & FAM49B &  &  &  &  & * & * &  &  & * &  &  &  &  \\ 
  58 & FAM98A &  &  & * &  &  &  &  &  &  &  &  &  &  \\ 
  59 & FBXL17 &  &  & * &  &  &  &  &  &  &  &  &  &  \\ 
  60 & FCGR2A &  &  &  &  &  & * &  &  & * &  &  &  & * \\ 
  61 & FGD4 & * &  &  &  &  &  & * & * & * &  & * & * & * \\ 
  62 & FIGN &  &  &  &  &  & * &  &  &  &  &  &  &  \\ 
  63 & FILIP1L &  &  &  &  &  &  & * &  &  &  &  &  &  \\ 
  64 & FMNL3 &  &  & * &  &  &  &  &  &  &  &  &  & * \\ 
  65 & FNDC3A &  &  & * &  &  &  &  &  &  &  &  &  &  \\ 
  66 & FOCAD &  &  &  & * &  &  &  &  &  &  &  &  &  \\ 
  67 & FOS &  &  &  &  &  &  & * & * & * &  & * & * & * \\ 
  68 & FRAS1 &  &  & * &  &  & * &  &  &  &  &  &  &  \\ 
  69 & FSD1 &  &  & * &  &  &  &  &  &  &  &  &  &  \\ 
  70 & FSTL5 &  &  &  & * &  &  &  &  &  &  &  &  &  \\ 
  71 & GABARAPL1 & * &  &  &  &  &  & * & * & * &  &  & * & * \\ 
  72 & GABRG3 &  &  & * &  &  &  &  &  &  &  &  &  &  \\ 
  73 & GALNT13 &  &  &  & * &  &  &  &  &  &  &  &  &  \\ 
  74 & GLIS3 &  &  &  & * &  &  &  &  &  &  &  &  &  \\ 
  75 & GNB4 &  &  &  &  &  &  &  &  & * &  &  &  &  \\ 
  76 & GPR183 &  &  & * &  &  &  &  &  &  &  &  &  &  \\ 
  77 & GRIA2 &  &  & * &  &  &  &  &  &  &  &  &  &  \\ 
  78 & HAGH &  &  & * &  &  &  &  &  &  &  &  &  &  \\ 
  79 & HAP1 &  &  & * &  &  &  &  &  &  &  &  &  &  \\ 
  80 & HDAC9 &  &  &  & * &  &  &  &  &  &  &  &  &  \\ 
  81 & HERPUD1 &  &  &  &  &  &  &  &  & * &  &  &  &  \\ 
  82 & HIF1A &  &  &  &  &  &  &  &  & * &  &  &  &  \\ 
  83 & HLA-DPA1 &  &  & * &  &  &  &  &  &  &  &  &  &  \\ 
  84 & HSPA2 &  &  & * &  &  &  &  &  &  &  &  &  &  \\ 
  85 & HSPA6 &  &  &  & * &  &  &  &  &  & * & * & * & * \\ 
  86 & HSPD1 &  &  &  &  &  &  &  &  &  &  &  &  & * \\ 
  87 & IDH3A &  &  & * &  &  &  &  &  &  &  &  &  &  \\ 
  88 & IFIT3 &  &  &  &  &  &  & * &  & * &  &  &  & * \\ 
  89 & IFRD1 & * &  &  &  & * &  & * & * & * &  &  &  & * \\ 
  90 & IQCA1 &  &  & * &  &  &  &  &  &  &  &  &  &  \\ 
  91 & ITM2B &  &  &  &  &  &  &  &  & * &  &  &  &  \\ 
  92 & JARID2 &  &  &  &  &  & * &  &  &  &  &  &  &  \\ 
  93 & JPH4 &  &  & * &  &  &  &  &  &  &  &  &  &  \\ 
  94 & KATNB1 &  &  & * &  &  &  &  &  &  &  &  &  &  \\ 
  95 & KBTBD8 &  &  & * &  &  &  &  &  &  &  &  &  &  \\ 
  96 & KCNJ6 &  &  & * &  &  &  &  &  &  &  &  &  &  \\ 
  97 & KHDC1 &  &  & * &  &  &  &  &  &  &  &  &  &  \\ 
  98 & KHDRBS3 &  &  & * &  &  &  &  &  &  &  &  &  &  \\ 
  99 & KIAA1958 &  &  &  &  &  &  &  &  &  &  & * &  & * \\ 
  100 & KLF6 &  &  &  &  &  &  &  &  & * &  &  &  &  \\ 
  101 & KLK6 &  &  & * &  &  &  &  &  &  &  &  &  &  \\ 
  102 & LAPTM5 &  &  & * &  &  &  &  &  &  &  &  &  &  \\ 
  103 & LIN7A &  &  &  &  &  &  &  &  & * &  &  &  & * \\ 
  104 & LINC01197 &  &  & * &  &  &  &  &  &  &  &  &  &  \\ 
  105 & LPAR6 &  &  & * &  &  &  &  &  &  &  &  &  &  \\ 
  106 & LPCAT2 &  &  &  &  &  &  &  &  & * &  &  &  &  \\ 
  107 & LRFN5 &  &  &  &  &  & * &  &  &  &  &  &  &  \\ 
  108 & LRRK2 &  &  &  &  &  &  &  &  & * &  &  &  &  \\ 
  109 & LRRN1 &  &  & * &  &  &  &  &  &  &  &  &  &  \\ 
  110 & LYRM1 &  &  &  &  &  &  &  &  & * &  &  &  & * \\ 
  111 & MAP2K4 &  &  &  &  &  &  &  &  & * &  &  &  &  \\ 
  112 & MAP4K5 &  &  & * &  &  &  &  &  &  &  &  &  &  \\ 
  113 & MARCKSL1 &  &  & * &  &  &  &  &  &  &  &  &  &  \\ 
  114 & MAST2 &  &  &  & * &  &  &  &  &  &  &  &  &  \\ 
  115 & MCTP1 &  &  &  &  &  &  & * &  & * &  &  &  & * \\ 
  116 & MFSD4A &  &  & * &  &  &  &  &  &  &  &  &  &  \\ 
  117 & MOCS2 &  &  & * &  &  &  &  &  &  &  &  &  &  \\ 
  118 & MRPL55 &  &  & * &  &  &  &  &  &  &  &  &  &  \\ 
  119 & MXD1 &  &  &  &  &  &  &  &  & * &  &  &  &  \\ 
  120 & NCALD &  &  & * &  &  &  &  &  &  &  &  &  &  \\ 
  121 & NKAIN2 &  &  & * &  &  &  &  &  &  &  &  &  &  \\ 
  122 & NPTN &  &  &  &  &  &  &  &  & * &  &  &  & * \\ 
  123 & NSF &  &  & * &  &  &  &  &  &  &  &  &  &  \\ 
  124 & NUTF2 &  &  & * &  &  &  &  &  &  &  &  &  &  \\ 
  125 & OR2W3 &  &  &  &  &  &  &  &  & * &  &  &  &  \\ 
  126 & ORMDL1 &  &  & * &  &  &  &  &  &  &  &  &  &  \\ 
  127 & PACSIN1 &  &  & * &  &  &  &  &  &  &  &  &  &  \\ 
  128 & PARP8 &  &  &  &  &  &  &  &  & * &  &  &  & * \\ 
  129 & PCSK5 &  &  &  & * &  &  &  &  &  &  &  &  &  \\ 
  130 & PDK2 &  &  &  &  &  &  &  &  & * &  &  &  &  \\ 
  131 & PDK4 &  &  &  &  &  &  &  &  &  &  &  &  & * \\ 
  132 & PELI1 & * &  &  &  &  &  & * & * & * &  &  &  & * \\ 
  133 & PGAM1 &  &  & * &  &  &  &  &  &  &  &  &  &  \\ 
  134 & PLEKHB1 &  &  & * &  &  &  &  &  &  &  &  &  &  \\ 
  135 & PLEKHB2 &  &  & * &  &  &  &  &  &  &  &  &  &  \\ 
  136 & PLXDC2 &  &  &  &  &  &  &  &  & * &  &  &  &  \\ 
  137 & PTPRD &  &  &  & * &  &  &  &  &  &  &  &  &  \\ 
  138 & PTPRG &  &  &  & * &  &  &  &  &  &  &  &  &  \\ 
  139 & PTPRN2 &  &  &  &  &  & * &  &  &  &  &  &  &  \\ 
  140 & PTPRR &  &  & * &  &  &  &  &  &  &  &  &  &  \\ 
  141 & RAMP1 &  &  & * &  &  &  &  &  &  &  &  &  &  \\ 
  142 & RANBP2 &  &  & * &  &  &  &  &  &  &  &  &  &  \\ 
  143 & RAP1GAP & * & * &  &  &  &  & * &  & * &  & * &  & * \\ 
  144 & RGS2 &  &  &  &  &  &  &  &  & * &  &  &  &  \\ 
  145 & RHOQ &  &  & * &  &  &  &  &  &  &  &  &  &  \\ 
  146 & RIN3 &  &  &  &  &  & * &  &  &  &  &  &  &  \\ 
  147 & RNF13 &  &  &  &  &  &  &  &  & * &  &  &  &  \\ 
  148 & RSRP1 &  &  &  &  &  &  &  &  & * &  &  &  & * \\ 
  149 & S100B &  &  & * &  &  &  &  &  &  &  &  &  &  \\ 
  150 & SCN1A &  &  & * &  &  &  &  &  &  &  &  &  &  \\ 
  151 & SCRN2 &  &  & * &  &  &  &  &  &  &  &  &  &  \\ 
  152 & SEMA3B &  &  & * &  &  &  &  &  &  &  &  &  &  \\ 
  153 & SERINC3 &  &  & * &  &  &  &  &  &  &  &  &  &  \\ 
  154 & SKAP2 &  &  &  &  &  &  &  &  & * &  &  &  &  \\ 
  155 & SLC16A4 &  &  &  &  &  &  & * &  &  &  &  &  &  \\ 
  156 & SLC24A2 &  &  &  & * &  &  &  &  &  &  &  &  &  \\ 
  157 & SLC25A29 &  &  & * &  &  &  &  &  &  &  &  &  &  \\ 
  158 & SLC31A2 &  &  &  &  &  &  &  &  & * &  &  &  &  \\ 
  159 & STRBP &  &  &  &  &  &  & * &  & * &  &  &  &  \\ 
  160 & SYT13 &  &  & * &  &  &  &  &  &  &  &  &  &  \\ 
  161 & SYT5 &  &  & * &  &  &  &  &  &  &  &  &  &  \\ 
  162 & SYTL3 &  &  &  &  & * &  & * &  & * &  & * &  & * \\ 
  163 & TIMM22 &  &  & * &  &  &  &  &  &  &  &  &  &  \\ 
  164 & TLR6 &  &  & * &  &  &  &  &  & * &  &  &  & * \\ 
  165 & TM2D3 &  &  &  &  &  &  &  &  & * &  &  &  &  \\ 
  166 & TMCC3 &  &  &  &  &  &  &  &  & * &  &  &  & * \\ 
  167 & TMEM125 &  &  & * &  &  &  &  &  &  &  &  &  &  \\ 
  168 & TMEM181 &  &  & * &  &  &  &  &  &  &  &  &  &  \\ 
  169 & TMEM246 &  &  & * &  &  &  &  &  &  &  &  &  &  \\ 
  170 & TMTC1 &  &  &  & * &  &  &  &  &  &  &  &  &  \\ 
  171 & TMX4 &  &  &  &  &  &  &  &  & * &  &  &  &  \\ 
  172 & TPI1 &  &  & * &  &  &  &  &  &  &  &  &  &  \\ 
  173 & TPST1 &  &  &  &  &  &  & * &  & * &  & * &  & * \\ 
  174 & TRANK1 &  &  &  &  &  &  &  &  & * &  &  &  &  \\ 
  175 & TRPS1 &  &  & * &  &  &  &  &  &  &  &  &  &  \\ 
  176 & TUFM &  &  & * &  &  &  &  &  &  &  &  &  &  \\ 
  177 & TYROBP &  &  &  &  &  &  &  &  & * &  &  &  &  \\ 
  178 & UQCR10 &  &  & * &  &  &  &  &  &  &  &  &  &  \\ 
  179 & UQCRC1 &  &  & * &  &  &  &  &  &  &  &  &  &  \\ 
  180 & VRK2 &  &  &  &  &  & * &  &  &  &  &  &  &  \\ 
  181 & WDR12 &  &  & * &  &  &  &  &  & * &  &  &  &  \\ 
  182 & XYLT1 &  &  &  & * &  &  &  &  &  &  &  &  &  \\ 
  183 & ZDHHC20 &  &  & * &  &  &  &  &  &  &  &  &  &  \\ 
   \hline
\hline
\caption{Genes identified as non-null in at least one of model. A gene is classified as non-null in all models with an “*”. The three groups (TG) models use the median model (cutoff at $P_{null} < 0.5$) except for the local GWAS model which included too many genes. We used a cutoff of 0.0001 for this GWAS model which still included 94 genes. All conventional models used a cutoff of 0.05.}
\label{tab:table_appendix_nonnull_all}
\end{longtable}
\end{spacing}}

\begin{landscape}

\section{11: Non-null genes and effect sizes from TG RNA only and GWAS only models.}
\label{sec:app_nonnull_each_branch}

{\footnotesize
\begin{spacing}{1.25}
\begin{longtable}{rl|rrrrrr|rrrrrr|llll|ll}
 & Gene & \begin{sideways}  $P_{null}$ RNA local \end{sideways} & \begin{sideways}  $P_{ben}$ RNA local \end{sideways} & \begin{sideways}  $P_{del}$ RNA local \end{sideways} & \begin{sideways}  $P_{null}$ RNA piMOM \end{sideways} & \begin{sideways}  $P_{ben}$ RNA piMOM \end{sideways} & \begin{sideways}  $P_{del}$ RNA piMOM \end{sideways} & \begin{sideways}  $P_{null}$ GWAS local \end{sideways} & \begin{sideways}  $P_{ben}$ GWAS local \end{sideways} & \begin{sideways}  $P_{del}$ GWAS local \end{sideways} & \begin{sideways}  $P_{null}$ GWAS piMOM \end{sideways} & \begin{sideways}  $P_{ben}$ GWAS piMOM \end{sideways} & \begin{sideways}  $P_{del}$ GWAS piMOM \end{sideways} & \begin{sideways} RNA effect local \end{sideways} & \begin{sideways} Dispersion local \end{sideways} & \begin{sideways} RNA effect piMOM \end{sideways} & \begin{sideways} Dispersion piMOM \end{sideways} & \begin{sideways} GWAS effect local \end{sideways} & \begin{sideways} GWAS effect piMOM \end{sideways} \\ 
  \hline
\endhead
\hline
\multicolumn{20}{l}{\footnotesize Continued on next page}
\endfoot
\endlastfoot
 \hline
 1 & ACTR10 & 1.00 & 0.00 & 0.00 & 1.00 & 0.00 & 0.00 & 0.00 & 0.00 & 1.00 & 1.00 & 0.00 & 0.00 & x & x & x & x & 179.34 & x \\ 
  2 & ADAMTS19 & 1.00 & 0.00 & 0.00 & 1.00 & 0.00 & 0.00 & 0.00 & 0.00 & 1.00 & 1.00 & 0.00 & 0.00 & x & x & x & x & 2.37 & x \\ 
  3 & ALOX5AP & 1.00 & 0.00 & 0.00 & 1.00 & 0.00 & 0.00 & 0.00 & 1.00 & 0.00 & 1.00 & 0.00 & 0.00 & x & x & x & x & 0.03 & x \\ 
  4 & ANK3 & 1.00 & 0.00 & 0.00 & 1.00 & 0.00 & 0.00 & 0.00 & 0.00 & 1.00 & 1.00 & 0.00 & 0.00 & x & x & x & x & 3.36 & x \\ 
  5 & ANKFN1 & 1.00 & 0.00 & 0.00 & 1.00 & 0.00 & 0.00 & 1.00 & 0.00 & 0.00 & 0.00 & 1.00 & 0.00 & x & x & x & x & x & 0.71 \\ 
  6 & AOAH & 1.00 & 0.00 & 0.00 & 1.00 & 0.00 & 0.00 & 0.00 & 0.00 & 1.00 & 1.00 & 0.00 & 0.00 & x & x & x & x & 2.92 & x \\ 
  7 & APOD & 0.22 & 0.10 & 0.68 & 1.00 & 0.00 & 0.00 & 1.00 & 0.00 & 0.00 & 1.00 & 0.00 & 0.00 & 0.45 & 5.16 & x & x & x & x \\ 
  8 & ARF1 & 1.00 & 0.00 & 0.00 & 1.00 & 0.00 & 0.00 & 0.00 & 0.00 & 1.00 & 1.00 & 0.00 & 0.00 & x & x & x & x & 20.39 & x \\ 
  9 & ARHGEF26 & 1.00 & 0.00 & 0.00 & 1.00 & 0.00 & 0.00 & 0.00 & 0.00 & 1.00 & 1.00 & 0.00 & 0.00 & x & x & x & x & 1.95 & x \\ 
  10 & ATP2B1 & 1.00 & 0.00 & 0.00 & 1.00 & 0.00 & 0.00 & 0.00 & 1.00 & 0.00 & 1.00 & 0.00 & 0.00 & x & x & x & x & 2.04 & x \\ 
  11 & ATP2C1 & 1.00 & 0.00 & 0.00 & 1.00 & 0.00 & 0.00 & 0.00 & 1.00 & 0.00 & 1.00 & 0.00 & 0.00 & x & x & x & x & 2.43 & x \\ 
  12 & ATP5A1 & 1.00 & 0.00 & 0.00 & 1.00 & 0.00 & 0.00 & 0.00 & 1.00 & 0.00 & 1.00 & 0.00 & 0.00 & x & x & x & x & 0.16 & x \\ 
  13 & AZI2 & 1.00 & 0.00 & 0.00 & 1.00 & 0.00 & 0.00 & 1.00 & 0.00 & 0.00 & 0.00 & 0.00 & 1.00 & x & x & x & x & x & 1.82 \\ 
  14 & BDH1 & 1.00 & 0.00 & 0.00 & 1.00 & 0.00 & 0.00 & 0.00 & 1.00 & 0.00 & 1.00 & 0.00 & 0.00 & x & x & x & x & 0.46 & x \\ 
  15 & C2CD5 & 1.00 & 0.00 & 0.00 & 1.00 & 0.00 & 0.00 & 0.00 & 1.00 & 0.00 & 1.00 & 0.00 & 0.00 & x & x & x & x & 0.08 & x \\ 
  16 & C6orf136 & 1.00 & 0.00 & 0.00 & 1.00 & 0.00 & 0.00 & 0.00 & 1.00 & 0.00 & 1.00 & 0.00 & 0.00 & x & x & x & x & 3.03 & x \\ 
  17 & CAMLG & 1.00 & 0.00 & 0.00 & 1.00 & 0.00 & 0.00 & 0.00 & 0.00 & 1.00 & 1.00 & 0.00 & 0.00 & x & x & x & x & 0.28 & x \\ 
  18 & CAMTA1 & 1.00 & 0.00 & 0.00 & 1.00 & 0.00 & 0.00 & 1.00 & 0.00 & 0.00 & 0.00 & 0.00 & 1.00 & x & x & x & x & x & 0.41 \\ 
  19 & CANX & 1.00 & 0.00 & 0.00 & 1.00 & 0.00 & 0.00 & 0.00 & 0.00 & 1.00 & 1.00 & 0.00 & 0.00 & x & x & x & x & 0.03 & x \\ 
  20 & CCDC136 & 1.00 & 0.00 & 0.00 & 1.00 & 0.00 & 0.00 & 0.00 & 1.00 & 0.00 & 1.00 & 0.00 & 0.00 & x & x & x & x & 0.25 & x \\ 
  21 & CCPG1 & 1.00 & 0.00 & 0.00 & 1.00 & 0.00 & 0.00 & 0.00 & 0.00 & 1.00 & 1.00 & 0.00 & 0.00 & x & x & x & x & 2.27 & x \\ 
  22 & CD200 & 0.42 & 0.46 & 0.12 & 0.44 & 0.49 & 0.07 & 1.00 & 0.00 & 0.00 & 1.00 & 0.00 & 0.00 & 0.77 & 0.43 & 0.75 & 0.43 & x & x \\ 
  23 & CD83 & 0.37 & 0.51 & 0.12 & 1.00 & 0.00 & 0.00 & 1.00 & 0.00 & 0.00 & 1.00 & 0.00 & 0.00 & 0.84 & 0.16 & x & x & x & x \\ 
  24 & CDH7 & 1.00 & 0.00 & 0.00 & 1.00 & 0.00 & 0.00 & 0.00 & 0.00 & 1.00 & 1.00 & 0.00 & 0.00 & x & x & x & x & 5.89 & x \\ 
  25 & CDIP1 & 1.00 & 0.00 & 0.00 & 1.00 & 0.00 & 0.00 & 1.00 & 0.00 & 0.00 & 0.00 & 1.00 & 0.00 & x & x & x & x & x & 0.49 \\ 
  26 & CHCHD3 & 1.00 & 0.00 & 0.00 & 1.00 & 0.00 & 0.00 & 0.00 & 0.00 & 1.00 & 1.00 & 0.00 & 0.00 & x & x & x & x & 2.26 & x \\ 
  27 & CHCHD6 & 1.00 & 0.00 & 0.00 & 1.00 & 0.00 & 0.00 & 0.00 & 0.00 & 1.00 & 0.00 & 1.00 & 0.00 & x & x & x & x & 0.41 & 0.8 \\ 
  28 & CHD6 & 1.00 & 0.00 & 0.00 & 1.00 & 0.00 & 0.00 & 0.00 & 0.00 & 1.00 & 1.00 & 0.00 & 0.00 & x & x & x & x & 2.71 & x \\ 
  29 & CNTNAP2 & 0.22 & 0.10 & 0.68 & 0.00 & 1.00 & 0.00 & 1.00 & 0.00 & 0.00 & 1.00 & 0.00 & 0.00 & 0.48 & 1.17 & 0.5 & 1.16 & x & x \\ 
  30 & CNTNAP5 & 1.00 & 0.00 & 0.00 & 1.00 & 0.00 & 0.00 & 0.00 & 1.00 & 0.00 & 0.39 & 0.14 & 0.47 & x & x & x & x & 5.58 & 0.61 \\ 
  31 & CSMD1 & 1.00 & 0.00 & 0.00 & 1.00 & 0.00 & 0.00 & 1.00 & 0.00 & 0.00 & 0.00 & 1.00 & 0.00 & x & x & x & x & x & 1.64 \\ 
  32 & CTSB & 1.00 & 0.00 & 0.00 & 1.00 & 0.00 & 0.00 & 0.00 & 0.00 & 1.00 & 1.00 & 0.00 & 0.00 & x & x & x & x & 5.36 & x \\ 
  33 & CYB5R1 & 1.00 & 0.00 & 0.00 & 1.00 & 0.00 & 0.00 & 0.00 & 1.00 & 0.00 & 1.00 & 0.00 & 0.00 & x & x & x & x & 5.62 & x \\ 
  34 & DACH1 & 1.00 & 0.00 & 0.00 & 1.00 & 0.00 & 0.00 & 1.00 & 0.00 & 0.00 & 0.00 & 1.00 & 0.00 & x & x & x & x & x & 1.53 \\ 
  35 & DDIT3 & 1.00 & 0.00 & 0.00 & 1.00 & 0.00 & 0.00 & 0.00 & 0.00 & 1.00 & 1.00 & 0.00 & 0.00 & x & x & x & x & 0.1 & x \\ 
  36 & DPYSL5 & 1.00 & 0.00 & 0.00 & 1.00 & 0.00 & 0.00 & 0.00 & 0.00 & 1.00 & 1.00 & 0.00 & 0.00 & x & x & x & x & 2.77 & x \\ 
  37 & DUSP1 & 0.22 & 0.10 & 0.68 & 1.00 & 0.00 & 0.00 & 0.00 & 0.00 & 1.00 & 1.00 & 0.00 & 0.00 & 1.18 & 0.09 & x & x & 6.12 & x \\ 
  38 & EDIL3 & 1.00 & 0.00 & 0.00 & 1.00 & 0.00 & 0.00 & 1.00 & 0.00 & 0.00 & 0.43 & 0.47 & 0.10 & x & x & x & x & x & 0.69 \\ 
  39 & EFEMP1 & 1.00 & 0.00 & 0.00 & 1.00 & 0.00 & 0.00 & 0.00 & 0.00 & 1.00 & 1.00 & 0.00 & 0.00 & x & x & x & x & 2.09 & x \\ 
  40 & ENPP2 & 1.00 & 0.00 & 0.00 & 1.00 & 0.00 & 0.00 & 0.00 & 1.00 & 0.00 & 1.00 & 0.00 & 0.00 & x & x & x & x & 17.02 & x \\ 
  41 & EVL & 1.00 & 0.00 & 0.00 & 1.00 & 0.00 & 0.00 & 0.00 & 0.00 & 1.00 & 1.00 & 0.00 & 0.00 & x & x & x & x & 0.17 & x \\ 
  42 & FAM98A & 1.00 & 0.00 & 0.00 & 1.00 & 0.00 & 0.00 & 0.00 & 0.00 & 1.00 & 1.00 & 0.00 & 0.00 & x & x & x & x & 3.85 & x \\ 
  43 & FBXL17 & 1.00 & 0.00 & 0.00 & 1.00 & 0.00 & 0.00 & 0.00 & 0.00 & 1.00 & 1.00 & 0.00 & 0.00 & x & x & x & x & 5.43 & x \\ 
  44 & FGD4 & 0.00 & 1.00 & 0.00 & 1.00 & 0.00 & 0.00 & 1.00 & 0.00 & 0.00 & 1.00 & 0.00 & 0.00 & 1.13 & 0.06 & x & x & x & x \\ 
  45 & FMNL3 & 1.00 & 0.00 & 0.00 & 1.00 & 0.00 & 0.00 & 0.00 & 0.00 & 1.00 & 1.00 & 0.00 & 0.00 & x & x & x & x & 0.06 & x \\ 
  46 & FNDC3A & 1.00 & 0.00 & 0.00 & 1.00 & 0.00 & 0.00 & 0.00 & 1.00 & 0.00 & 1.00 & 0.00 & 0.00 & x & x & x & x & 5.18 & x \\ 
  47 & FOCAD & 1.00 & 0.00 & 0.00 & 1.00 & 0.00 & 0.00 & 1.00 & 0.00 & 0.00 & 0.00 & 1.00 & 0.00 & x & x & x & x & x & 0.7 \\ 
  48 & FRAS1 & 1.00 & 0.00 & 0.00 & 1.00 & 0.00 & 0.00 & 0.00 & 0.00 & 1.00 & 1.00 & 0.00 & 0.00 & x & x & x & x & 4.4 & x \\ 
  49 & FSD1 & 1.00 & 0.00 & 0.00 & 1.00 & 0.00 & 0.00 & 0.00 & 0.00 & 1.00 & 1.00 & 0.00 & 0.00 & x & x & x & x & 2.13 & x \\ 
  50 & FSTL5 & 1.00 & 0.00 & 0.00 & 1.00 & 0.00 & 0.00 & 1.00 & 0.00 & 0.00 & 0.00 & 1.00 & 0.00 & x & x & x & x & x & 0.71 \\ 
  51 & GABARAPL1 & 0.49 & 0.38 & 0.13 & 1.00 & 0.00 & 0.00 & 1.00 & 0.00 & 0.00 & 1.00 & 0.00 & 0.00 & 1.11 & 0.04 & x & x & x & x \\ 
  52 & GABRG3 & 1.00 & 0.00 & 0.00 & 1.00 & 0.00 & 0.00 & 0.00 & 0.00 & 1.00 & 1.00 & 0.00 & 0.00 & x & x & x & x & 4.58 & x \\ 
  53 & GALNT13 & 1.00 & 0.00 & 0.00 & 1.00 & 0.00 & 0.00 & 1.00 & 0.00 & 0.00 & 0.00 & 1.00 & 0.00 & x & x & x & x & x & 0.7 \\ 
  54 & GLIS3 & 1.00 & 0.00 & 0.00 & 1.00 & 0.00 & 0.00 & 1.00 & 0.00 & 0.00 & 0.00 & 1.00 & 0.00 & x & x & x & x & x & 1.53 \\ 
  55 & GPR183 & 1.00 & 0.00 & 0.00 & 1.00 & 0.00 & 0.00 & 0.00 & 1.00 & 0.00 & 1.00 & 0.00 & 0.00 & x & x & x & x & 0.02 & x \\ 
  56 & GRIA2 & 1.00 & 0.00 & 0.00 & 1.00 & 0.00 & 0.00 & 0.00 & 0.00 & 1.00 & 1.00 & 0.00 & 0.00 & x & x & x & x & 2.08 & x \\ 
  57 & HAGH & 1.00 & 0.00 & 0.00 & 1.00 & 0.00 & 0.00 & 0.00 & 0.00 & 1.00 & 1.00 & 0.00 & 0.00 & x & x & x & x & 0.15 & x \\ 
  58 & HAP1 & 1.00 & 0.00 & 0.00 & 1.00 & 0.00 & 0.00 & 0.00 & 1.00 & 0.00 & 1.00 & 0.00 & 0.00 & x & x & x & x & 4.58 & x \\ 
  59 & HDAC9 & 1.00 & 0.00 & 0.00 & 1.00 & 0.00 & 0.00 & 1.00 & 0.00 & 0.00 & 0.14 & 0.81 & 0.05 & x & x & x & x & x & 0.74 \\ 
  60 & HLA-DPA1 & 1.00 & 0.00 & 0.00 & 1.00 & 0.00 & 0.00 & 0.00 & 0.00 & 1.00 & 1.00 & 0.00 & 0.00 & x & x & x & x & 4.21 & x \\ 
  61 & HSPA2 & 1.00 & 0.00 & 0.00 & 1.00 & 0.00 & 0.00 & 0.00 & 0.00 & 1.00 & 1.00 & 0.00 & 0.00 & x & x & x & x & 33.82 & x \\ 
  62 & HSPA6 & 1.00 & 0.00 & 0.00 & 1.00 & 0.00 & 0.00 & 1.00 & 0.00 & 0.00 & 0.00 & 0.00 & 1.00 & x & x & x & x & x & 0.6 \\ 
  63 & IDH3A & 1.00 & 0.00 & 0.00 & 1.00 & 0.00 & 0.00 & 0.00 & 0.00 & 1.00 & 1.00 & 0.00 & 0.00 & x & x & x & x & 0.22 & x \\ 
  64 & IFRD1 & 0.00 & 1.00 & 0.00 & 1.00 & 0.00 & 0.00 & 1.00 & 0.00 & 0.00 & 1.00 & 0.00 & 0.00 & 1.13 & 0.07 & x & x & x & x \\ 
  65 & IQCA1 & 1.00 & 0.00 & 0.00 & 1.00 & 0.00 & 0.00 & 0.00 & 0.00 & 1.00 & 1.00 & 0.00 & 0.00 & x & x & x & x & 2.34 & x \\ 
  66 & JPH4 & 1.00 & 0.00 & 0.00 & 1.00 & 0.00 & 0.00 & 0.00 & 0.00 & 1.00 & 1.00 & 0.00 & 0.00 & x & x & x & x & 0.01 & x \\ 
  67 & KATNB1 & 1.00 & 0.00 & 0.00 & 1.00 & 0.00 & 0.00 & 0.00 & 0.00 & 1.00 & 1.00 & 0.00 & 0.00 & x & x & x & x & 3.68 & x \\ 
  68 & KBTBD8 & 1.00 & 0.00 & 0.00 & 1.00 & 0.00 & 0.00 & 0.00 & 0.00 & 1.00 & 1.00 & 0.00 & 0.00 & x & x & x & x & 1.59 & x \\ 
  69 & KCNJ6 & 1.00 & 0.00 & 0.00 & 1.00 & 0.00 & 0.00 & 0.00 & 0.00 & 1.00 & 1.00 & 0.00 & 0.00 & x & x & x & x & 3.6 & x \\ 
  70 & KHDC1 & 1.00 & 0.00 & 0.00 & 1.00 & 0.00 & 0.00 & 0.00 & 0.00 & 1.00 & 1.00 & 0.00 & 0.00 & x & x & x & x & 6.61 & x \\ 
  71 & KHDRBS3 & 1.00 & 0.00 & 0.00 & 1.00 & 0.00 & 0.00 & 0.00 & 0.00 & 1.00 & 1.00 & 0.00 & 0.00 & x & x & x & x & 3.33 & x \\ 
  72 & KLK6 & 1.00 & 0.00 & 0.00 & 1.00 & 0.00 & 0.00 & 0.00 & 0.00 & 1.00 & 1.00 & 0.00 & 0.00 & x & x & x & x & 44.74 & x \\ 
  73 & LAPTM5 & 1.00 & 0.00 & 0.00 & 1.00 & 0.00 & 0.00 & 0.00 & 1.00 & 0.00 & 1.00 & 0.00 & 0.00 & x & x & x & x & 2.14 & x \\ 
  74 & LINC01197 & 1.00 & 0.00 & 0.00 & 1.00 & 0.00 & 0.00 & 0.00 & 0.00 & 1.00 & 1.00 & 0.00 & 0.00 & x & x & x & x & 0.45 & x \\ 
  75 & LPAR6 & 1.00 & 0.00 & 0.00 & 1.00 & 0.00 & 0.00 & 0.00 & 1.00 & 0.00 & 1.00 & 0.00 & 0.00 & x & x & x & x & 0.13 & x \\ 
  76 & LRRN1 & 1.00 & 0.00 & 0.00 & 1.00 & 0.00 & 0.00 & 0.00 & 0.00 & 1.00 & 1.00 & 0.00 & 0.00 & x & x & x & x & 3.8 & x \\ 
  77 & MAP4K5 & 1.00 & 0.00 & 0.00 & 1.00 & 0.00 & 0.00 & 0.00 & 0.00 & 1.00 & 1.00 & 0.00 & 0.00 & x & x & x & x & 1.7 & x \\ 
  78 & MARCKSL1 & 1.00 & 0.00 & 0.00 & 1.00 & 0.00 & 0.00 & 0.00 & 0.00 & 1.00 & 1.00 & 0.00 & 0.00 & x & x & x & x & 0.04 & x \\ 
  79 & MAST2 & 1.00 & 0.00 & 0.00 & 1.00 & 0.00 & 0.00 & 1.00 & 0.00 & 0.00 & 0.00 & 0.00 & 1.00 & x & x & x & x & x & 0.68 \\ 
  80 & MFSD4A & 1.00 & 0.00 & 0.00 & 1.00 & 0.00 & 0.00 & 0.00 & 1.00 & 0.00 & 1.00 & 0.00 & 0.00 & x & x & x & x & 0.25 & x \\ 
  81 & MOCS2 & 1.00 & 0.00 & 0.00 & 1.00 & 0.00 & 0.00 & 0.00 & 0.00 & 1.00 & 1.00 & 0.00 & 0.00 & x & x & x & x & 2.6 & x \\ 
  82 & MRPL55 & 1.00 & 0.00 & 0.00 & 1.00 & 0.00 & 0.00 & 0.00 & 0.00 & 1.00 & 1.00 & 0.00 & 0.00 & x & x & x & x & 25.93 & x \\ 
  83 & NCALD & 1.00 & 0.00 & 0.00 & 1.00 & 0.00 & 0.00 & 0.00 & 0.00 & 1.00 & 1.00 & 0.00 & 0.00 & x & x & x & x & 0.42 & x \\ 
  84 & NKAIN2 & 1.00 & 0.00 & 0.00 & 1.00 & 0.00 & 0.00 & 0.00 & 0.00 & 1.00 & 1.00 & 0.00 & 0.00 & x & x & x & x & 6.08 & x \\ 
  85 & NSF & 1.00 & 0.00 & 0.00 & 1.00 & 0.00 & 0.00 & 0.00 & 1.00 & 0.00 & 1.00 & 0.00 & 0.00 & x & x & x & x & 0.48 & x \\ 
  86 & NUTF2 & 1.00 & 0.00 & 0.00 & 1.00 & 0.00 & 0.00 & 0.00 & 1.00 & 0.00 & 1.00 & 0.00 & 0.00 & x & x & x & x & 4.29 & x \\ 
  87 & ORMDL1 & 1.00 & 0.00 & 0.00 & 1.00 & 0.00 & 0.00 & 0.00 & 1.00 & 0.00 & 1.00 & 0.00 & 0.00 & x & x & x & x & 55.44 & x \\ 
  88 & PACSIN1 & 1.00 & 0.00 & 0.00 & 1.00 & 0.00 & 0.00 & 0.00 & 0.00 & 1.00 & 1.00 & 0.00 & 0.00 & x & x & x & x & 3.01 & x \\ 
  89 & PCSK5 & 1.00 & 0.00 & 0.00 & 1.00 & 0.00 & 0.00 & 1.00 & 0.00 & 0.00 & 0.00 & 1.00 & 0.00 & x & x & x & x & x & 2.22 \\ 
  90 & PELI1 & 0.12 & 0.83 & 0.05 & 1.00 & 0.00 & 0.00 & 1.00 & 0.00 & 0.00 & 1.00 & 0.00 & 0.00 & 1.11 & 0.05 & x & x & x & x \\ 
  91 & PGAM1 & 1.00 & 0.00 & 0.00 & 1.00 & 0.00 & 0.00 & 0.00 & 0.00 & 1.00 & 1.00 & 0.00 & 0.00 & x & x & x & x & 68.49 & x \\ 
  92 & PLEKHB1 & 1.00 & 0.00 & 0.00 & 1.00 & 0.00 & 0.00 & 0.00 & 1.00 & 0.00 & 1.00 & 0.00 & 0.00 & x & x & x & x & 2.32 & x \\ 
  93 & PLEKHB2 & 1.00 & 0.00 & 0.00 & 1.00 & 0.00 & 0.00 & 0.00 & 0.00 & 1.00 & 1.00 & 0.00 & 0.00 & x & x & x & x & 30.29 & x \\ 
  94 & PTPRD & 1.00 & 0.00 & 0.00 & 1.00 & 0.00 & 0.00 & 1.00 & 0.00 & 0.00 & 0.00 & 1.00 & 0.00 & x & x & x & x & x & 1.49 \\ 
  95 & PTPRG & 1.00 & 0.00 & 0.00 & 1.00 & 0.00 & 0.00 & 1.00 & 0.00 & 0.00 & 0.00 & 0.00 & 1.00 & x & x & x & x & x & 2.18 \\ 
  96 & PTPRR & 1.00 & 0.00 & 0.00 & 1.00 & 0.00 & 0.00 & 0.00 & 0.00 & 1.00 & 1.00 & 0.00 & 0.00 & x & x & x & x & 1.56 & x \\ 
  97 & RAMP1 & 1.00 & 0.00 & 0.00 & 1.00 & 0.00 & 0.00 & 0.00 & 1.00 & 0.00 & 1.00 & 0.00 & 0.00 & x & x & x & x & 35.86 & x \\ 
  98 & RANBP2 & 1.00 & 0.00 & 0.00 & 1.00 & 0.00 & 0.00 & 0.00 & 0.00 & 1.00 & 1.00 & 0.00 & 0.00 & x & x & x & x & 0.27 & x \\ 
  99 & RAP1GAP & 0.22 & 0.10 & 0.68 \relax\relax& 0.15 & 0.83 & 0.02 & 1.00 & 0.00 & 0.00 & 1.00 & 0.00 & 0.00 & 0.58 & 1.5 & 0.6 & 1.49 & x & x \\ 
  100 & RHOQ & 1.00 & 0.00 & 0.00 & 1.00 & 0.00 & 0.00 & 0.00 & 0.00 & 1.00 & 1.00 & 0.00 & 0.00 & x & x & x & x & 8.8 & x \\ 
  101 & S100B & 1.00 & 0.00 & 0.00 & 1.00 & 0.00 & 0.00 & 0.00 & 0.00 & 1.00 & 1.00 & 0.00 & 0.00 & x & x & x & x & 0.47 & x \\ 
  102 & SCN1A & 1.00 & 0.00 & 0.00 & 1.00 & 0.00 & 0.00 & 0.00 & 0.00 & 1.00 & 1.00 & 0.00 & 0.00 & x & x & x & x & 2.85 & x \\ 
  103 & SCRN2 & 1.00 & 0.00 & 0.00 & 1.00 & 0.00 & 0.00 & 0.00 & 0.00 & 1.00 & 1.00 & 0.00 & 0.00 & x & x & x & x & 2.57 & x \\ 
  104 & SEMA3B & 1.00 & 0.00 & 0.00 & 1.00 & 0.00 & 0.00 & 0.00 & 1.00 & 0.00 & 1.00 & 0.00 & 0.00 & x & x & x & x & 8.22 & x \\ 
  105 & SERINC3 & 1.00 & 0.00 & 0.00 & 1.00 & 0.00 & 0.00 & 0.00 & 1.00 & 0.00 & 1.00 & 0.00 & 0.00 & x & x & x & x & 0.1 & x \\ 
  106 & SLC24A2 & 1.00 & 0.00 & 0.00 & 1.00 & 0.00 & 0.00 & 1.00 & 0.00 & 0.00 & 0.00 & 0.00 & 1.00 & x & x & x & x & x & 0.67 \\ 
  107 & SLC25A29 & 1.00 & 0.00 & 0.00 & 1.00 & 0.00 & 0.00 & 0.00 & 0.00 & 1.00 & 1.00 & 0.00 & 0.00 & x & x & x & x & 0.03 & x \\ 
  108 & SYT13 & 1.00 & 0.00 & 0.00 & 1.00 & 0.00 & 0.00 & 0.00 & 1.00 & 0.00 & 1.00 & 0.00 & 0.00 & x & x & x & x & 2.34 & x \\ 
  109 & SYT5 & 1.00 & 0.00 & 0.00 & 1.00 & 0.00 & 0.00 & 0.00 & 1.00 & 0.00 & 1.00 & 0.00 & 0.00 & x & x & x & x & 25.12 & x \\ 
  110 & TIMM22 & 1.00 & 0.00 & 0.00 & 1.00 & 0.00 & 0.00 & 0.00 & 1.00 & 0.00 & 1.00 & 0.00 & 0.00 & x & x & x & x & 69.37 & x \\ 
  111 & TLR6 & 1.00 & 0.00 & 0.00 & 1.00 & 0.00 & 0.00 & 0.00 & 0.00 & 1.00 & 1.00 & 0.00 & 0.00 & x & x & x & x & 4.82 & x \\ 
  112 & TMEM125 & 1.00 & 0.00 & 0.00 & 1.00 & 0.00 & 0.00 & 0.00 & 0.00 & 1.00 & 1.00 & 0.00 & 0.00 & x & x & x & x & 0.17 & x \\ 
  113 & TMEM181 & 1.00 & 0.00 & 0.00 & 1.00 & 0.00 & 0.00 & 0.00 & 0.00 & 1.00 & 1.00 & 0.00 & 0.00 & x & x & x & x & 0.17 & x \\ 
  114 & TMEM246 & 1.00 & 0.00 & 0.00 & 1.00 & 0.00 & 0.00 & 0.00 & 1.00 & 0.00 & 1.00 & 0.00 & 0.00 & x & x & x & x & 0.54 & x \\ 
  115 & TMTC1 & 1.00 & 0.00 & 0.00 & 1.00 & 0.00 & 0.00 & 1.00 & 0.00 & 0.00 & 0.00 & 1.00 & 0.00 & x & x & x & x & x & 1.33 \\ 
  116 & TPI1 & 1.00 & 0.00 & 0.00 & 1.00 & 0.00 & 0.00 & 0.00 & 0.00 & 1.00 & 1.00 & 0.00 & 0.00 & x & x & x & x & 15.83 & x \\ 
  117 & TRPS1 & 1.00 & 0.00 & 0.00 & 1.00 & 0.00 & 0.00 & 0.00 & 0.00 & 1.00 & 1.00 & 0.00 & 0.00 & x & x & x & x & 3.56 & x \\ 
  118 & TUFM & 1.00 & 0.00 & 0.00 & 1.00 & 0.00 & 0.00 & 0.00 & 0.00 & 1.00 & 1.00 & 0.00 & 0.00 & x & x & x & x & 25.96 & x \\ 
  119 & UQCR10 & 1.00 & 0.00 & 0.00 & 1.00 & 0.00 & 0.00 & 0.00 & 1.00 & 0.00 & 1.00 & 0.00 & 0.00 & x & x & x & x & 2.07 & x \\ 
  120 & UQCRC1 & 1.00 & 0.00 & 0.00 & 1.00 & 0.00 & 0.00 & 0.00 & 0.00 & 1.00 & 1.00 & 0.00 & 0.00 & x & x & x & x & 0.39 & x \\ 
  121 & WDR12 & 1.00 & 0.00 & 0.00 & 1.00 & 0.00 & 0.00 & 0.00 & 0.00 & 1.00 & 1.00 & 0.00 & 0.00 & x & x & x & x & 1.77 & x \\ 
  122 & XYLT1 & 1.00 & 0.00 & 0.00 & 1.00 & 0.00 & 0.00 & 1.00 & 0.00 & 0.00 & 0.00 & 1.00 & 0.00 & x & x & x & x & x & 1.36 \\ 
  123 & ZDHHC20 & 1.00 & 0.00 & 0.00 & 1.00 & 0.00 & 0.00 & 0.00 & 0.00 & 1.00 & 1.00 & 0.00 & 0.00 & x & x & x & x & 0.22 & x \\ 
   \hline
   \caption{Genes identified as non-null in either the TG RNA only model or GWAS only model with proportion of time spent in each group and effect sizes as in Manuscript Table 1. The top 94 genes in the GWAS only local model are reported (a cutoff of 0.9999) as opposed to all genes in the median model which is used for the other three models.}
\label{tab:table_appendix_GWASRNAonly}
\end{longtable}
\end{spacing}}
\end{landscape}

\end{document}